\documentclass[12pt,a4paper]{article}
\usepackage[utf8]{inputenc}
\usepackage[english]{babel}
\usepackage{amsmath}
\usepackage{amsfonts}
\usepackage{amssymb}
\usepackage{graphicx}
\usepackage{cancel}
\usepackage{soul}
\usepackage[usenames,dvipsnames]{color}
\usepackage{slashed, epsf, latexsym}
\usepackage{epsfig}
\usepackage{graphics}
\usepackage{jheppub}

\begin{document}

	\preprint{TIFR/TH/17-15} 
	\title{The large $D$ black hole dynamics in AdS/dS backgrounds}
	\author[a]{Sayantani Bhattacharyya} 
	\author[a]{, Parthajit Biswas}
	\author[b,c]{, Bidisha Chakrabarty}
	\author[d]{, Yogesh Dandekar}
	\author[a,e]{, Anirban Dinda}
	\affiliation[a]{Indian Institute of Technology Kanpur, Kanpur 208016, India.}
	\affiliation[b]{Institute of Physics, Sachivalaya Marg, Bhubaneswar 751005, Odisha, India}
	\affiliation[c]{Homi Bhabha National Institute, Training School Complex,
	Anushakti Nagar, Mumbai 400085, India}
	\affiliation[d]{Tata Institute of Fundamental Research, Mumbai 400005, India}
	\affiliation[e]{S N Bose National Centre for Basic Sciences, JD Block, Sector-III, Salt Lake City, Kolkata 700106, India}
	\emailAdd{sayanta@niser.ac.in}
	\emailAdd{parthajit.biswas@niser.ac.in}
	\emailAdd{bidisha@iopb.res.in}
	\emailAdd{yogesh@theory.tifr.res.in}
	\emailAdd{dindaanirban@gmail.com}
	
	\abstract{We have constructed a class of  perturbative dynamical black hole solutions in presence of cosmological constant. We have done our calculation in large number of dimensions. The inverse power of dimension has been used as the perturbation parameter and our calculation is valid upto the first subleading order.  The solutions are in one to one correspondence with a dynamical membrane and a velocity field   embedded in the asymptotic geometry.
Our method is manifestly covariant with respect to the asymptotic geometry. One single calculation and the  same universal result works for both dS and AdS geometry or in case of AdS for both  global AdS and Poincare patch. We have checked our final answer with various known exact solutions and the known spectrum of Quasi Normal modes in AdS/dS. }

\maketitle


\section{Introduction}\label{sec:intro}
Recently it has been shown (\cite{EmparanCoupling}) that at infinite dimension limit,  a finite number of quasinormal modes around any black hole solution are effectively confined in the near horizon region and are decoupled from the rest of the infinite tower of QNMs that extend upto asymptotic infinity. The spectrum also develops an infinite gap. In such cases it is natural  that the decouple modes will have a closed dynamics among themselves even at nonlinear levels. The non-linear dynamics of these decoupled QNMs can be constructed in a power series expansion around infinite dimension (i.e., a series in $\left(\frac{1}{D}\right)$). This gives an algorithm to construct a new class of approximate solutions to Einstein equations with a dynamical event horizon. It turns out that there is a one-to-one correspondence between these nonlinear gravity solutions and a codimension one membrane, embedded in the asymptotic flat space, and whose dynamics is governed by a very particular equation, also determined in a power series in $\left(\frac{1}{D}\right)$ \cite{membrane, Chmembrane,yogesh1}, see also \cite{Emparan:2015hwa}.  

See papers \cite{Emparan:2013moa,Emparan:2013xia,Emparan:2013oza,Emparan:2014cia,Emparan:2014jca,Giribet:2013wia,Prester:2013gxa} for initial development of the subject of gravity at large $D$ and inverse dimensional expansion.

There have been many papers recently, that use the technique of $\frac{1}{D}$ expansion, see \cite{Suzuki:2015iha,Emparan:2015gva,Tanabe:2015hda,EmparanHydro,Suzuki:2015axa,Tanabe:2016pjr,Tanabe:2016opw,Sadhu:2016ynd,Herzog:2016hob,Rozali:2016yhw,Chen:2015fuf,Chen:2016fuy,Chen:2017hwm,Chen:2017wpf}.  

In all the above cases, since the dynamics is confined in the near horizon region, it does not care much about the asymptotic geometry. This, in effect, implies that this whole `large $D$' technique of solving gravity equations could be easily extended to situation where the asymptotic geometry is not exactly flat. We expect that the membrane-gravity correspondence will still hold for such cases, but now the membrane will be embedded in whatever non-flat asymptotic geometry (which we shall refer to here as `background') we are interested in. In particular, similar construction should be possible in presence of cosmological constant, \cite{EmparanHydro}. \\
 In \cite{Chmembrane} and \cite{yogesh1}, though the analysis is strictly applicable to asymptotically flat space, the final answer has been presented in this `background-covariant' form. In \cite{yogesh1}, such covariance has been implemented even in the complicated intermediate steps for the derivation of the final metric upto the second  subleading order.

In this paper, we have extended the formalism of \cite{membrane, Chmembrane}, in a way so that this covariant dependence on the `background' is manifest in every step. In addition we have also included cosmological constant, which could have any sign.

One of the key motivation to include the cosmological constant is the following.
We know that in presence of negative cosmological constant, there exists another class of approximate solutions with dynamical event horizons which are dual to conformal fluids, living in a space with dimension one less than that of the  bulk \cite{nonlinfluid}. This duality holds in any dimension and therefore in large $D$ as well. We would eventually like to see how the large $D$ limit of these fluid modes map to the large $D$ decoupled modes in presence of cosmological constant (see \cite{EmparanHydro}). This paper is the first step towards this goal.\\

The organization of this paper is as follows.\\
In section (\ref{sec:setup}) we have described the initial set-up of the problem, the main  equation that we would like to solve for and the scheme of our perturbation technique. In section (\ref{sec:dscale}) we described how in our scheme, different quantities scale with the dimension $D$, the perturbation parameter. In section (\ref{sec:LeadAnsatz}) we have described how we could guess the leading ansatz. Next in a small section (\ref{sec:backcov}) we described how our approach becomes manifestly covariant with respect to the embedding geometry of the membrane. In section ({\ref{sec:strategy}) we briefly explained the algorithm we used to solve for the first subleading correction. In section (\ref{sec:detail1st}) and section (\ref{sec:finalresult}) we have derived and presented the first subleading correction to the metric and the equation governing the dual membrane and the velocity field.
Then in section (\ref{sec:checks}) we have performed several checks on our ansatz. We have matched our solution with Schwarzschild AdS/dS black hole/brane and then with rotating black hole solution upto the required order in an expansion in $\left(\frac{1}{D}\right)$. We have linearized our membrane equation and reproduced the known spectrum of quasi-normal modes . In an appropriate scaling limit of our equation (and after some linear field redefinition) we reproduced the effective hydrodynamic equation of \cite{EmparanHydro}. Finally in section (\ref{sec:conclude}) we concluded.\\
We have several appendices with the details of all computation.
%

\section{Set-up}\label{sec:setup}
In this section we shall describe the basic set-up of our problem and the final goal  in terms of equations. We shall also present the schematic form of the solution that we will eventually determine. 


We are working with the following two derivative action of gravity.
\begin{equation}\label{eq:action}
\begin{split}
{\cal S} = \int\sqrt{-G} \left[R - \Lambda\right]
\end{split}
\end{equation}
where for $\Lambda$ we have assumed the following scaling with dimension $D$.~\footnote{See section (\ref{sec:strategy}) for the motivation of this choice}.
\begin{equation}\label{eq:lamdaScale}
\Lambda =\left[ (D-1)(D-2)\right]\lambda,~~~~~\lambda\sim{\cal O}(1)
\end{equation}
Varying \eqref{eq:action} with respect to the metric, we get the equation of motion.
\begin{equation}\label{eq:eom1}
E_{AB}\equiv R_{AB} -\left(\frac{R-\Lambda}{2} \right)G_{AB}=0
\end{equation}
As mentioned before, our goal is to find new `black-hole type' solutions (i.e. metric with an event horizon) of equation \eqref{eq:eom1} in a power series expansion around $D\rightarrow\infty$. Schematically the solution will have the form
\begin{equation}\label{eq:expansion}
G_{AB} = g_{AB}+\sum_{k=0}^\infty \left(\frac{1}{D}\right)^kG^{(k)}_{AB} 
\end{equation}
\footnote{For our analysis in the later sections we shall often use a notation $G_{AB}^{[k]}$ to denote the metric corrected upto order ${\cal O}\left({1\over D^k}\right)$
$$G^{[k]}_{AB} = g_{AB} + \sum_{m=0}^{m=k} \frac{1}{D^m}G_{AB}^{(m)}$$}
Here $g_{AB}$ is a smooth metric that also solves the same equation eq.\eqref{eq:eom1}. It is the metric that we have referred to as the `background' in the previous section. On the other hand, the $G^{(k)}_{AB}$'s are not smooth and their forms are such that the full metric $G_{AB}$ would necessarily possess an event horizon, and might have singularities behind it. $G^{(k)}_{AB}$'s together will capture the nonlinear dynamics of the `decoupled quasi-normal modes'. Since the decoupled modes are confined within a thin region around the horizon, the $G^{(k)}_{AB}$'s should vanish fast as we go away from the horizon, implying that the $g_{AB}$ is the asymptotic form of the metric.\\
 As explained in \cite{membrane, Chmembrane, yogesh1}, our final solution for $G^{(k)}_{AB}$'s will be parametrized by a codimension-one time-like membrane, embedded in the background space-time with metric  $g_{AB}$, with a velocity field along it. However the curvature of this membrane and the velocity field are not completely independent data. We can solve for $G^{(k)}_{AB}$'s consistently provided the extrinsic curvature of the  membrane as embedded in $g_{AB}$ and the velocity field on it together satisfy some integrability condition which follows from the constraint equations of gravity. We would be viewing this integrability condition as a dynamical equation on the coupled system of the membrane and the velocity. This leads to a `membrane-gravity' duality in the sense that for every solution of this membrane equation we shall be able to find a metric solution for equation \eqref{eq:eom1} in an expansion in $\left(\frac{1}{D}\right)$.
 
 Following \cite{Chmembrane,yogesh1} we shall determine $G^{(k)}_{AB}$'s in a way so that if we view the same membrane as a hypersurface embedded in the full space-time with metric $G_{AB}$,  it becomes the event horizon and the velocity field on it reduces to its null generators.

 \section{Scaling with $D$}\label{sec:dscale}
 Roughly speaking, gravity equation \eqref{eq:eom1} in $D$ dimensions is a collection of $\frac{D(D+1)}{2}$ equations that are used to determine the $\frac{D(D+1)}{2}$ components of the metric (modulo the coordinate redefinition freedom). So a naive large $D$ limit, simply means that both the number of equations and the number of variables we would like to solve for, are blowing up along with our perturbation parameter. 
 
 To get rid of this complication, we shall implicitly assume that a large part of the geometry is fixed by  some symmetry and the metric is dynamical only along some finite number of directions. In other words, the metric  will be of the form \cite{membrane}
 \begin{equation}\label{eq:jadarkar}
 \begin{split}
 dS^2=G_{AB}~dX^A dX^B = \tilde G_{ab}(\{x^a\})~dx^a dx^b + f(\{x^a\}) d\Omega^2
 \end{split}
 \end{equation}
 where $\tilde G_{ab}(\{x^a\}),~\{a,b\}=\{0,1,\cdots,p\}$ is a dynamical finite ($p+1$) dimensional metric and $d\Omega^2$ is the line element on the infinite $(D-p-1)$ dimensional symmetric space. $f(\{x^a\})$ is an arbitrary function of $\{x^a\}$ which is not constant.\\
 Since the metric is dual to a membrane embedded in the background $g_{AB}$ with a velocity field along it, the symmetry of the metric must  be there in the membrane, the velocity field  and the background as well. This will imply that the dual membrane is  dynamical only along the $x^a$ directions and simply wrap the symmetric space (with metric $\Omega_{AB}$). Similarly the dual velocity field must have components only along the $x^a$ directions and also the nonzero components should not depend on the coordinates along the $\Omega$-space. As a consequence, the same feature (i.e., no components along the symmetry directions and all nonzero components are functions of $x^a$'s only) would be true of any vector constructed out of the membrane data. Similarly for tensor structures the components, where all indices are along the isometry directions, must be proportional to the metric in the symmetric space,  $\Omega_{ab}$.
 
%

 In such cases we could easily see that for any generic vector or one form, the order of its divergence will always be $D$ times larger than the order of the form itself \cite{membrane,Chmembrane,yogesh1}. \\
In fact such a rule will apply to  tensors with any number of indices. If $T_{A_1 A_2\cdots A_n}$ is a generic tensor of order ${\cal O}\left(\frac{1}{D}\right)^k$ maintaining the symmetry of equation \eqref{eq:jadarkar},  then its divergence is of order $ {\cal O}\left(\frac{1}{D}\right)^{k-1}$.
\begin{equation}\label{eq:dscale}
T_{A_1 A_2\cdots A_n}\sim {\cal O}\left(\frac{1}{D}\right)^k
\Rightarrow g^{A_p A_q}\nabla_{A_p} T_{A_1,A_2,\cdots A_q\cdots}\sim {\cal O}\left(\frac{1}{D}\right)^{k-1}
\end{equation}
In the same way we could see that if  the background metric $g_{AB}$  admits a decomposition of the form \eqref{eq:jadarkar}, then the Riemann tensor evaluated on $g_{AB}$ will be of order ${\cal O}(1)$ and Ricci tensor and Ricci scalar will be order ${\cal O}(D)$ and ${\cal O}(D^2)$ respectively.
\begin{equation}\label{eq:dscale2}
R_{ABCD}\vert_{\text{on}~g_{AB}}\sim{\cal O}(1),~~~R_{AB}\vert_{\text{on}~g_{AB}}\sim{\cal O}(D),~~~R\vert_{\text{on}~g_{AB}}\sim{\cal O}(D^2)
\end{equation}
It follows that the Einstein tensor evaluated on $g_{AB}$ would be of ${\cal O}(D^2)$\footnote{Such scaling is true for a generic case. It is always possible to have special background where equation \eqref{eq:dscale2} is not true. A different choice for the $D$ dependence of $\Lambda$ would have led to such `non-generic' background.} and since we want $g_{AB}$ to satisfy equation \eqref{eq:eom1}, it justifies our choice of $D$ scaling for the cosmological constant as given in equation \eqref{eq:lamdaScale}.

However as explained in \cite{yogesh1}, for our calculation  we do not need any details of the decomposition as given in equation \eqref{eq:jadarkar} or the symmetry itself. The only aspect of it that will be used is the scaling law \eqref{eq:dscale} and the $D$ dependence of the curvatures, evaluated on $g_{AB}$. \\

\section{Leading Ansatz }\label{sec:LeadAnsatz}
In our calculation $G^{(0)}_{AB}$ is the starting ansatz that captures the nonlinear dynamics of the decoupled modes at the very leading order. Any perturbation theory works provided we have a clear guess for what should be the leading answer. In this sense, we can carry on with our program only if we know the appropriate form of $G^{(0)}_{AB}$ that solves the equation \eqref{eq:eom1} at leading order in $\left(\frac{1}{D}\right)$ expansion. So we shall first describe how we could guess the form of this leading ansatz.
\subsection{The form of the leading ansatz}
 As mentioned before,  the two parameters that will characterize our solution are a codimension-one hypersurface, embedded in the background space-time with metric $g_{AB}$, and  a unit normalized velocity field $u_\mu$ defined along the membrane \footnote{Throughout this paper, we shall use Greek indices to denote indices along the world volume of the membrane as embedded in $g_{AB}$, whereas capital Latin indices will denote full space-time indices.\\
 The velocity field $u_\mu$ is unit normalized with respect to the induced metric along the membrane (denoted as $h_{\mu\nu}$) .
 $$u_\mu u_\nu h^{\mu\nu}=-1$$} .
 
 We shall first construct a smooth function  $\psi$ of the background coordinates  $\{X^A\}$ such that $(\psi=1)$ gives the equation for the membrane.  Next we shall construct a smooth one-form field $(O= O_A ~dX^A)$, defined everywhere in the background space-time, such that the  projection of $(-O^A)$ along the membrane reduces to the velocity vector field $u^\mu$.
We shall determine our final solution in terms of this bulk function $\psi$ and bulk one-form field $O$. Note, at this stage there is a huge ambiguity in the construction of $\psi$ and $O$. The condition, they have to reduce to something specific on $\psi=1$, is certainly not enough to fix them completely. We shall fix this ambiguity with certain convenient choice (see subsection (\ref{subsec:subsidiary}) for a more detailed discussion on this point).\\
At this stage the simplest structure that we could imagine for $G^{(0)}_{AB}$ (that does not involve any derivative of the two basic fields $\psi$ and $O$) is the following
\begin{equation}\label{eq:guess1}
G^{(0)}_{AB}= F~ O_A O_B\Rightarrow G_{AB} = g_{AB} + F~ O_A O_B + {\cal O}\left(\frac{1}{D}\right)
\end{equation}
where $F$ is any arbitrary scalar function of $\psi$ and $(O\cdot O)~~$\footnote{Throughout this paper `$\cdot$' denotes contraction with respect to the metric $g_{AB}$}. \\
From equation \eqref{eq:guess1} it follows 
\begin{equation}\label{eq:inv0}
G^{AB} = g^{AB}  -\left(\frac{F}{1+F (O\cdot O)}\right) O^A O^B + {\cal O}\left(\frac{1}{D}\right)
\end{equation}
Here all raising/lowering and contractions are with respect to the background metric $g_{AB}$.\\
Now firstly we want $\psi=1$ surface to be the horizon when embedded in the metric $G_{AB}$. This implies that $(\partial_A\psi)(\partial_B\psi) G^{AB}=0$ on $\psi =1$. Now we shall impose this condition order by order in $\left(\frac{1}{D}\right)$ expansion. At leading order it implies
\begin{equation}\label{eq:cond0}
\begin{split}
&\bigg[d\psi\cdot d\psi - \left(\frac{F}{1+F (O\cdot O)}\right)(O\cdot d\psi)^2\bigg]_{\psi=1} ={\cal O}\left(\frac{1}{D}\right)\\
&\left[\frac{F}{1+F (O\cdot O)}\right]_{\psi=1}= \left[\frac{1}{O\cdot n}\right]^2_{\psi=1}+{\cal O}\left(\frac{1}{D}\right)\\
&\\
&\text{where} ~~n_A =\frac{\partial_A\psi}{\sqrt{d\psi\cdot d\psi}}
\end{split}
\end{equation}

Secondly we want the velocity vector field to be the null generator of the horizon, which is given by 
$$t^A= G^{AB}n_A\vert_{\psi=1}$$
Also by definition, the projection of $(-O^A)$ along the membrane will give the velocity field. This in turn implies
\begin{equation}\label{eq:cond1}
\begin{split}
&\left[\Pi^A_B O^B+G^{AB} n_B\right]_{\psi=1} = 0,~~~\text{where}~~ \Pi^A_B =\text{projector} = \delta^A_B -n^An_B\\
\Rightarrow~&\left[O^A - (O\cdot n) n^A+n^A-O^A \left(\frac{F}{1+F (O\cdot O)}\right)(O\cdot n)\right]_{\psi=1} = {\cal O}\left(\frac{1}{D}\right)\\
\Rightarrow~&\left[\left(1-\frac{1}{O\cdot n}\right)\left(O^A - (O\cdot n) n^A\right)\right]_{\psi=1} ={\cal O}\left(\frac{1}{D}\right)\\
\end{split}
\end{equation}
In equation \eqref{eq:cond1} to go from second to the third line we have used equation \eqref{eq:cond0}. From equation \eqref{eq:cond1} it follows that 
\begin{equation}\label{eq:cond3}
(O\cdot n)\vert_{\psi=1} = 1+{\cal O}\left(\frac{1}{D}\right)
\end{equation}
On the other hand using the fact that velocity vector field on the membrane (viewed as a hypersurface embedded in the background $g_{AB}$) is normalized to minus one, we see
\begin{equation}\label{eq:cond4}
\Pi^{AB} O_A O_B = -1
\end{equation}
From equations \eqref{eq:cond3} and \eqref{eq:cond4} it follows that at leading order in $\left(\frac{1}{D}\right)$ expansion,  $O$ is a null one-form  with respect to the background metric $g_{AB}$.
\begin{equation}\label{eq:cond5}
g^{AB} O_A O_B ={\cal O}\left(\frac{1}{D}\right)
\end{equation}
 We shall often express $O_A$ as
\begin{equation}\label{eq:reO}
\begin{split}
&O_A = n_A - u_A\\
\text{where}~~&u_A \equiv -\Pi_A^B O_B,~~\Pi^A_B = \text{projector}= \delta^A_B - n^A n_B
\end{split}
\end{equation}
By construction $u_A$ is always along the membrane and it will be the velocity field $u_\mu$ , if expressed in terms of the intrinsic coordinate of the membrane.
 So far from our analysis we could see that the simplest form of $G_{AB}^{(0)} $ is the following.
 $$G_{AB}^{(0)} = F~ O_A O_B= F~ (n_A -u_A)(n_B -u_B)$$
 Since $(O\cdot O)$ is zero at leading order, $F$ could only be a function of $\psi$, if we do not want any derivative at zeroth order. We also want $F$ to be vanishing outside a thin region of thickness of order ${\cal O}\left(\frac{1}{D}\right)$ around $\psi=1$. This would be ensured provided $F(\psi)\propto \psi^{-D}$.\\
  Now in equation \eqref{eq:cond0} if we substitute the fact $O$ is null at leading order, we find 
 $$F\vert_{\psi=1} = 1 + {\cal O}\left(\frac{1}{D}\right)$$
 This fixes the proportionality constant in $F$ to be one.\\
 So finally we are lead to the following expression for our leading ansatz\footnote{We would like to emphasize that what we have presented here is {\it not} a derivation for the leading ansatz. In the end this is a `guess' and our perturbation program is developed around this starting point. This guess could also be motivated from the fact that our final solution, in a very small patch of size of the order of ${\cal O}\left(\frac{1}{D}\right)$, looks like a $D$ dimensional Schwarzschild black hole with a local radius and boost velocity. See \cite{Chmembrane},\cite{yogesh1} for a more detailed discussion on this}
 \begin{equation}\label{exg0}
 G^{(0)}_{AB} = \psi^{-D} O_A O_B
 \end{equation}

  It turns out that this form will solve equation \eqref{eq:eom1} at leading order provided the following conditions are satisfied (see \cite{membrane,Chmembrane} for a more detailed discussion)
\begin{equation}\label{eq:g0ab}
\begin{split}
&G^{(0)}_{AB} = \psi^{-D} O_A O_B~~~~~\text{such that}\\
&K\equiv\text{Trace of  extrinsic curvature of the membrane}= {\cal O}(D)\\
&\sqrt{g^{AB}(\partial_A\psi)(\partial_B\psi )}\vert_{\psi=1}=\frac{K}{D} + {\cal O}\left(\frac{1}{D}\right)\\
&g^{AB}\nabla_A O_B=K+ {\cal O}(1)
\end{split}
\end{equation}
As before, the membrane $\psi=1$ is viewed as a hypersurface embedded in the background space-time with metric $g_{AB}$ and $\nabla_A$ denotes covariant derivative with respect to $g_{AB}$.

 \subsection{When ansatz solves the leading equation}\label{subsubsec:leading}
  Now we shall demonstrate that given the $D$ scaling law  \eqref{eq:dscale} and \eqref{eq:dscale2}, how $G^{(0)}_{AB}$ as given in equation \eqref{eq:g0ab} indeed satisfies the equation \eqref{eq:eom1} at leading order.\\
We shall simply evaluate the gravity equations on the metric $G^{(0)}_{AB}$ plus the background $g_{AB}$ and we shall see that  leading order (which will turn out to be of order ${\cal O}(D^2)$) piece vanishes provided the conditions as mentioned in equation \eqref{eq:g0ab} are satisfied.
 
 Before getting into any details, we shall first simplify the equation \eqref{eq:eom1} a bit, by subtracting the trace part of the equation.
 \begin{equation}\label{eq:eom2}
 \begin{split}
 &R_{AB}- \left(\frac{R}{2} \right)G_{AB} = -\left[\frac{(D-2)(D-1)\lambda}{2}\right] G_{AB}\\
 \Rightarrow~&R = D(D-1)\lambda\\
  \Rightarrow~&{\cal E}_{AB}\equiv R_{AB}- (D-1)\lambda~G_{AB}=0\\
 \end{split}
 \end{equation}
Now we shall evaluate $R_{AB}$ on the metric $G_{AB}^{[0]} = g_{AB} + G^{(0)}_{AB}$.
 Details of the calculation are presented in appendix (\ref{app:calsource}). Here we are simply quoting the final result.
\begin{equation}\label{eq:decomp}
 \begin{split}
 R_{AB}\vert_{G_{AB}^{[0]}} = ~&\psi^{-D}\left(\frac{DN}{2}\right)\bigg\{[DN-(\nabla\cdot O)]\left(n_A O_B + n_B O_A\right)+( K -DN)O_A O_B \bigg\} \\
 &+\left(\frac{\psi^{-2D}}{2}\right)\bigg\{DN\left[DN-(\nabla\cdot O)\right]O_AO_B
 \bigg\}\\
&+\tilde R_{AB}   +{\cal O}\left(D\right)
 \end{split}
 \end{equation} 
 where
 \begin{itemize}
 \item  $\tilde R_{AB}$  is the Ricci tensor evaluated on the background $g_{AB}$
  \item $\nabla_A$ denotes the covariant derivative with respect to background metric $g_{AB}$
 \item $K$ is the extrinsic curvature of the membrane as embedded in the background:~~~
 $K\equiv \nabla_A ~n^A $
 \item $N$ is the norm of the one form $d\psi$:~~~
 $N \equiv \sqrt{(\partial_A\psi)(\partial_B\psi) g^{AB}}$
 \end{itemize}

%
%
%

Now from equation \eqref{eq:dscale2} it follows that $\tilde R_{AB}\sim{\cal O}(D)$. Therefore the leading equation reduces to
\begin{equation}\label{eq:leadeq}
\begin{split}
&\left[\psi^{-D}(K -DN)  + \psi^{-2D}(DN-\nabla\cdot O)\right]O_AO_B\\
& +\psi^{-D}(DN-\nabla\cdot O)(n_A O_B +n_B O_A)= {\cal O}(1)
\end{split}
\end{equation}
Since $O_A$ and $n_A$ are two independent vector directions in the background spacetime, equation \eqref{eq:leadeq} finally implies
\begin{equation}\label{eq:leadeq2}
\begin{split}
&(\nabla\cdot O - DN)_{\psi=1} = {\cal O}(1)\\
&(K - DN)_{\psi=1} = {\cal O}(1)
\end{split}
\end{equation}
Equation \eqref{eq:leadeq} is simply the conditions mentioned in equation \eqref{eq:g0ab}. Note also that at leading order the RHS of equation \eqref{eq:eom1}, which captures the effect of cosmological constant, does not contribute. \\
Also note that the two equations in \eqref{eq:leadeq} together imply that
\begin{equation}\label{sclrcon}
\left(\nabla\cdot u\right)_{\psi=1}={\cal O}(1)
\end{equation}
Here $u$ is defined in equation \eqref{eq:reO}\footnote{$\left(\nabla\cdot u\right)$ should have been of ${\cal O}(D)$ if we naively use the rules for counting the order in $\left(\frac{1}{D}\right)$ expansion as explained in section(\ref{sec:dscale})}

\section{Covariance w.r.t.  `background' metric}\label{sec:backcov}
In this section  we shall discuss how we could recast all subsequent calculation in a manifestly covariant form  with respect to the background $g_{AB}$. \\
In fact this feature has already appeared in the previous section (see equation \eqref{eq:decomp}).
As we know, the computation of $R_{AB}$, to begin with, involves partial derivatives of the metric components $G_{AB}^{[0]}$. However,  the expressions appearing in \eqref{eq:decomp} have only covariant derivatives with respect to the background $g_{AB}$.
In \cite{yogesh1} this point has been argued  from a physical point of view and has been used extensively.

Here we shall see how it follows algebraically.
This is a consequence of the fact that though Christoffel symbols are not covariant tensors, their differences are and therefore the Christoffel symbols on the full metric $G_{AB}$ could always be written as the Christoffel symbols evaluated on the background $g_{AB}$ plus a correction which will have a  form of a covariant tensor with respect to the background. Then this feature could easily be extended to the construction of the Riemann tensor and the Ricci tensor for the full background.

The general form of our metric
$$G_{AB} = g_{AB} + \chi_{AB}$$
Let us denote the Christoffel symbols corresponding to $G_{AB}$ as $\Gamma^A_{BC}$, whereas $\hat\Gamma^A_{BC}$ denotes the Christoffel symbols corresponding to the background $g_{AB}$.
\begin{equation}\label{eq:gamma}
\begin{split}
\Gamma^A_{BC}=&~ \frac{1}{2}G^{AC'} \bigg(\partial_{C}~ G_{C'B} + \partial_{B} ~G_{C'C} - \partial_{C'}~ G_{BC}\bigg)\\
=&~ \hat\Gamma^A_{BC}+ \frac{1}{2}G^{AC'} \bigg(\nabla_{C}~ \chi_{C'B} + \nabla_{B} ~\chi_{C'C} - \nabla_{C'}~ \chi_{BC}\bigg) \\
\end{split}
\end{equation}
where $\nabla_A$ denotes the covariant derivative with respect to $g_{AB}$. In deriving equation \eqref{eq:gamma} we have used the fact that Christoffel symbols are symmetric in its lower two indices.\\
Now in our convention, Ricci Tensor, $R_{AB}$, of the full metric is given by the following expression.
$$R_{AB} = \partial_k \Gamma^k_{AB} - \partial_B\Gamma^k_{Ak} + \Gamma^k_{km} \Gamma^m_{AB} - \Gamma^k_{Bm} \Gamma^m_{Ak}$$
Using equation \eqref{eq:gamma} we could easily rewrite it in a covariant form.
\begin{equation}\label{eq:covricci}
\begin{split}
R_{AB} &= \bar R_{AB} + \nabla_k\left[\delta\Gamma^k_{AB}\right] - \nabla_B\left[\delta\Gamma^k_{Ak}\right] + \left[\delta\Gamma^k_{km}\right]\left[\delta\Gamma^m_{AB}\right] - \left[\delta\Gamma^k_{Bm}\right]\left[\delta\Gamma^m_{Ak}\right]
\end{split}
\end{equation}
where $\bar R_{AB}$ is the Ricci Tensor evaluated on the background and $\left[\delta\Gamma^A_{BC}\right]$ is the covariant tensor appearing in the second term of equation \eqref{eq:gamma}
\begin{equation}\label{eq:gamma2}
\left[\delta\Gamma^A_{BC}\right] = \frac{1}{2}G^{AC'} \bigg(\nabla_{C}~ \chi_{C'B} + \nabla_{B} ~\chi_{C'C} - \nabla_{C'}~ \chi_{BC}\bigg)
\end{equation}
Equations\eqref{eq:gamma} and  \eqref{eq:covricci} are the key equations that we shall use to determine  the subleading corrections to the metric in a manifestly covariant fashion.
\section{General strategy for the first subleading correction} \label{sec:strategy}
 Once the leading ansatz $G^{(0)}_{AB}$, the function $\psi$ and the one-form $O$ are well-defined everywhere in the  background with metric $g_{AB}$, we can describe the strategy to determine the subleading corrections to the metric i.e., the $G_{AB}^{(k)}$ s for $k>0$. In this paper our goal is to determine $G^{(1)}_{AB}$.  Our method is essentially  same as the one described in  \cite{yogesh1}. 
 The purpose of this section is to mainly set up the notation and convention. We shall omit any detailed justification or `all order proof', for the statements. Interested reader should refer to \cite{yogesh1} for a thorough discussion. 
 
\subsection{Summary of the algorithm}
 We already know that if we evaluate Ricci tensor on $G^{[0]}_{AB}= g_{AB} + G^{(0)}_{AB}$, the leading piece is of order ${\cal O}(D^2)$. This leading piece vanishes provided  $O_A$ and $\psi$ satisfy equations \eqref{eq:leadeq2}. Clearly after imposing equation \eqref{eq:leadeq2},  the leading non-vanishing piece in $R_{AB}$  would be of order ${\cal O}(D)$. To cancel this piece upto corrections of order ${\cal O}(1)$ we add the new terms in the metric - $\left(\frac{1}{D}\right) G^{(1)}_{AB}$ . Therefore to begin with $\left(\frac{1}{D}\right) G^{(1)}_{AB}$ will have  the most general  form that could contribute to  the equation of motion \eqref{eq:eom2} at order ${\cal O}(D)$. Also  any term in equation of motion that involves product of two components of $G^{(1)}_{AB}$ (i.e., non-linear in $G^{(1)}_{AB}$) will contribute at most at order ${\cal O}(1)$. Since in this paper we are interested only at order ${\cal O}(D)$, we have to treat $G^{(1)}_{AB}$ simply as a linear perturbation on $G^{[0]}_{AB}$.
Then at order ${\cal O}\left(D\right)$, the equation of motion \eqref{eq:eom2}  will have two pieces. One piece will take the form of a linear differential operator acting on different (and so far unknown) components $G^{(1)}_{AB}$ and the second piece will involve the ${\cal O}(D)$ piece coming from $G^{[0]}_{AB}$.  The first piece will have an universal structure at all orders and we shall call it as `homogeneous piece' or $H_{AB}$. The second part will be termed as `source' ($S_{AB}$) .
Schematically
$${\cal E}_{AB}\sim H_{AB} +S_{AB}$$
 Our solution procedure will essentially be an `inversion' of the universal differential operator  in $H_{AB}$.

 We shall determine $G^{(1)}_{AB}$  completely in terms of the function $\psi$ and the one-form $O$, that are directly related to the basic data of our construction - the membrane and the velocity field. One advantage of our formalism is that  we never need to choose any specific coordinate system on the membrane or for the background $g_{AB}$.

 \subsection{Subsidiary condition}\label{subsec:subsidiary}
 Note that so far all the conditions on $\psi$ and $O$ are imposed only along the membrane. We want $\psi$ to be one on the membrane hypersurface and the projection of $O$ onto the membrane to reduce to the velocity field $u_\mu$. The gravity equation \eqref{eq:eom1} at leading order (see equation \eqref{eq:leadeq2}) imposes some more constraints on $\psi$ and $O$, but still they needed to be satisfied only at $(\psi=1)$. Therefore there is a large ambiguity in the construction of the function $\psi$ and the one-form $O$. In this subsection we shall fix this ambiguity with a certain convenient choice, which, following \cite{membrane},\cite{Chmembrane},\cite{yogesh1}, we shall refer to as `subsidiary conditions'.\footnote{The subsidiary conditions we have chosen in this paper are different from what has been used in \cite{membrane},\cite{Chmembrane} or \cite{yogesh1}.  We found this choice most convenient because the  metric correction at the first subleading order takes the simplest form. As we shall see, with this subsidiary condition, it simply vanishes and  the first non-trivial correction appears only at the second subleading order.}.\\
%

 Subsidiary condition on $\psi$ is chosen as follows.
 \begin{equation}\label{eq:SubsidiaryPsi}
 \begin{split}
 \nabla^2\psi^{-D} =0~~~~\text{everywhere}\\
 \end{split}
 \end{equation}
It could be shown that equation \eqref{eq:SubsidiaryPsi} is  enough to determine $\psi$ in an expansion in $\left(\frac{1}{D}\right)$ around the membrane $(\psi=1)$\cite{radiation}. Also we could easily see that \eqref{eq:SubsidiaryPsi} is consistent with the second equation \eqref{eq:leadeq2}(See appendix (\ref{app:Identity})).\\
Now we shall describe how we fixed the ambiguity in the definition of $O_A$. Unlike $\psi$, since $O_A$ is a vector in the background with $D$ components, we need $D$ equations to fix it completely. From the construction of $G^{(0)}_{AB}$ we know that on the membrane, $O^A$ is a null vector and $O\cdot n=1$, where $n_A$ in the unit normal to the membrane. Firstly note that once we have imposed equation \eqref{eq:SubsidiaryPsi}, $(\psi=\text{constant})$ surfaces and therefore the unit normal to them are well-defined everywhere. Therefore we could easily lift these two conditions on $O$, which are initially imposed only on the membrane,  to everywhere in the background. In terms of equation what we mean is the following 
 \begin{equation}\label{eq:SubsidiaryO1}
 \begin{split}
O\cdot O =0~~\text{and}~~O\cdot n = 1~~~\text{everywhere}
 \end{split}
 \end{equation}
Equation \eqref{eq:SubsidiaryO1} gives two scalar conditions on $O$. 
 We still need $(D-2)$ equations through which we would be able to  determine the remaining $(D-2)$ components $O_A$, everywhere in the background . To fix them  we use the following differential equation.
 \begin{equation}\label{eq:SubsidiaryO2}
 \begin{split}
&P_A^B(O\cdot\nabla)O^A=0~~~\text{everywhere}\\
\text{where}~& P_A^B \equiv \delta_A^B - n_A O^B -O_An ^B + O_A O^B,
 \end{split}
 \end{equation}
 Since $P^A_B$ is the projector to the subspace orthogonal to both $n$ and $O$, equation \eqref{eq:SubsidiaryO2} is effectively a collection of $(D-2)$ equations as required\footnote{ Because of equation \eqref{eq:SubsidiaryO1} $O_A(O\cdot\nabla)O^A$  and $n_A(O\cdot\nabla)O^A$ are already  determined.
 $$O_A(O\cdot\nabla)O^A=0,~~n_A(O\cdot\nabla)O^A=-O_A(O\cdot\nabla)n^A$$}.
Equations \eqref{eq:SubsidiaryO1} and \eqref{eq:SubsidiaryO2} together fix the ambiguities in all components $O$, everywhere in the background.
%
 
 It is possible to rewrite the subsidiary condition on $O$ in a more geometric form. Equation \eqref{eq:SubsidiaryO1} and \eqref{eq:SubsidiaryO2} it follows that 
  \begin{equation}\label{eq:SubsidiaryO3}
 \begin{split}
&(O\cdot\nabla)O^A= \left[n_B(O\cdot\nabla)O^B\right]O^A~~~\text{everywhere}
 \end{split}
 \end{equation}
Equation \eqref{eq:SubsidiaryO3} simply implies that throughout the background geometry, $O^A$s are the tangent vectors to the null geodesics passing through the membrane.
 
  In course of analysis we shall often define a  $u_A$ field everywhere in the background\footnote{Equation \eqref{eq:reOeve} apparently looks very similar to equation \eqref{eq:reO}. However the main difference is that equation \eqref{eq:reOeve} is true for any constant $\psi$ slices whereas equation \eqref{eq:reO} was specifically applied to the membrane i.e., $(\psi=1)$.}.
\begin{equation}\label{eq:reOeve}
u_A \equiv - \Pi_A^B O_B~~\text{where}~~\Pi^A_B \equiv \text{Projector on constant $\psi$ slices} = \delta^A_B -n^A n_B
\end{equation}
Note that as a consequence of equation \eqref{eq:SubsidiaryO1}, $u_A$ turns out to be a unit normalized time-like vector, which is orthogonal to $n_A$ by construction.
$$g^{AB} u_A u_B =-1,~~g^{AB} u_A n_B =0$$
From equation \eqref{eq:SubsidiaryO1} it follows that $O\cdot n =O\cdot u =1$  or $O_A = n_A -u_A$ everywhere. Also the projector $P_{AB}$ of equation \eqref{eq:SubsidiaryO2} is actually a projector orthogonal to both $n_A$ and $u_A$ and therefore could equivalently be expressed as
 $$P_{AB} = g_{AB} -n_A n_B + u_A u_B$$

 \subsection{Choice of gauge}
 We shall choose a gauge such that 
\begin{equation}\label{eq:gaugechoice} 
 O^A G_{AB}^{(1)} =0
 \end{equation}
 Note that our leading ansatz also satisfies this same gauge.

After imposing equation \eqref{eq:gaugechoice} the most general structure for $G_{AB}^{(1)}$ is the following
 \begin{equation}\label{eq:ggenstu}
 \begin{split}
 G_{AB}^{(1)}=&~{\cal S}_1O_A O_B +\left(\frac{1}{D}\right){\cal S}_2P_{AB}+ \left[O_A {\cal V}_B +O_B {\cal V}_A\right] +  {\cal T}_{AB}\\
 \text{where}~~&\\
 &u^A {\cal V}_A = n^A{\cal V}_A =0;~~~u^A {\cal T}_{AB}= n^A{\cal T}_{AB} =0;~~g^{AB}{\cal T}_{AB}^{(1)}=0 
 \end{split}
 \end{equation}
Here the unknown scalar, vector and the tensors, $\left[{\cal S}_i,~~i=\{1,2\}\right]$, ${\cal V}_A$, ${\cal T}_{AB}$ are all of order ${\cal O}(1)$ and have explicit dependence on $\psi$ as well as the derivatives of $\psi$ and $O$.\\
Note the extra factor of $\left(\frac{1}{D}\right)$ in the term proportional to $P_{AB}$. This is because, by definition,  $G^{(1)}_{AB}$ is the collection of  those terms in the metric that contribute to the gravity equation at order ${\cal O}(D)$. As we shall see below, the term proportional to $P_{AB}$ will contribute one extra factor of $D$ in some terms of the gravity equation  (the ones that involve a trace of the metric tensor). In other words, unless we suppress this term by an extra factor of $\left(\frac{1}{D}\right)$, it will contribute and mess-up the matching and solving of the equations at order ${\cal O}(D^2)$.

\subsection{The form of explicit $\psi$ dependence}
We know that within the region where the metric correction is nontrivial, $(\psi-1)$ is of order ${\cal O}\left(\frac{1}{D}\right)$. Therefore we would define a new order ${\cal O}(1)$ variable $R\equiv D(\psi-1)$ to parametrize the explicit $\psi$ dependence of the unknown scalar, vector and the tensor functions in equation \eqref{eq:ggenstu}. In terms of equation we mean the following.
\begin{equation}\label{eq:psidepend}
\begin{split}
&{\cal S}_1 = \sum_n f_n(R)~ {\mathfrak s}_n,~~~
{\cal S}_2 = \sum_n h_n(R)~ {\mathfrak s}_n\\
&{\cal V}_A = \sum_n v_n(R) ~\left[{\mathfrak v}_n\right]_A~~~{\cal T}_{AB} = \sum_n t_n(R) ~\left[{\mathfrak t}_n\right]_{AB}\\
&R \equiv D(\psi-1)
\end{split}
\end{equation}
Here $f_n(R),~v_n(R),~t_n(R)$ and $h_n(R)$ are functions that do not involve any explicit factors of $D$. The other expressions, ${\mathfrak s}_n$, $\left[{\mathfrak v}_n\right]_A$, $\left[{\mathfrak t}_n\right]_{AB}$ are the different scalar, vector and the tensor structures of order ${\cal O}(1)$, involving the derivatives of $n_A$ and $O_A$  that could appear at  order ${\cal O}(1)$. The upper limit for the sum over $n$ will generically be different in scalar, vector and tensor sector.   These structures, by construction will not have any explicit dependence on $\psi$, since all such explicit dependence at this order will be captured by  the function $f_n,~v_n,~t_n$ and $h_n$.
However these structures will depend on $\psi$ implicitly through the derivatives of $n_A$ and $O_A$. But note that this will be a `slow' dependence in $\left(\frac{1}{D}\right)$ expansion. More precisely if we compute of the variations of ${\mathfrak s}_n$, ${\mathfrak v}_n$ or ${\mathfrak t}_n$ in the direction of $\partial_A\psi$ it will always be of ${\cal O}(1)$, whereas the variations of $  f_n(R)$, $h_n(R)$, $ v_n(R)$ and $ t_n(R)$, will be of order ${\cal O}(D)$.
This is the reason, we could treat these structures, ${\mathfrak s}_n$, $[{\mathfrak v}_n]_A$ and $[{\mathfrak t}_n]_{AB}$ effectively as constants when we are doing the leading order computation with $G^{(1)}_{AB}$. See the next subsection for details.

\subsection{Structure of `Homogeneous piece'}
In this subsection we shall list  the detailed form of the homogeneous piece. As mentioned before, the homogeneous piece could be computed by simply linearizing the gravity equations \eqref{eq:eom2} around  as $G^{[0]}_{AB}$, where the gauge-fixed form of the linear perturbation is given by $G^{(1)}_{AB}$. (See appendix \ref{app:homo} for the details of the computation)\\
 For convenience, we shall decompose the homogeneous piece into three parts.
\begin{equation}\label{eq:homogeneous}
\begin{split}
&H_{AB} = H_{AB}^{scalar} +H_{AB}^{vector} +H_{AB}^{tensor} + H_{AB}^{trace} \\
\end{split}
\end{equation}
where
\begin{equation}\label{eq:homoscalar}
\begin{split}
H_{AB}^{scalar}
&=\left(\frac{D N^2}{2}\right)\sum_n{\mathfrak s}_n\left( f_n^{\prime\prime}+  f^{\prime}_n\right)\bigg[n_B O_A + n_A O_B -\left(1-\psi^{-D}\right)O_BO_A\bigg]
\end{split}
\end{equation}
\begin{equation}\label{eq:homovector}
\begin{split}
H_{AB}^{vector}
=&~
\left(\frac{N}{2}\right)\sum_n \left(\nabla\cdot{\mathfrak v}_n\right)\bigg[ v^{\prime}_n\left(n_AO_B + n_B O_A\right) -\psi^{-D}v_nO_BO_A\bigg]\\
&~+\left(\frac{D N^2}{2}\right)\sum_n\left( v^{\prime\prime}_n+  v^{\prime}_n\right)\bigg\{\bigg(u_B\left[{\mathfrak v}_n\right]_A+u_A\left[{\mathfrak v}_n\right]_B\bigg)\\
&~~~~~~~~~~~~~~~~~~~~~~~~~~~~~~~~~~~~+\psi^{-D}\bigg(O_B\left[{\mathfrak v}_n\right]_A+O_A\left[{\mathfrak v}_n\right]_B\bigg)\bigg\}\\
\end{split}
\end{equation}
\begin{equation}\label{eq:homotensor}
\begin{split}
H_{AB}^{tensor}=&~- \left(\frac{D N^2}{2}\right)\sum_n\left[ t^{\prime\prime}_n(1-\psi^{-D}) +  t^{\prime}_n\right]\left[{\mathfrak t}_n\right]_{AB}\\
&+\left(\frac{N}{2}\right)\sum_n  t^{\prime}_n\bigg( n_B~(\nabla_C\left[{\mathfrak t}_n\right]^C_A)+A\leftrightarrow B\bigg)\\
\end{split}
\end{equation}\\
\begin{equation}\label{eq:homoTrace}
\begin{split}
H^{trace}_{AB} &= 
-\left(\frac{D N^2}{4}\right)\sum_n{\mathfrak s}_n\bigg\{2  h^{\prime\prime}_n ~n_A n_B + h^{\prime}_n \left[\psi^{-D}(n_A n_B -u_Au_B)+\psi^{-2D}O_BO_A\right]\bigg\}\\
\end{split}
\end{equation}
Here $ X^{\prime}$ for any function $X(R)$ denotes $\frac{dX}{dR}$.\\

From the explicit expressions of $H_{AB}$ it follows that 
\begin{equation}\label{eq:trvanish}
\left(\frac{ 1}{D}\right)\Pi^{AB}H_{AB}= {\cal O}(1)
\end{equation}
where $\Pi_{AB}$ is the projector perpendicular to $(\psi=1)$ hypersurface as embedded in the background.

It turns out that we could easily decouple these homogeneous parts of the ${\cal E}_{AB}$ by taking the following linear combination of the components.\\
\begin{equation}\label{eq:homot}
\begin{split}
~&P^A_C H_{AB} P^{B}_{C'}-\frac{ P_{CC'}}{D}\left( P^{AB}H_{AB}\right)\\
=&-\left(\frac{DN^2}{2}\right)\sum[\mathfrak{t}_n]_{CC'}\left[t^{\prime\prime}_n\left(1-\psi^{-D}\right)+t^{\prime}_n\right]
\end{split}
\end{equation}

%

\begin{equation}\label{eq:homov}
\begin{split}
u^A H_{AB} P^B_C=-\left(\frac{DN^2}{2}\right) \sum_n(1-\psi^{-D})(v^{\prime}_n+v^{\prime\prime}_n)[{\mathfrak v}_n]_C
\end{split}
\end{equation}

\begin{equation}\label{eq:homof}
\begin{split}
u^A H_{AB} u^B&=-\left(\frac{DN^2 }{2}\right)(1-\psi^{-D})\sum_n{\mathfrak s}_n\left[ f^{\prime\prime}_n + f^{\prime}_n -\left(\frac{\psi^{-D}}{2}\right) h^{\prime}_n\right]\\
&-\left(\frac{N}{2}\right)\psi^{-D}\sum_{n} v_{n} (\nabla\cdot {\mathfrak{v}}_n)
\end{split}
\end{equation}

\begin{equation}\label{eq:homoh}
\begin{split}
O^A O^B{H}_{AB} =  -\frac{DN^2}{2}\sum_n h^{\prime\prime}_n{\mathfrak s}_n 
\end{split}
\end{equation}

Note that given equation \eqref{eq:trvanish},  equations \eqref{eq:homot}, \eqref{eq:homov} and \eqref{eq:homof} are simply the different components of $\bigg(\Pi_A^{A'}\Pi^{B'}_B H_{A'B'}\bigg)$ at leading non-trivial order in $(1/D)$ expansion.

\subsection{Structure of `Source'}\label{subsec:source}
In general the source $S_{AB}$ will depend on all the coordinates, through some explicit dependence on $\psi$ and also through different derivatives of $O_A$ and $n_A$. As before, we can classify the $\psi$ dependence of $S_{AB}$ as `slow' and `fast'. The `fast'  pieces are those whose derivatives in the directions of increasing will have a factor of $D$, (i.e., the dependence on $\psi$ is through $R \equiv D(\psi-1)$). These are the parts which have been treated exactly at a given order. All other variations of the source terms, both along and away from the membrane hypersurface, are `slow' (i.e., the derivatives are suppressed by a factor of $\left(\frac{1}{D}\right)$ compared to the `fast' dependence) and therefore could effectively be treated as constants while solving for the next correction to the metric i.e, $G^{(1)}_{AB}$.
This is why we simply invert the homogeneous piece $H_{AB}$ assuming it to be an ordinary differential operator in the `fast' variable $R$. See \cite{membrane} and \cite{Chmembrane} for a more detailed explanation.\\

As we have seen in the previous subsection, the projected components of the homogeneous piece ($\Pi^{A'}_A \Pi^{B'}_B H_{A'B'}$) could be viewed as ordinary second order differential operator in the `fast' variable $R$, acting on the unknown functions appearing in the metric correction. It follows that to determine the unknown functions $f(R),~ v(R)$ and $t(R)$, it is enough to solve the projected components the gravity equations \eqref{eq:eom2}  or
$$\Pi^{A'}_A \Pi^{B'}_B {\cal E}_{A'B'}=0$$
The traceless piece of the projected ${\cal E}_{AB}$ leads to the following set of second order inhomogeneous differential equations for three sets of the unknown functions, $f_n(R),~v_n(R)$ and $t_n(R)$.
\begin{equation}\label{eq:fulleq}
\begin{split}
&\sum_n\frac{d}{dR}\left[\left(e^{R}-1\right)t^{\prime}_n\right]~[\mathfrak{t}_n]_{AB}
=\left(\frac{2~e^R}{DN^2}\right)\left[P^{C}_A P^{C'}_B-P_{AB}\left(\frac{ P^{CC'}}{D}\right) \right]S_{CC'} \\
\\
&(1-e^{-R})\sum_n~\frac{d}{dR}\left[e^{R}v^{\prime}_n\right]~[{\mathfrak v}_n]_A=\left(\frac{2~e^R}{DN^2}\right)\left[u^B S_{BC} P^C_A\right]\\
\\
&(1-e^{-R})\sum_n\frac{d}{dR}\left[e^{R} f^{\prime}_n -\frac{ h_n}{2}\right]~{\mathfrak s}_n
=\left(\frac{2~e^R}{DN^2}\right)\left(u^A S_{AB} u^B\right)
-\sum_{n} v_{n} \left(\frac{\nabla\cdot {\mathfrak{v}}_n}{DN}\right)
\end{split}
\end{equation}
In equation \eqref{eq:fulleq} we have also used the fact that $\left[\psi^{-D}= e^{-R} + {\cal O}\left(\frac{1}{D}\right)\right]$.\\
The equation for $h(R)$  is given by the ${\cal E}_{AB}$ with both indices projected in the direction of $O$.
\begin{equation}\label{eq:decouph}
\begin{split}
O^A O^B{\cal E}_{AB} =0\Rightarrow  \sum_n h^{\prime\prime}_n{\mathfrak s}_n =\left(\frac{2}{DN^2}\right)\left[ O^A~ S_{AB}~ O^B\right]
\end{split}
\end{equation}
%
%
%
%
%

Note that the last two equations in \eqref{eq:fulleq} will admit regular solutions at $\psi=1$ only if 
\begin{equation}\label{eq:reg}
\begin{split}
&\left[u^B S_{BC} P^C_A\right]_{R=0} =0\\
&\bigg[\left(\frac{2}{DN^2}\right)\left(u^A S_{AB} u^B\right)-\sum_{n} v_{n} \left(\frac{\nabla\cdot {\mathfrak{v}}_n}{DN}\right)\bigg]_{R=0}=0
\end{split}
\end{equation}
We shall see that both of these conditions will be true as a consequence of our membrane equation. In fact in \cite{membrane} this is the regularity condition that has been used to determine the membrane equation.

\subsection{Boundary condition}\label{subsec:boundary}
Since our differential operator (in $R$) is second order, we need two sets of boundary conditions to fix the integration constants. \\
One of these is the `normalizability' .
In our construction it must be true that the metric is non-trivial only in a thin region of thickness ${\cal O}\left(\frac{1}{D}\right)$ around the membrane $\psi=1$. This defines the normalizability conditions on the metric functions $f_n(R), ~v_n(R),~t_n(R)$ and $h_n(R)$; in $R$ coordinates they must vanish exponentially as $R\rightarrow\infty$ (recall $R= D~(\psi-1)$), so that outside the `membrane region' the metric is that of the background. This `normalizability' fixes one integration constant in each of the three differential equations in  \eqref{eq:fulleq}. It turns out that for equation \eqref{eq:decouph} both the zero modes are non-normalizable or in other words in this case the `normalizability' condition is enough to fix $h_n(R)$.

The other integration constant is fixed by the condition on the horizon.\\
For $f_n$ and $v_n$ it is fixed by our definition of the horizon itself.
We want  $\psi=1$ to be the exact equation for the horizon of this geometry and $u^A$ to be the null generator of the horizon. This implies that the following `all order'  equation on the horizon
\begin{equation}\label{eq:bc1}
u^AG_{AB} \vert_{\psi=1}= n_B
\end{equation}
Note that by construction at any order the metric will take the form
$$G_{AB} = g_{AB} + f~ O_A O_B  + (V_A~ O_B + V_B~O_A) + h~ P_{AB} + t_{AB}$$
where $O_A = n_A - u_A,~~V\cdot O = V\cdot n=0,~~O^A t_{AB} = n^A t_{AB} =0,~P^{AB} t_{AB}=0$
Contracting this metric with $u^A$ we find
$$u^A G_{AB} = u_B + f~O_B + V_B$$
Now \eqref{eq:bc1}  fixes the values of $f$ and $V_A$ on $\psi=1$ or equivalently $R=0$.
\begin{equation}\label{eq:bc2}
\begin{split}
&f~\vert_{\psi=1} =1~\Rightarrow ~f_n(R=0) =0,\\
&V_A~\vert_{\psi=1} = 0~\Rightarrow~v_n(R=0) =0
\end{split}
\end{equation}
For the tensor sector i.e., the function $t_n(R)$, the other integration constant could be fixed by demanding the solution is regular at the horizon.\\

\subsection{Solution in the form of integral}
Once the boundary conditions are fixed, we can explicitly invert the differential operators and could write the solutions for $f_n(R),~v_n(R),~t_n(R),$ and $h_n(R)$ in terms of some definite integrals of the source. In this subsection we shall present these formulas explicitly.\\
As mentioned before in subsection (\ref{subsec:source}) we could always rewrite source $S_{AB}$ at any given order as some functions of `fast' variable $R$ multiplied by the `slowly' varying scalar, vector or tensor structures relevant for that order. In other words the RHS of the three equations in \eqref{eq:fulleq} could be expressed as 
\begin{equation}\label{eq:resource}
\begin{split}
\text{RHS of 1st eqn} &=\left(\frac{2e^{R}}{N^2}\right)\sum_n \left[{\mathfrak t}_n\right]_{AB}S^\text{tensor}_{n}(R)\\
\text{RHS of  2nd eqn} &=\left(\frac{2e^{R}}{N^2}\right)\sum_n \left[{\mathfrak v}_n\right]_{A}S^\text{vector}_{n}(R)\\
\text{RHS of  3rd eqn} &=\left(\frac{2e^{R}}{N^2}\right)\sum_n \left[{\mathfrak s}_n\right]S^\text{scalar}_{n}(R)
-\left(\frac{1}{DN}\right)\sum_{n} v_{n}(R) \left(\nabla\cdot {\mathfrak v}_n\right)
\end{split}
\end{equation}
Similarly RHS of \eqref{eq:decouph} could be written as
\begin{equation}\label{eq:resource2}
\begin{split}
RHS &=\left(\frac{2}{N^2}\right)\sum_n S^\text{trace}_{n}(R)~{\mathfrak s}_n
\end{split}
\end{equation}
Now we can explicitly write the solution for $ G_{AB}^{(1)}$ in terms of definite integral of the source.
\begin{equation}\label{eq:ggensturep}
 \begin{split}
 G_{AB}^{(1)}=&~{\cal S}_1 O_A O_B +\left(\frac{1}{D}\right){\cal S}_2P_{AB}+ \left[O_A {\cal V}_B +O_B {\cal V}_A\right] +  {\cal T}_{AB}\\
 \text{where}~~&\\
 &u^A {\cal V}_A = n^A{\cal V}_A =0;~~~u^A {\cal T}_{AB} = n^A{\cal T}_{AB} =0;~~g^{AB}{\cal T}_{AB}=0 
 \end{split}
 \end{equation}
 where
 \begin{equation}\label{eq:explicit}
 \begin{split}
{\cal T}_{AB} =&-\left(\frac{2}{N^2}\right)\sum_n \left[{\mathfrak t}_n\right]_{AB}\int_R^\infty \left(\frac{dy}{e^{y}-1}\right)\bigg( \int_0^y dx\left[e^x ~S^\text{tensor}_{n}(x)\right]\bigg)\\
{\cal V}_{A}  =&-\left(\frac{2}{N^2}\right)\sum_n \left[{\mathfrak v}_n\right]_{A} \int_R^\infty dy ~e^{-y}\bigg[\int_0^y dx\left(\frac{e^{2x} }{e^x -1}\right)S^\text{vector}_{n}(x)\bigg]+e^{-R}{\cal K}^\text{vector}_A \\
{\cal S}_2 =& \left(\frac{2}{N^2}\right)\sum_n{\mathfrak s}_n\int_R^\infty dy\bigg[\int_y^\infty dx~S^\text{trace}_{n}(x)\bigg]\\
{\cal S}_1= &-\left(\frac{2}{N^2}\right)\sum_n{\mathfrak s}_n \int_R^\infty dy ~e^{-y}\bigg[\int_0^y dx\left(\frac{e^{2x} }{e^x -1}\right)S^\text{scalar}_{n}(x)\bigg]\\
&+\left(\frac{1}{2}\right)\int_R^\infty dz ~e^{-z}\bigg[-{\cal S}_2+ 2\int_0^z\left(\frac{dx}{1-e^{-x}}\right)\left(\frac{\nabla\cdot {\cal V}}{DN}\right)\bigg]+e^{-R}{\cal K}^\text{scalar} \\
 \end{split}
 \end{equation}
Here ${\cal K}_s$ and ${\cal K}_v$ are two constants added so that $S_1\vert_{R=0} = {\cal V}_{A}\vert_{R=0} =0$
\begin{equation}\label{eq:kskv}
\begin{split}
{\cal K}^\text{scalar} &= \left(\frac{2}{N^2}\right)\sum_n{\mathfrak s}_n \int_0^\infty dy ~e^{-y}\bigg[\int_0^y dx\left(\frac{e^{2x} }{e^x -1}\right)S^\text{scalar}_{n}(x)\bigg]\\
&-\left(\frac{1}{2}\right)\int_0^\infty dz ~e^{-z}\bigg[-{\cal S}_2+ 2\int_0^z\left(\frac{dx}{1-e^{-x}}\right)\left(\frac{\nabla\cdot {\cal V}}{DN}\right)\bigg]\\
{\cal K}^\text{vector}_A &= \left(\frac{2}{N^2}\right)\sum_n \left[{\mathfrak v}_n\right]_{A} \int_0^\infty dy ~e^{-y}\bigg[\int_0^y dx\left(\frac{e^{2x} }{e^x -1}\right)S^\text{vector}_{n}(x)\bigg]
\end{split}
\end{equation}

\subsection{Constraint and membrane equation}
 Consider the following combinations of different components of $H_{AB}$. 
\begin{enumerate}
\item $(n^B- \psi^{-D} O^B) ~H_{BC}~P^C_A = \frac{N}{2} (1-e^{-R}) \sum_{n} \nabla^B  ({\mathfrak t}_n)_{BA}~ t^{\prime}_n$
\item $(n^B- \psi^{-D} O^B)~ H_{BC}~u^C=\left(\frac{DN}{2}\right)\sum_n\left(\frac{\nabla\cdot{\mathfrak v}_n}{D}\right) \left[ v^{\prime}_n(1-e^{-R})-v_ne^{-R}\right]$
\end{enumerate}
Note that the above combinations have at most one $R$ derivative of the unknown functions. Clearly the same feature would be true if we take the above combinations on the components of ${\cal E}_{AB}$, since the source $S_{AB}$ does not involve any of the unknown functions. Hence these combinations could be viewed as equations that restrict the `initial conditions' (defined on any constant $R$ slice) for the second order differential equations (see \eqref{eq:fulleq}) controlling the `$R$-evolution' of the unknown functions. It follows that the `constraint' equations in our case has the following form
\begin{equation}\label{eq:combine2}
\begin{split}
&{\cal C}\equiv (n^B- \psi^{-D} O^B)~{\cal E}_{BC}~u^C\\
&C_A \equiv (n^B- \psi^{-D} O^B) ~{\cal E}_{BC}~P^C_A\\
\end{split}
\end{equation}
In terms of source $S_{AB}$ and the unknown metric functions,  the above two constraints will take the following structure \footnote{We know that given the foliation of the space-time with $\psi =const$ hypersurfaces,  the equations of gravity could be decomposed into dynamical and constraint equations \cite{WaldBook}.  The constraint equations are the ones where one of the indices of the Einstein equation is projected along the normal to the foliating hypersurfaces. In \cite{yogesh1}, this theory has been used and explained in detail the context  of our large $D$ expansion. Along with the two combinations we mentioned in equations \eqref{eq:combine2} one more constraint equation appears in \cite{yogesh1}, whose abstract form is the following
$${\cal A} \equiv (1-\psi^{-D})~ O^A O^B ~{\cal E}_{AB} - P^{AB}~ {\cal E}_{AB}$$
However,  we shall not analyze this combination here since it will not be required to obtain the final gravity solution and the membrane equations
}
	 \begin{equation}\label{eq:combine3}
	\begin{split}
	&{\cal C}= (n^B- e^{-R} O^B)~S_{BC}~u^C +\left(\frac{DN}{2}\right)\sum_n\left(\frac{\nabla\cdot{\mathfrak v}_n}{D}\right) \left[ v^{\prime}_n(1-e^{-R})-v_ne^{-R}\right]\\
	&C_A = (n^B- e^{-R}O^B) ~S_{BC}~P^C_A + \frac{N}{2} (1-e^{-R}) \sum_{n} \nabla^B  ({\mathfrak t}_n)_{BA}~ t^{\prime}_n \\
	\\
	\end{split}
	\end{equation}
	Now it is known  that  if the constraint is satisfied along one slice and the dynamical equations are satisfied everywhere, then the constraint is automatically satisfied along all hypersurfaces\cite{WaldBook}. In \cite{yogesh1}, this theorem has been explicitly verified for the constraint equations listed above in equations \eqref{eq:combine3}.
Because  of this theorem, we are allowed to impose the constraints \eqref{eq:combine3} only on $\psi=1$ hypersurface and do not worry about how these equations are solved away from the membrane. So at order ${\cal O}(D)$,  the final form of the membrane equations
\begin{equation}\label{eq:memfinal}
\begin{split}
&{\cal C}\vert_{R=0}= u^B~S_{BC}~u^C\vert_{R=0} \\
&C_A\vert_{R=0} = u^B ~S_{BC}~P^C_A\vert_{R=0}\\
\end{split}
\end{equation}
In deriving equation \eqref{eq:memfinal} we have used the fact that $O^A = n^A -u^A$ and $v_n(R=0)=0$ because of our boundary condition. We also used the fact that ${\cal T}_{AB}^{(1)}$ is regular at $R=0$ due to choice of integration limits (see equation \eqref{eq:explicit}) and thus the term involving unknown tensor metric correction in $C_A$ vanishes at $R=0$.\\
Equations \eqref{eq:memfinal} are the genuine membrane equations that do not involve any of the unknown functions and therefore only constrain our membrane data. Also note that these are the combinations that appear in the RHS of the first two equations in \eqref{eq:fulleq} and the regularity of the solutions also demand the vanishing of these constraints on $R=0$.

The fact that given a solution to these constraint  equations along the membrane, we can always solve the other dynamical equations (i.e. the other components of the ${\cal E}_{AB}$), by inverting the linear differential operator appearing in $H_{AB}$, establishes the `membrane-gravity duality' that we have mentioned in the introduction.

\section{The first subleading correction: $G^{(1)}_{AB}$}\label{sec:detail1st}
In this section we shall describe how we calculate the first subleading correction to the metric along with the coupled  equations of motion for the membrane and the velocity field along it. As described in the previous section,  at this order the source $S_{AB}$ will be simply be determined by evaluating the Ricci Tensor $R_{AB}$ on the metric $G^{[0]}_{AB} = g_{AB} + G^{(0)}_{AB} =  g_{AB} + \psi^{-D} O_A O_B$. The details of the computation of the Ricci Tensor are presented in  the appendix (\ref{app:calsource}). Here we simply take the appropriate combinations of the different components equation of motion.
For convenience we quote the final answer for the source at first subleading order.
\begin{equation}\label{eq:S^1ABmntxt}
\begin{split}
S_{AB}&=e^{-R}\left(\frac{K}{2}\right)\Bigg[e^{-R} \left(\left(\tilde{\nabla}\cdot u\right)_{R=0}-\frac{R}{K}\left(\tilde{\nabla}\cdot E\right)_{R=0}\right) O_BO_A\\
&~~~~~~~+(n_A O_B+n_B O_A)\left(\left(\tilde{\nabla}\cdot u\right)_{R=0}-\frac{R}{K}\left(\tilde{\nabla}\cdot E\right)\right)_{R=0}\\
&+(O_BP^C_A+O_AP^C_B)\left(\frac{\tilde{\nabla}^2 u_C}{K}-\frac{\nabla_C K}{K}+u^D K_{DC}-(u\cdot\nabla)u_C\right)_{R=0}\Bigg]\\
\end{split}
\end{equation}
Where, $\tilde{\nabla}$ is defined as follows, for any general  tensor with $n$ indices $W_{A_1A_2\cdots A_n}$
\begin{equation}\label{tildedef}
\tilde{\nabla}_A W_{A_1A_2\cdots A_n}=\Pi_A^C~\Pi_{A_1}^{C_1}\Pi_{A_2}^{C_2}\cdots \Pi_{A_n}^{C_n}\left(\nabla_C W_{C_1C_2\cdots C_n}\right)
\end{equation}\\

\subsection{Constraint equation}
In the previous section we have  described how we could determine the constraint equations on the membrane by taking appropriate combination of the components of the source terms evaluated at $\psi=1$. In this subsection, we shall first  evaluate those combinations on $S_{AB}$ and determine the constraints on the membrane data at the first subleading order. Note that at leading order there was only one scalar constraint on the membrane data
$$\nabla\cdot u\sim {\cal O}(1)$$
It turns out that at first subleading order we shall have one scalar and one vector equation. This matches with the number of free data we have on the membrane: the shape of the membrane (scalar function) and the unit normalized velocity field on it (the vector function).
\subsubsection{Constraint in the vector sector}
First we shall describe the constraint equation in the direction perpendicular to $u_A$. We shall refer to this as `Vector constraint'. 
\begin{equation}\label{eq:vecconst1}
\begin{split}
&u^B S_{BC}P^C_A={\cal O}(1)\\
\Rightarrow~~&~{K\over 2}P^C_A\left[\frac{\tilde{\nabla}^2 u_C}{K}-\frac{\nabla_C K}{K}+u^D K_{DC}-(u\cdot\nabla)u_C\right]={\cal O}(1)\\
\Rightarrow~~&P^C_A\left[\frac{\tilde{\nabla}^2 u_C}{K}-\frac{\nabla_C K}{K}+u^D K_{DC}-(u\cdot\nabla)u_C\right]={\cal O}\left(1\over D\right)
\end{split}
\end{equation}
Note that  in equation \eqref{eq:vecconst1}, all derivatives and the all the indices (both contracted and free) are projected along the hypersurface $(\psi=1)$. Now it is easy to rewrite the constraint equation as an equation intrinsic to the membrane.
\begin{equation}\label{eq:vecconst1final}
\begin{split}
{\cal{P}}^\nu_\mu\left[ \frac{\hat\nabla^2 u_\nu}{\cal K} -\frac{\hat\nabla_\nu{\cal K}}{\cal K}+  u^\alpha {\cal K}_{\alpha\nu} - (u\cdot\hat\nabla)u_\nu \right] = {\cal O}\left(1\over D\right)
\end{split}
\end{equation}
Here $\hat\nabla$ denotes the covariant derivative with respect to the intrinsic metric of the membrane. ${\cal K}_{\mu\nu}$ is the extrinsic curvature of the membrane, viewed as a tensor intrinsic to the hypersurface and ${\cal K} $ is the trace of ${\cal K}_{\mu\nu}$.

\subsubsection{Constraint in the scalar sector}
Now we shall describe the constraint equation in the scalar sector, i.e.,the constraint in the 
direction of $u_A$.
\begin{equation}\label{eq:scalconst1}
\begin{split}
0=&~u^B S_{BC}~u^C  ={K\over 2}\left[\tilde\nabla\cdot  u\right]\\
\end{split}
\end{equation}
  As before, this equation also could be written purely in terms of the intrinsic data of the membrane.
\begin{equation}
\begin{split}
\left[\tilde\nabla\cdot u\right]_{\psi=1} = \hat\nabla\cdot u
\end{split}
\end{equation}
where $\hat\nabla$ denotes the covariant derivative with respect to the intrinsic metric of the hypersurface $(\psi=1)$ viewed as a membrane embedded in the background.

We finally find

\begin{equation}\label{eq:nablau}
\tilde\nabla\cdot u \sim {\cal O}\left(\frac{1}{D}\right)
\end{equation}
\subsection{Dynamical equation}
In this section we shall give details of the dynamical equations. It turns out that given our subsidiary condition and  after imposing the scalar and vector constraint equations, the sources for all dynamical equation simply vanish leading to the vanishing of $G^{(1)}_{AB}$.

\subsubsection{Tensor sector}
From the first equation of \eqref{eq:fulleq} we get the relevant differential equation for the `tensor-type' correction at the first subleading order.
\begin{equation}\label{eq:tensoreq}
\begin{split}
&D~\sum_n\left[\left(1-e^{-R}\right){t}_n^{\prime\prime}+t_n^{\prime}\right]~[\mathfrak{t}_n]_{AB}
=\left(\frac{2}{N^2}\right)\left[P^{C}_A P^{C'}_B-P_{AB}\left(\frac{ P^{CC'}}{D}\right) \right]S_{CC'} 
\end{split}
\end{equation}
 But from equation \eqref{eq:S^1ABmntxt} we could simply see that
$$P^C_A P^{C'}_B S_{CC'} =0$$
In the language of equation\eqref{eq:resource} it implies that $S^{tensor}_{n}(R)$ vanishes for all $(n)$. Substituting this in the first equation of \eqref{eq:explicit} we find ${\cal T}^{(1)}_{AB}$ is zero.
\subsubsection{Vector sector}
From the second equation of \eqref{eq:fulleq} we get the relevant differential equation for the `vector-type' correction at the first subleading order.
\begin{equation}\label{eq:vectoreq}
\begin{split}
&De^{-R}(1-e^{-R})\sum_n~\frac{d}{dR}\left[e^{R}v_n^{\prime}\right]~[{\mathfrak v}_n]_A=\left(\frac{2}{N^2}\right)\left[u^B S_{BC} P^C_A\right]\\
\end{split}
\end{equation}
Note that the RHS of equation \eqref{eq:vectoreq} implicitly  depends on $\psi$. However the dependence is `slow', in the sense as one goes away from $(\psi=1)$ hypersurface, the variation of the RHS is suppressed by a factor of $\left(\frac{1}{D}\right)$. Thus, at this order, we need to evaluate the RHS only at $(\psi =1)$ hypersurface.\\
Now from equations \eqref{eq:vecconst1} and \eqref{eq:vecconst1final} it follows that 
$$\left[u^B S_{BC} P^C_A\right]_{\psi=1} =0$$
In the language of equation\eqref{eq:resource} it implies that $S^{vector}_{n}(R)$ vanishes for all $(n)$. Substituting this in the second equation of \eqref{eq:explicit} we find ${\cal V}^{(1)}_{A}$ is zero.
\subsubsection{Scalar sector}
In the scalar sector there are two unknown functions $h(R)$ and $f(R)$ and therefore we need two equations. Clearly equation \eqref{eq:decouph} and the last equation of \eqref{eq:fulleq}  are the relevant equations here.
\begin{equation}\label{eq:scalareq}
\begin{split}
&
O^A O^B{\cal E}_{AB} =0\Rightarrow  \sum_n h_n^{\prime\prime}{\mathfrak s}_n =\left(\frac{2}{DN^2}\right)\left[ O^A~ S_{AB}~ O^B\right]\\
&\text{and}\\
&De^{-R}(1-e^{-R})\sum_n\frac{d}{dR}\left[e^{R} f_n^{\prime} -\frac{ h_n}{2}\right]~{\mathfrak s}_n\\
=&~\left(\frac{2}{N^2}\right)\left(u^A S_{AB} u^B\right)
-e^{-R}\sum_{n} v_{n} \left(\frac{\nabla\cdot {\mathfrak{v}}_n}{N}\right)
\end{split}
\end{equation}

Now since $P^{AB}S_{AB}$  vanishes, the boundary conditions (see section (\ref{subsec:boundary}) ensure that $h_n(R)$ is zero for every $n$. Given that $h_n(R)$ is zero and there is no correction in the vector sector (implying $v_n^{(1)}(R)$ is zero for every $n$) the second equation of \eqref{eq:scalareq} reduces to
\begin{equation}\label{eq:scalareqf}
\begin{split}
&De^{-R}(1-e^{-R})\sum_n\frac{d}{dR}\left[e^{R} f_n^{\prime} \right]~{\mathfrak s}_n
=~\left(\frac{2}{N^2}\right)\left(u^A S_{AB} u^B\right)
\end{split}
\end{equation}
Now, following the same logic as we have used in `Vector sector' ,  the RHS of equation \eqref{eq:scalareqf} is simply the scalar constraint equation and therefore vanishes.  Now the boundary conditions ensures that $f_n(R)=0$ for every $n$.

\section{Final metric and membrane equation}\label{sec:finalresult}
In this section we shall simply summarize our final result i.e., the metric and the membrane equation of motion upto the first subleading order. As we have seen in the previous section, given our subsidiary condition, the next to leading correction to the metric vanishes.
\begin{equation}\label{eq:finalMetric}
G_{AB} = g_{AB} + \psi^{-D} O_A O_B + {\cal O}\left(\frac{1}{D}\right)^2
\end{equation}
where the scalar function $\psi$ and $O_A$ are defined everywhere in the background (with metric $g_{AB}$) through the following equations
\begin{equation}\label{eq:subdef}
\begin{split}
&O^C\partial_C\psi =\sqrt{\partial\psi\cdot\partial\psi},~~~~O\cdot O =0\\
&\nabla^2\psi^{-D} =0\\
&(O\cdot\nabla)O_A = \left[\left(\frac{\partial_C\psi }{\sqrt{\partial\psi\cdot\partial\psi}}\right)(O\cdot\nabla) O^C\right]O_A\\
\end{split}
\end{equation}
Clearly the asymptotic form of the full space-time is given by the metric $g_{AB}$, which we have referred to as `background'. $\nabla$ is the covariant derivative with respect to $g_{AB}$.\\
The equations \eqref{eq:subdef} are enough to fix $\psi$ and $O$ everywhere provided the shape of the $(\psi=1)$ hypersurface and the one form field  $O_A$ on $(\psi=1)$ hypersurface are given. We have referred to these two pieces of information as `membrane data' . It turns out that \eqref{eq:finalMetric} is a solution of the gravity equation provided the membrane data satisfy the following equation of motion

\begin{equation}\label{eq:finalMembrane}
\begin{split}
&{\cal P}^\nu_\mu\left\{ \hat\nabla^2 u_\nu -\hat\nabla_\nu {\cal K}+{\cal K}\left[  u_\alpha{\cal K}^\alpha_\nu - (u_\alpha\hat\nabla^\alpha)u_\nu\right] \right\} = {\cal O}\left(1\right)\\
&\hat\nabla_\alpha u^\alpha= {\cal O}\left(\frac{1}{D}\right)
\end{split}
\end{equation}
Equation \eqref{eq:finalMembrane} is an equation intrinsic to the membrane, in the sense that all raising and lowering of indices and the covariant derivatives are defined with respect to the induced metric on the membrane - a hypersurface embedded in the background $g_{AB}$.  All the indices now can take $(D-1)$ values.  ${\cal K}_{\mu\nu}$ is the extrinsic curvature tensor ,  viewed as a tensor structure defined on the membrane only. ${\cal K}$  is the trace of ${\cal K}_{\mu\nu}$. The velocity field $u_\mu$ is the projection of the one form $O_A$ along the hypersurface. And ${\cal P}^\nu_\mu$ is the projector perpendicular to the velocity field $u_\mu$. Like the extrinsic curvature tensor, this projector is also defined only along  the membrane worldvolume.

Equations \eqref{eq:finalMetric}, \eqref{eq:subdef} and \eqref{eq:finalMembrane} together are the final result of this paper.

\section{Checks}\label{sec:checks}
In this section we shall perform several checks on our solution for the metric and the equation of motion for the membrane. 
\subsection{Matching with known exact solution}
We know of few exact  static and stationary black hole / brane solutions of the equation \eqref{eq:eom1} in arbitrary dimension. Now our effective membrane equation \eqref{eq:finalMembrane} and the metric \eqref{eq:finalMetric} are valid as long as the number of dimensions is very large. Clearly static and stationary exact solutions are special cases which must solve our equation and must match with our metric in the appropriate limit. In this subsection we shall show this matching explicitly for three different exact solutions in Asymptotically AdS space.
\subsubsection{Schwarzschild Black-Brane in AdS}
In Kerr-Schild form AdS-black brane is given by
\begin{equation}\label{bbmetric}
\begin{split}
&dS^2
=dS^2_{\text{Poincare}}+r^{-(D-3)}\left(dt+\frac{dr}{r^2}\right)^2\\
\end{split}
\end{equation}
where $dS^2_\text{Poincare}$ is the line element in Poincare patch AdS space.
\begin{equation}\label{eq:AdsPoincare}
\begin{split}
&dS^2_\text{Poincare} =\frac{dr^2}{r^2}-r^2 dt^2+r^2 d\vec{x}^2_{D-2}
\end{split}
\end{equation}
For the black-brane geometry \eqref{bbmetric}, the hypersurface $r=1$ is the horizon and the null generator of the horizon is given by
$$l^A\partial_A = ~~\partial_t$$
It follows that the dual membrane is given by the same surface  $r=1$, however viewed as a hypersurface embedded in the AdS space with metric $dS^2_\text{Poincare}$ and the velocity field along the horizon is simply $u = -dt$.
The induced metric on the membrane
$$ dS^2_{\text{induced}}=-~dt^2~+~d\vec{x}^2_{D-2}$$
We can easily see that this velocity field $u$ is divergence free along the membrane.
It is very easy to compute the  extrinsic curvature tensor for this configuration. The non-zero components of extrinsic curvature and trace of extrinsic curvature are given by
\begin{equation}\label{braneextrinsic}
{\cal{K}}_{ij}=\delta_{ij},~~ {\cal{K}}_{tt}=-1,~~ {\cal{K}}=D-1\\
\text{~~Where \{$i~= 1,...,~D~-~2$\}}
\end{equation}
All the components of the derivatives of the velocity field on the membrane vanishes
\begin{equation}\label{branederiu}
\hat{\nabla}_{\mu} u_{\nu}=0,~~ \{\mu={t,i}\}
\end{equation}
Substituting equations \eqref{braneextrinsic} and \eqref{branederiu} in the membrane equation \eqref{eq:finalMembrane} and using the fact that $P^t_t=P^t_i=0$, we see that it is satisfied upto the required order.

Next we shall match the form of the metric.  For this we need
 to read off $\psi$ and $u_A$ in such a way that 
\begin{enumerate}
\item $\psi=1$ surface is same as the $r=1$ surface. In other words if we consider $\psi$ as a function of $r$, then $\psi(r=1) =1$.
\item $u^A\vert_{r=1} = l^A$
\item Both $\psi$ and the $u^A$ satisfy the subsidiary conditions \eqref{eq:SubsidiaryPsi} and \eqref{eq:SubsidiaryO3}.
\end{enumerate}
The normalized form of $u$ is easy to guess.
\begin{equation}\label{velsolbrane}
u_A ~dx^A = -r ~dt
\end{equation}
Translation symmetry  in $t$ and all $i$ directions guarantees  that $\psi$ must be a function of $r$ alone and it follows that  the subsidiary condition on $u$ is trivially satisfied (since any vector in the space perpendicular to $n\sim dr$ and $u\sim dt$  must vanish because of the symmetry).
Now we shall  solve for $\psi$ in an expansion in $\left(\frac{1}{D}\right)$. 
Let us start by expanding $\psi$ around the horizon $r=1$.
\begin{equation}\label{psiexp}
\begin{split}
\psi(r) &= 1  + \left(a_{10} +\frac{a_{11}}{D}\right)(r-1) + a_{20} (r-1)^2 + {\cal O}\left(\frac{1}{D}\right)^3\\
\end{split}
\end{equation}
Here $a_{10},~a_{11},~a_{20}$ are constants  (to be determined by solving the subsidiary condition \eqref{eq:SubsidiaryPsi}) and we have also used the fact that within the `membrane region'  $(r-1)\sim{\cal O}\left(\frac{1}{D}\right)$.
Substituting \eqref{psiexp} in \eqref{eq:SubsidiaryPsi}) and solving order by order 
 we find
\begin{equation}\label{psisolbrane}
\begin{split}
\psi(r)&=1+\left(1-\frac{1}{D}\right)(r-1)+ {\cal O}\left(\frac{1}{D}\right)^3\\
&= r - \left(\frac{r-1}{D}\right)+ {\cal O}\left(\frac{1}{D}\right)^3\\
\end{split}
\end{equation}
Note that equations \eqref{psisolbrane} and \eqref{velsolbrane} imply that   in the `membrane region'
\begin{equation}\label{psidifbrane}
\begin{split}
\frac{dr }{r} - r~dt &= O_A dx^A\\
\frac{r^{-(D-3)}}{r^2}&=\psi^{-D} + \left(\frac{1}{D}\right)^2\\
\end{split}
\end{equation}
From equations \eqref{psidifbrane} it follows that the metric of  AdS Schwarzschild black-brane is same as the one we determined in equation \eqref{eq:finalMetric} upto correction of order ${\cal O}\left(\frac{1}{D}\right)^2$.

\subsubsection{Schwarzschild Black-Hole in Global  AdS}
In Kerr-Schild form of global AdS black holes are given as
\begin{equation}\label{globalSt}
\begin{split}
dS^2
&=dS_{Global}^2 +\left(\frac{ r^{-(D-3)}}{1+r^2}\right) \left(\sqrt{1+r^2}~dt + \frac{dr}{\sqrt{1+r^2}}\right)^2\\
\end{split}
\end{equation}
where $dS_{Global}^2$ is given by
\begin{equation}\label{eq:dsglobal}
dS_{Global}^2=\frac{dr^2}{1+r^2} - (1+r^2)dt^2 + r^2d\Omega_{D-2}^2 
\end{equation}
Horizon of this black hole space time (\eqref{globalSt}) is located at the zero of the function 
$f(r)= 1 +r^2 - r^{-(D-3)}$,
$$ \text{Horizon is at $r=r_0$}\Rightarrow f(r_0) =0,~~r_0\neq1$$
The null generator of the horizon is given by $$l^A\partial_A = ~\frac{1}{\sqrt{1+r_0^2}}\partial_t$$

It follows that our membrane is given by the hypersurface $r=r_0$ embedded in the AdS space with metric as given by $dS^2_\text{Global}$ and the velocity field along the horizon is simply $u = -\sqrt{1+r_0^2} ~dt$.  The induced metric on the membrane
$$ds^2_{\text{induced}}=- (1+r_0^2)dt^2 + r_0^2d\Omega_{D-2}^2$$
We can easily see that this velocity field $u$ is divergence free along the membrane.
It is very easy to compute the  extrinsic curvature tensor for this configuration. The non-zero component of the extrinsic curvature and the trace of extrinsic curvature are given by
\begin{equation}\label{holeextrinsic}
\begin{split}
&{\cal{K}}_{tt}=-~\sqrt{2},~~ {\cal{K}}_{ab}=\sqrt{2}~\Omega_{ab},~~{\cal{K}}=\frac{1}{\sqrt{2}}+(D-2)\sqrt{2}\\
\text{Where~$\Omega_{ab}$}&\text{ is the metric on $(D-2)$ dimensional unit sphere}
\end{split}
\end{equation}
All the components of the derivatives of the velocity field on the membrane vanishes
\begin{equation}\label{branederiu2}
\hat{\nabla}_{\mu} u_{\nu}=0,~~ \{\mu={t,a}\}
\end{equation}
Substituting equations \eqref{braneextrinsic} and \eqref{branederiu} in the membrane equation \eqref{eq:finalMembrane} and using the fact that $P^t_t=P^t_a=0$,  we see that it is satisfied upto the required order.

Next we shall match the form of the metric. As in previous subsubsection we have to read off appropriate $\psi$ and $u_A$ defined everywhere in Global AdS space.
 \begin{itemize}
 \item Since the space-time is static and also maintains spherical  symmetry,  $\psi$ must be a function of $r$ only.
 This implies $n_A dx^A \propto dr$. \\
 After  normalization $n_A dx^A  = \frac{dr}{\sqrt{1+r^2}}$.
 \item It follows that the normalized $u$ has the form
 \begin{equation}\label{ua}
 u_A~ dx^A = -~\sqrt{1+r^2}~dt~~\text{or}~~O_A~ dx^A = \left(\sqrt{1+r^2}~dt + \frac{dr}{\sqrt{1+r^2}}\right)
 \end{equation}
 It is easy to see that this $u$ will satisfy all the subsidiary condition as a consequence of the symmetry.
 \item Now $\psi$ has to satisfy the subsidiary condition, 
 \begin{equation}\label{subpsi}
 \nabla^2\psi^{-D} =0
 \end{equation}
  \end{itemize}
  To solve the equation \eqref{subpsi} we have to repeat the same procedure as we have done in the previous subsubsection. Now the only difference is that the background is not AdS-Poincare but global AdS and the covariant derivatives are also modified accordingly. This calculation is a bit complicated and the details are given in appendix (\ref{app:AdsBHMatch})
  \begin{equation}\label{psiglobal}
  \begin{split}
  \psi(r)&=1+\frac{\text{log 2}}{D}+\left(\frac{1}{D}\right)^2\left[\frac{(\text{log 2})^2}{2}\right]+\left(1+\frac{\text{log 2~-~2}}{D}\right)(r-1)+{\cal O}\left(\frac{1}{D}\right)^3\\
  &=r\left(1+\frac{\text{log 2}}{D}\right)-(r-1)\frac{2}{D}+\left(\frac{1}{D}\right)^2\frac{(\text{log 2})^2}{2}+{\cal O}\left(\frac{1}{D}\right)^3
  \end{split}
  \end{equation}
Here also \eqref{psiglobal} imply that in the `membrane region'
\begin{equation}\label{psidifglobal}
\frac{r^{-(D-3)}}{1+r^2} = \psi^{-D}  +{\cal O}\left(\frac{1}{D}\right)^2
\end{equation}
As in the previous subsection from equations \eqref{psidifbrane} and \eqref{ua} it follows that the metric of  AdS Schwarzschild black-hole is same as the one we determined in equation \eqref{eq:finalMetric} upto correction of order ${\cal O}\left(\frac{1}{D}\right)^2$.

\subsubsection{Rotating Black Hole in AdS}
The explicit form of Kerr de-Sitter metric in $D~=~2n~+~1$ dimensions(\cite{gibbons}, \cite{myersperry}) in Kerr-Schild form is given by 
\begin{equation}\label{KerrRot}
\begin{split}
dS^2&=d\bar{S}^2_{\text{AdS}}+\frac{2M}{U}(k_A dx^A)^2\\
g_{AB}dx^{A}dx^{B}&=\bar{g}_{AB}dx^{A}dx^{B}+\frac{2M}{U}k_{A}k_{B}dx^{A}dx^{B}
\end{split}
\end{equation}
where
\begin{equation}\label{rotback}
\begin{split}
d\bar{S}^2&=-W(1+r^2) dt^2 + F dr^2+\sum_{i=1}^{n}\frac{r^2+a_i^2}{1-a_i^2}(d\mu_i^2+\mu_i^2 d\phi_i^2)\\
&~~~~~~~~~~~~~~-\frac{1}{W(1+r^2)}\bigg(\sum_{i=1}^n\frac{(r^2+a_i^2)\mu_i d\mu_i}{1-a_i^2}\bigg)^2
\end{split}
\end{equation}
\begin{equation}
\begin{split}
W=\sum_{i=1}^n\frac{\mu_i^2}{1-a_i^2};\quad
F=\frac{r^2}{1+r^2}\sum_{i=1}^n\frac{\mu_i^2}{r^2+a_i^2};\quad
U=\sum_{i=1}^n\frac{\mu_i^2}{r^2+a_i^2}\prod_{j=1}^n(r^2+a_j^2)
\end{split}
\end{equation}
\begin{equation}
\begin{split}
k_{\mu} dx^{\mu}=W dt+Fdr-\sum_{i=1}^n\frac{a_i \mu_i^2}{1-a_i^2}d\phi_i
\end{split}
\end{equation}
$\bar{g}_{\mu\nu}$ is actually the metric of global AdS, but written in some rotating coordinate. The coordinate transformation that will bring it back to standard form (the one presented in  equation \eqref{globalSt}) is given in \cite{gibbons}.  However we shall continue to work in the coordinates as given in equation \eqref{rotback}. One of the advantage of using these coordinates is that the horizon of the black hole space time in these rotating coordinates  is given by constant $r$ slices, where the value of the constant is determined from the zero of the following function.
\begin{equation}\label{rothor}
\begin{split}
\frac{U}{F}-2M=0
\end{split}
\end{equation}
 For convenience of computation we shall scale the parameter $M$ in the following way $$M=\prod_{i=1}^{n}(1+a_i^2)$$ so that the horizon lies at $r=1$, which would be the equation of our membrane.  The induced metric on the membrane
 \begin{equation}\label{indrot}
 \begin{split}
 dS^2_{\text{induced}}&=-2~W~dt^2 +\sum_{i=1}^{n}\frac{1+a_i^2}{1-a_i^2}(d\mu_i^2+\mu_i^2 d\phi_i^2)-\frac{1}{2~W}\bigg(\sum_{i=1}^n\frac{(1+a_i^2)\mu_i d\mu_i}{1-a_i^2}\bigg)^2
 \end{split}
 \end{equation}
  It turns out that $k_{\mu}$ is null with respect to both the metric $g_{\mu\nu}$ and $\bar{g}_{\mu\nu}$. The null generator of the horizon is given by
\begin{equation}\label{rotgen}
\begin{split}
l^A\partial_A = \frac{1}{\sqrt{2}}\left(\sum_{j=1}^{n}\frac{\mu_j^2}{1+a_j^2}\right)^{-\frac{1}{2}}\left(\partial_t+2\sum_{i=1}^{n}\frac{a_i}{1+a_i^2}~\partial_{\phi_i}\right)
\end{split}
\end{equation}  
From here it follows that the velocity field along the horizon is given by 
\begin{equation}\label{uAmem}
\begin{split}
&u_A dx^A=-\sqrt{2}\left(\sum_{j=1}^{n}\frac{\mu_j^2}{1+a_j^2}\right)^{-\frac{1}{2}}\left(\sum_{i=1}^n\frac{\mu_i^2}{1-a_i^2}(dt-a_i~d\phi_i)\right)
\end{split}
\end{equation}
Once we have  the explicit form of the membrane equation and the velocity field, each term of \eqref{eq:memfinal} are computable.\\
Now as we have explained before, our $\left(\frac{1}{D}\right)$ expansion is valid provided the space-time satisfies some large symmetry and is dynamical or non-trivial only in a finite number of dimensions. The metric in \eqref{KerrRot} will belong to this class, if only a finite number rotation parameters $a_i$'s are non-zero.  But if we turn on arbitrary (though finite) number of $a_i$' s,  it turns out that explicit computation is very tedious for this complicated metric. So we have used Mathematica (version 9.0) here and to be explicit we have used two non-zero rotation parameters.
We have first checked that this velocity field and the extrinsic curvature of the membrane do satisfy our membrane equation \eqref{eq:finalMembrane} upto the required order.

 The next job is to check whether the space-time metric \eqref{KerrRot} matches with equation \eqref{eq:finalMetric} upto correction of order ${\cal O}\left(\frac{1}{D}\right)^2$. Now we know that $k_A$ is exactly null with respect to $\bar g_{AB}$. Clearly $k_A$ is the most natural candidate for the null vector $O_A$ we have in our metric. \\
 Suppose
 $$k_A = {\cal A}~ O_A$$
 where ${\cal A}$ is some unknown function of $r$ at the moment.
Now note that the metric \eqref{KerrRot} will be precisely of the form \eqref{eq:finalMetric} provided we identify \\
$${{\cal A}^2}\left(\frac{2M}{U}\right)\rightarrow\bigg[\psi^{-D} + {\cal O}\left(\frac{1}{D}\right)^2\bigg]$$
The above equation along with the fact that ${\cal A}$ is a function of $r$ , will imply that $\psi$ also depends only on $r$. The unit normal to $\psi = $ constant slices is then given by
$$n_A dx^A = \sqrt{F}~dr$$
 Now  $O_A ~n^A =1$  implies $k_A n^A = {\cal A}$. Therefore once we know the explicit expression of $n_A$, we can fix ${\cal A}$.
 It turns out 
 $${\cal A} = \sqrt{F} $$
However just identifying  $\bigg[{{\cal A}^2}\left(\frac{2M}{U}\right)\bigg]$ with $\psi^{-D}$ is not enough for the matching of the two metrics. We also have to see whether these $\psi$ and $O_A$ satisfy our subsidiary conditions.
The above identification will be consistent with our subsidiary condition \eqref{eq:SubsidiaryPsi} provided  
\begin{equation}\label{eqcond}
\begin{split}
&\nabla^2\left[{{\cal A}^2}\left(\frac{2M}{U}\right)\right]= {\cal O}(1)\\
&(k\cdot\nabla)k_A \propto \bigg[k_A + {\cal O}\left(\frac{1}{D}\right)\bigg],~~~n^A k_A =1
\end{split}
\end{equation} 
\footnote{We already know that $k_A k^A =0$. Now as along as $n_A k^A=1$, we could always  express $k_A$ as $k_A = n_A -u_A$ such that $u\cdot u= -1$ and $n\cdot u = 0$ everywhere}
Here $\nabla$ is defined with respect to the background metric $\bar g_{AB}$ and all raising and lowering of indices have been done using $\bar g_{AB}$. In Mathematica we have explicitly verified this condition for two nonzero rotation parameters.

\subsection{Quasinormal Modes for Schwarzschild black hole in background AdS/dS spacetime}
In this subsection, using our membrane equations we shall compute the spectrum of light quasi normal modes for Schwarzschild Black hole with horizon topology $S^{D-2}\times \text{R}$ in background AdS/dS spacetimes. We will find that the QNM frequencies that we get match exactly with the earlier obtained result from purely gravitational analysis done in \cite{Emparan:2015rva}. 

We do this calculation in Global AdS/dS coordinates. The background AdS/dS in global coordinates can be written as
\begin{equation}\label{globmet}
 ds^2_{(bgd)} = g_{AB} dX^A dX^B= -\left( 1-\sigma\frac{r^2}{L^2} \right)dt^2 + \frac
 {dr^2}{\left( 1-\sigma\frac{r^2}{L^2} \right)} + r^2 d\Omega^2_{D-2}.
\end{equation}
Where
\begin{equation} 
\begin{split}
\Lambda &= \frac{\sigma}{L^2} (D-1)(D-2)\\
L&=\text{AdS/dS radius}\\
 \sigma &= 0 \quad \text{for Flat} \\
 &= 1 \quad \text{for dS} \\
 &= -1 \quad \text{for AdS}
 \end{split}
\end{equation}
The black hole solution in this coordinate system has the following form
\begin{equation}\label{eq:blackstatic}
ds^2_{(BH)} =  -\left( 1-\sigma\frac{r^2}{L^2} - \left( \frac{r_0}{r}\right)^{D-3}  \right)dt^2 + \frac
 {dr^2}{\left( 1-\sigma\frac{r^2}{L^2}- \left( \frac{r_0}{r}\right)^{D-3} \right)} + r^2 d\Omega^2_{D-2}.
\end{equation}
In equation \eqref{eq:blackstatic} $r_0$ is an arbitrary constant.
From now on we choose $r_0=1$ for convenience. After we find the answers for QNM frequencies we can reinstate these factors easily from dimensional analysis. 

As we have seen in the previous subsection, a static black hole corresponds to a spherical membrane with a velocity field purely in the time direction. Here we introduce a small fluctuation around this spherical membrane along with a small fluctuation in the velocity field. The amplitude of the fluctuation will be denoted by $\epsilon$ - the linearization parameter for our analysis.
\begin{equation}\label{bhflu}
 \begin{split}
  r &= 1 + \epsilon~ \delta r(t,a) \\
  u &= u_0 ~dt + \epsilon ~\delta u_\mu (t,a) dx^\mu
 \end{split}
\end{equation}
Here $a$ indices denote the angle coordinates along the $(D-2)$ dimensional sphere and the coordinates along the membrane (i.e., time $t$ and angles $a$) are denoted by $\mu$ indices.
The induced metric on the membrane worldvolume upto linear order in $\epsilon$ is given by (with the components denoted by $g_{\mu\nu}^{(ind)}$)
\begin{equation}\label{lind}
 ds^2_{(ind)} =g_{\mu\nu}^{(ind)}dy^{\mu}dy^{\nu}= -\left( 1-\sigma\frac{1+2\epsilon \delta r}{L^2} \right)dt^2 + (1+2\epsilon \delta r) d\Omega^2_{D-2}
\end{equation}
Normalization of the velocity field $( u_\mu g^{\mu\nu}_{(ind)} u_\nu=-1)$ implies
\begin{equation}
 \begin{split}
& u_0 = -\left( 1-\frac{\sigma}{L^2} \right)^{\frac{1}{2}} ~~\text{and}~~~
  \delta u_t(t,a) = \left( 1-\frac{\sigma}{L^2} \right)^{-\frac{1}{2}}\left(\frac{\sigma}{L^2}\right)\delta r(t,a)
 \end{split}
\end{equation}

The membrane equations have the following form
\begin{equation}\label{membrane}
\begin{split}
\hat{\nabla}\cdot u=0,~~~~&{\cal P}^{\nu}_\mu\left\{\frac{\hat{\nabla}^2 u_{\nu}}{\cal{K}}-\frac{\hat{\nabla}_{\nu}{\cal{K}}}{\cal{K}}+u^{\alpha}{\cal K}_{\alpha\nu}-(u\cdot\hat{\nabla})u_{\nu}\right\}=0\\
\end{split}
\end{equation}
Here $\hat\nabla$ denotes the covariant derivative with respect to the metric \eqref{lind}. ${\cal K}_{\mu\nu}$ is the extrinsic curvature viewed as a symmetric tensor along the membrane world volume. ${\cal K}$ is the trace of ${\cal K}_{\mu\nu}$ and the projector (again as a tensor along the membrane world volume)  perpendicular to $u_\mu$ is denoted as $$ {\cal{P}}_{\mu\nu}\equiv g^{(ind)}_{\mu\nu}+u_{\mu} u_{\nu}$$ 
Different components of this projector are given by
\begin{equation}
 {\mathcal P}^t_t = 0,\quad {\mathcal P}^a_t = -\left( 1-\frac{\sigma }{L^2} \right)^{\frac{1}{2}}(\epsilon \delta u^a),\quad {\mathcal P}^t_a = \left( 1-\frac{\sigma }{L^2} \right)^{-\frac{1}{2}}(\epsilon \delta u_a),\quad {\mathcal P}^a_b = \delta^a_b
\end{equation}
Now, for convenience we rewrite the vector membrane equation (second equation in \eqref{membrane}) as
 $$E^{tot}_{\mu} \equiv \mathcal{P}^\nu_\mu E_{\nu}$$ where,
 $$E_\mu \equiv \frac{\hat{\nabla}^2u_\mu}{\mathcal{K}}-\frac{\hat{\nabla}_\mu {\mathcal K}}{\mathcal{K}}+u^\nu {\cal K}_{\nu\mu}-u^\nu \hat{\nabla}_{\nu}u_{\mu} $$ 
 Hence we have
 \begin{equation}
\begin{split}
 E^{tot}_t &= E_t {\mathcal P}^t_t +  E_b {\mathcal P}^b_t \\
 E^{tot}_a &= E_t {\mathcal P}^t_a +  E_b {\mathcal P}^b_a
\end{split}
\end{equation}
 Note that because of the spherical symmetry of the background $E_a$(remember `$a$' denotes the angle coordinates along the sphere) will be nonzero only if fluctuations are present or in other words $E_a\sim{\cal O}(\epsilon)$.
 Also ${\mathcal P}^t_t= 0$ and  ${\mathcal P}^a_t \sim {\mathcal O}(\epsilon)$. It follows that the time component $E^{tot}_t$ identically vanishes at the linear order. Now since ${\mathcal P}^t_a = {\mathcal O}(\epsilon)$, we see that only ${\mathcal O}(\epsilon^0)$ pieces of $ E_t$ matter in the computation of  $E^{tot}_a$. We keep these facts in mind and evaluate only those terms in $E_\mu$ that will be important for our linearized analysis.\\

First note that ${\cal K}_{\mu\nu}$ is generically not equal to the $(\mu\nu)$ component of  $K_{AB}$, the extrinsic curvature viewed as a tensor in the full background space-time.
${\cal K}_{\mu\nu}$ is given by the pullback of the extrinsic curvature on the membrane surface as
\begin{equation}\label{eq:pullback}
 {\cal K}_{\mu\nu} = \left( \frac{\partial X^M}{\partial y^\mu} \right) \left( \frac{\partial X^N}{\partial y^\nu} \right) K_{MN} \vert_{r= 1 + \epsilon \delta r}
 \end{equation}
 where we have denoted the set $(r,t,\theta^a)$ by $X^M$ and the set $(t,\theta^a)$ by $y^\mu$.
The space-time form of extrinsic curvature $K_{AB}$  is given by 
\begin{equation}
\begin{split}
K_{AB}&=\Pi_{A}^{C} {\nabla}_C n_B,~~~
\text{where } \Pi_{AC}=g_{AC} -n_A n_C
\end{split}
\end{equation}
Now if we apply equation \eqref{eq:pullback} for our case, we find
\begin{equation}\label{eq:pullapply}
\begin{split}
 {\cal K}_{\mu\nu} =\epsilon (\partial_\mu\delta r ) K_{r\nu} +\epsilon (\partial_\nu\delta r ) K_{r\mu} + K_{\mu\nu} + {\cal O}(\epsilon^2)
 \end{split}
\end{equation}
From explicit computation we know (see appendix \ref{QNM AdS Global}) that $K_{rN}={\mathcal O}(\epsilon)$. It follows that for this linearized analysis ${\cal K}_{\mu\nu}$ is just the `truncation' of the $K_{MN}$ evaluated on the membrane surface and is given by (see appendix \ref{QNM AdS Global} for details).
 \begin{equation}
  \begin{split}
   {\cal K}_{tt} &=  \left( 1-\frac{\sigma}{L^2} \right)^{-\frac{1}{2}} (-\epsilon \partial^2_t \delta r) + \left( 1-\frac{\sigma}{L^2} \right)^{\frac{1}{2}} \left(\frac{\sigma}{L^2}\right) \left( 1 + \epsilon\delta r - \frac{\sigma \epsilon \delta r}{L^2-\sigma} \right) \\
    {\cal K}_{ta} &=  \left( 1-\frac{\sigma}{L^2} \right)^{-\frac{1}{2}} (-\epsilon \partial_t \bar\nabla_a \delta r) \\
     {\cal K}_{ab} &=  \left( 1-\frac{\sigma}{L^2} \right)^{-\frac{1}{2}} (-\epsilon \bar\nabla_a \bar\nabla_b \delta r) + \left( 1-\frac{\sigma}{L^2} \right)^{\frac{1}{2}}  \left( 1 + \epsilon\delta r - \frac{\sigma \epsilon \delta r}{L^2-\sigma} \right)\hat{g}_{ab} \\
  \end{split}
 \end{equation}
 Here $\bar\nabla_a$ denotes the covariant derivative with respect to the metric on a $(D-2)$ dimensional unit sphere.\\
The trace of the Extrinsic curvature is given by
\begin{equation}
 \begin{split}
  {\mathcal K} &= \left( 1-\frac{\sigma}{L^2} \right)^{-\frac{3}{2}} (\epsilon \partial^2_t \delta r) - \left( 1-\frac{\sigma}{L^2} \right)^{-\frac{1}{2}} \left(\frac{\sigma}{L^2}\right) \left( 1 + \frac{\epsilon L^2 \delta r}{L^2-\sigma} \right) \\ & + \left( 1-\frac{\sigma}{L^2} \right)^{-\frac{1}{2}} (-\epsilon \bar\nabla^2 \delta r) + \left( 1-\frac{\sigma}{L^2} \right)^{\frac{1}{2}}  \left( 1 - \frac{\epsilon L^2 \delta r}{L^2-\sigma} \right) (D-2)
 \end{split}
\end{equation}
Thus the  components that would be relevant for the linearized membrane equation are  given by
 \begin{equation}\label{eq:list}
  \begin{split}
   u^\nu {\cal K}_{\nu t} &= \frac{\sigma}{L^2} + {\mathcal O}(\epsilon) \\
   u^\nu {\cal K}_{\nu a} &= \left( 1-\frac{\sigma}{L^2} \right)^{-1} (-\epsilon \partial_t \bar\nabla_a \delta r) + \left( 1-\frac{\sigma}{L^2} \right)^{\frac{1}{2}}(\epsilon \delta u_a) \\
   u^\nu \hat{\nabla}_{\nu}u_{t} &= 0\\
   u^\nu \hat{\nabla}_{\nu}u_{a} &= \left( 1-\frac{\sigma}{L^2} \right)^{-\frac{1}{2}}(\epsilon \partial_t \delta u_a)-\left( 1-\frac{\sigma}{L^2} \right)^{-1}\frac{\sigma}{L^2}(\epsilon \bar\nabla_a \delta r) \\
  \hat{\nabla}_t {\mathcal K} &=  {\mathcal O}(\epsilon) \\
\hat{\nabla}_a {\mathcal K} &=  \left( 1-\frac{\sigma}{L^2} \right)^{-\frac{3}{2}}(\epsilon \partial^2_t \bar\nabla_a \delta r)-\left( 1-\frac{\sigma}{L^2} \right)^{-\frac{3}{2}}\frac{\sigma}{L^2}(\epsilon \bar\nabla_a \delta r) \\
   &+ \left( 1-\frac{\sigma}{L^2} \right)^{-\frac{1}{2}}(-\epsilon \bar\nabla_a \bar\nabla^2 \delta r) - (D-2) \left( 1-\frac{\sigma}{L^2} \right)^{-\frac{1}{2}}(\epsilon \bar\nabla_a \delta r) \\
  \hat{\nabla}^2u_t &= {\mathcal O}(\epsilon) \\
  \hat{\nabla}^2u_a &= -\left( 1-\frac{\sigma}{L^2} \right)^{-1}(\epsilon \partial_t^2 \delta u_a) + \left( 1-\frac{\sigma}{L^2} \right)^{-\frac{3}{2}}\frac{\sigma}{L^2} (\epsilon \partial_t \bar\nabla_a \delta r) \\
   &+ \epsilon \bar\nabla^2 \delta u_a + \left( 1-\frac{\sigma}{L^2} \right)^{-\frac{1}{2}} (\epsilon \partial_t \bar\nabla_a \delta r)
  \end{split}
 \end{equation}
As before in equations \eqref{eq:list} $\hat\nabla$ denotes the covariant derivative with respect to the the induced metric as given in equation \eqref{lind} and $\bar\nabla$ denotes the covariant derivative with respect to the metric on a $(D-2)$ dimensional unit sphere.

Using equations \eqref{eq:list} the linearized vector membrane equation in the angular directions evaluates to
\begin{equation}\label{lineq}
 \begin{split}
  E_a^{tot} &\equiv \left( \frac{\sigma}{L^2} \right)\left( 1-\frac{\sigma}{L^2} \right)^{-\frac{1}{2}}(\epsilon \delta u_a) + \left( 1-\frac{\sigma}{L^2} \right)^{-\frac{1}{2}}\epsilon \frac{\bar\nabla^2\delta u_a}{D-2} + \left( 1-\frac{\sigma}{L^2} \right)^{-1} \epsilon \frac{\bar\nabla_a\bar\nabla^2\delta r}{D-2} \\
  &+ \left( 1-\frac{\sigma}{L^2} \right)^{-1} (\epsilon \bar\nabla_a \delta r) + \left( 1-\frac{\sigma}{L^2} \right)^{-1} (-\epsilon \partial_t\bar\nabla_a \delta r) + \left( 1-\frac{\sigma}{L^2} \right)^{\frac{1}{2}} (\epsilon \delta u_a) \\ &- \left( 1-\frac{\sigma}{L^2} \right)^{-\frac{1}{2}} (\epsilon \partial_t\delta u_a) +\left( 1-\frac{\sigma}{L^2} \right)^{-1}\left(\frac{\sigma}{L^2}\right) (\epsilon \bar\nabla_a\delta r)
 \end{split}
\end{equation}
In writing \eqref{lineq} we have also neglected the terms which are subleading in $1/D$. \\
We also need to process the first equation of \eqref{membrane}. We have to evaluate  the divergence of the velocity field. It comes out to be
\begin{equation}\label{veldiv}
 \hat{\nabla}.u = 0 = \epsilon \bar\nabla^a \delta u_a + \epsilon \left( 1-\frac{\sigma}{L^2} \right)^{-\frac{1}{2}} (\partial_t \delta r)(D-2)
\end{equation}
Now, similar to calculation done in Section (5) of \cite{Chmembrane} we divide the fluctuation $\delta u_a$ in two parts
\begin{equation}\label{velcomp}
 \delta u_a = \delta v_a + \bar\nabla_a \Phi~,\quad \text{with}\quad\bar\nabla^a \delta v_a =0 
\end{equation}
Substituting \eqref{velcomp} into \eqref{veldiv} we get 
\begin{equation}\label{paru}
 \bar\nabla^2\Phi = -\left( 1-\frac{\sigma}{L^2} \right)^{-\frac{1}{2}} (\partial_t \delta r)(D-2) 
\end{equation}
Now consider $\bar\nabla^aE_a^{tot}$. Using the identity $\bar\nabla^a\bar\nabla^2V_a = ((D-2)+\bar\nabla^2)\bar\nabla^aV_a$ and simplifying we get
\begin{equation}\label{eqdiv}
 \begin{split}
  -2(D-2) \partial_t \delta r - 2\partial_t\bar\nabla^2 \delta r + \frac{\bar\nabla^2 \bar\nabla^2\delta r}{D-2} + \bar\nabla^2\delta r  + (D-2) \partial^2_t \delta r + \frac{\sigma}{L^2}\bar\nabla^2\delta r =0
 \end{split}
\end{equation}
Note that compared to the flat case (refer to Eq. (5.16) in \cite{Chmembrane}), it is easy to see that the only term extra in equation \eqref{eqdiv} is the last term which is crucial.

We expand the fluctuation as
\begin{equation}\label{rsph}
\delta r= \sum_{l,m}a_{lm} Y_{lm} e^{-i \omega^s_l t}
\end{equation} 
where the scalar spherical harmonics $Y_{lm}$ on $S^{D-2}$ obey 
\begin{equation}
\bar\nabla^2 Y_{lm} = -l(D+l-3) Y_{lm}.
\end{equation} 
After substituting \eqref{rsph} in \eqref{eqdiv} and solving we get the scalar QNM frequencies
\begin{equation}
 \omega^s = \pm \sqrt{l\left(1-\frac{\sigma}{L^2}\right)-1} - i(l-1)
\end{equation}
Reinstating the factors of $r_0$ we get
\begin{equation}
 \omega^s r_0 = \pm \sqrt{l\left(1-\frac{\sigma r_0^2}{L^2}\right)-1} - i(l-1)
\end{equation}
Upto the required order, this answer agrees  with the corresponding answer given in expressions (D.3),(D.4) of \cite{Emparan:2015rva}.


Now we find the vector QNM frequencies. Since we have solved \eqref{eqdiv} the $\delta r$ and $\Phi$ terms in \eqref{lineq} drop out and the equation reduces to

\begin{equation}\label{eqvel}
 \frac{\bar\nabla^2\delta v_a}{D-2} + \delta v_a - \partial_t \delta v_a = 0
\end{equation}

We expand the fluctuation as
\begin{equation}\label{usph}
\delta v_a = \sum_{l,m} b_{lm} Y_a^{lm} e^{-i \omega^v_l t}
\end{equation}
where the vector spherical harmonics $Y_a^{lm}$ on $S^{D-2}$ obey
\begin{equation}
\bar\nabla^2 Y_a^{lm}=- [(D+l-3)l-1]Y_a^{lm}
\end{equation}

Substituting \eqref{usph} in \eqref{eqvel} and solving we get the vector QNM frequency as

\begin{equation}
 \omega^v = -i(l-1) 
\end{equation}
Reinstating the factors of $r_0$ we have
\begin{equation}
 \omega^v r_0 = -i(l-1) 
\end{equation}
Upto the required order, this answer agrees  with the corresponding answer given in expression (D.2) of \cite{Emparan:2015rva}.

\subsection{Nonlinear effective equations for AdS Black brane dynamics from scaled membrane equations}

In a parallel development, the authors of \cite{EmparanHydro} have developed an effective theory for black brane dynamics in background AdS spacetime. They focus on the length scales of order $\frac{1}{\sqrt{D}}$ and derive a pair of nonlinear effective differential equations that govern the dynamics of fluctuations which are suppressed for large $D$ by appropriate inverse powers of $D$ given in \cite{EmparanHydro}. In this section we show that by doing appropriate scalings in our membrane equations, we are able to reduce our membrane equations to the form that matches the effective equations given in \cite{EmparanHydro} under appropriate field redefinition. This analysis is very similar to the one done for the case of black p-brane in flat spacetime in \cite{yogesh2}. In this section, we first do the linearized analysis of fluctuations without any scalings and get idea about how the various quantities need to be rescaled. Then we do the nonlinear analysis by applying the scalings and show the correspondence to effective equations of \cite{EmparanHydro}. 

\subsubsection{Linearized fluctuation analysis and hints for scalings}

 Now we do the linearized fluctuation analysis for a planar membrane in AdS. This membrane corresponds to a Schwarzschild black brane in AdS with horizon topology $R^{D-2}\times R$ in Poincare patch metric. We will consider the fluctuations in shape and velocity field in time plus all the $D-2$ brane directions.
 
 
 The background metric in Poincare patch (with AdS radius $L=1$) is given by
 \begin{equation}\label{ppold}
  ds^2 = -\hat{r}^2 d\hat{t}^2 + \frac{d\hat{r}^2}{\hat{r}^2} + \hat{r}^2 d\hat{x}^ad\hat{x}_a
 \end{equation}
 Where the indices $a,b$ take $D-2$ number of values.\\
 Let $\hat{r}=r_0$ be the position of static unperturbed membrane in Poincare patch coordinates. We choose to scale the coordinates with $r_0$ in the following way 
 \begin{equation}
  \hat{r} = r_0 r, \quad \hat{t} = \frac{t}{r_0}, \quad  \hat{x}^a = \frac{x^a}{r_0}
 \end{equation}
In these scaled coordinates the background metric takes the following form
 \begin{equation}\label{ppmet}
  ds^2_{(bgd)} = g_{AB} dX^AdX^B = -r^2 dt^2 + \frac{dr^2}{r^2} + r^2 dx^adx_a
 \end{equation}
 And also the position of the membrane is now at $r=1$.\\
  When we introduce the fluctuations on this membrane we will consider the time dependence as $$ e^{-i\hat{\omega}\hat{t}}=e^{-i\omega t} , \quad \text{with}~~\hat{\omega}=r_0 \omega $$ We work with this choice from now on. 
 Note that with this choice all the new (non-hatted) coordinates are dimensionless.

As mentioned before, in this section we shall consider small fluctuations around a static membrane solution. The fluctuations will be of the form
 \begin{equation}\label{bbflu}
  \begin{split}
   r &= 1 + \epsilon \delta r(t,a) \\
   u &= u_0 dt + \epsilon \delta u_t(t,a) dt + \epsilon \delta u_b(t,a) dx^{b} 
  \end{split}
 \end{equation}
 $\epsilon$ is the linearization parameter.
 
 To the leading order in $\epsilon$, the induced metric on the membrane worldvolume is
 \begin{equation}\label{ppind}
  ds^2 = g_{\mu\nu}^{(ind)} dy^\mu dy^\nu = -(1 + 2 \epsilon \delta r) dt^2 + (1 + 2 \epsilon \delta r) dx^adx_a 
 \end{equation}

As before,  we use the notation $\hat{\nabla}$ for denoting the covariant derivative constructed from the induced metric \eqref{ppind} and $\nabla$ for denoting the covariant derivative constructed from the background metric \eqref{ppmet}. In this notation the membrane equation will have the same form in as in equation \eqref{membrane}, where ${\cal P}_{\mu\nu}$, ${\cal K}_{\mu\nu}$  and ${\cal K}$ are the projector, extrinsic curvature and its trace  exactly as in previous section.\\
${\cal K}_{\mu\nu}$ is given by the pullback of the space-time extrinsic curvature $K_{MN}$ on the membrane. Here also using the same reasoning as in the previous subsection one could show that ${\cal K}_{\mu\nu}$ is just the `truncation' of the $K_{MN}$ evaluated on the membrane surface.
 The nonzero components of ${\cal K}_{\mu\nu}$ are given by
 \begin{equation}
    {\cal  K}_{tt} = -\epsilon \partial^2_t \delta r - (1 + 2\epsilon \delta r),\quad {\cal  K}_{ta} = -\epsilon \partial_t \partial_a \delta r, \quad{\cal  K}_{ab} = -\epsilon \partial_a \partial_b \delta r + (1 + 2\epsilon \delta r) \delta_{ab}
 \end{equation}
 Thus the trace of Extrinsic curvature becomes
 \begin{equation}\label{kap}
  \mathcal{K} = (D-1) + \epsilon \partial^2_t \delta r
  - \epsilon \partial_a \partial^a \delta r
 \end{equation}
where the index $a$ in \eqref{kap} is raised with $\delta^{ab}$.

%
%
%
%
%
%
%
%
%
%
%
Normalization of the velocity field fixes $u_0$ and $\delta u_t$, defined in \eqref{bbflu}, in terms of the radial fluctuation
\begin{equation}\label{eq:utexp}
 u_t = u_0 +\epsilon \delta u_t = -(1+\epsilon\delta r)
\end{equation}
Given  the velocity field, different components of the projectors  $\mathcal{P}^\mu_\nu = \delta^\mu_\nu + u^\mu u_\nu$ are given by
\begin{equation}
  \mathcal{P}^a_b = \delta^a_b, \quad \mathcal{P}^t_t = 0, \quad \mathcal{P}^t_a = \epsilon \delta u_a, \quad  \mathcal{P}^a_t = -\epsilon \delta u_a
\end{equation}
Now, following the same trick as  in the previous subsection we denote the vector membrane equation (the 2nd equation in \eqref{membrane}) as
 $$E^{tot}_{\mu} \equiv \mathcal{P}^\nu_\mu E_{\nu}$$ where,
 $$E_\mu \equiv \frac{\hat{\nabla}^2u_\mu}{\mathcal{K}}-\frac{\hat{\nabla}_\mu {\mathcal K}}{\mathcal{K}}+u^\nu {\cal K}_{\nu\mu}-u^\nu \hat{\nabla}_{\nu}u_{\mu} $$ 
Then we have
\begin{equation}\label{eomdiv}
\begin{split}
 E^{tot}_t &= E_t {\mathcal P}^t_t +  E_b {\mathcal P}^b_t \\
 E^{tot}_a &= E_t {\mathcal P}^t_a +  E_b {\mathcal P}^b_a 
\end{split}
\end{equation}
 
The background has a translational symmetry along the  $x_a$ directions that is broken by the fluctuations. Hence $E_b\sim{\mathcal O}(\epsilon)$.
Now using the facts that  ${\mathcal P}^t_t = 0$,  ${\mathcal P}^a_t\sim {\mathcal O}(\epsilon)$ and $E_b\sim{\mathcal O}(\epsilon)$ we can see that the time component $E^{tot}_t$ vanishes at the linear order. 
Similarly for $E_a^{tot}$, since ${\mathcal P}^t_a \sim {\mathcal O}(\epsilon)$, we see that only ${\mathcal O}(\epsilon^0)$ pieces of $E_t$ contributes.
 We keep these facts in mind and calculate only those terms that are important.
 \begin{equation}
   \frac{\hat{\nabla}_t \mathcal{K}}{\mathcal{K}} = \mathcal{O}(\epsilon), \quad
   \frac{\hat{\nabla}^2 u_t}{\mathcal{K}} = \mathcal{O}(\epsilon), \quad
   u^\mu{\cal K}_{\mu t} = -1, \quad
  (u\cdot\hat\nabla) u_t = \mathcal{O}(\epsilon)
 \end{equation}
and (with the notation $\partial^2=\partial_a\partial^a$)
 \begin{equation}
  \begin{split}
   \frac{\hat{\nabla}_a \mathcal{K}}{\mathcal{K}} &= \frac{\epsilon \partial_a\partial_t^2 \delta r}{D} - \frac{\epsilon\partial_a \partial^b\partial_b \delta r}{D}, \quad
   \frac{\hat{\nabla}^2 u_a}{\mathcal{K}} = \frac{-\epsilon \partial_t^2 \delta u_a}{D} - \frac{\epsilon\partial_a \partial_t \delta r}{D} + \frac{\epsilon \partial^b\partial_b \delta u_a}{D} + \frac{\epsilon \partial_a \partial_t \delta r}{D} ,\\ 
   u^\mu{\cal K}_{\mu a} &= -\epsilon \partial_t\partial_a\delta r+\epsilon \delta u_a, \quad
  (u\cdot\hat\nabla)u_a = \epsilon \partial_t \delta u_a + \epsilon \partial_a \delta r
  \end{split}
 \end{equation}
 Using equation \eqref{eomdiv} the expression for  linearized $E_a^{tot}$ is given by
  \begin{equation}\label{eq:eomPrelim}
  -\partial_a \delta r - \partial_t\partial_a\delta r -\partial_t \delta u_a+\frac{1}{D} \left(-\partial_a\partial_t^2 \delta r -\partial_t^2 \delta u_a\right) + \frac{1}{D}( \partial^b\partial_b \delta u_a + \partial^b\partial_b\partial_a \delta r) = 0
 \end{equation}
 The $\hat{\nabla}.u=0$ equation becomes
 \begin{equation}\label{delu}
  \hat{\nabla}.u=0= \epsilon \partial^a \delta u_a + \epsilon (D-2) \partial_t \delta r 
 \end{equation}
 Note that if we assume all spatial and temporal frequencies are of order ${\cal O}(1)$, then in  equation \eqref{eq:eomPrelim} the last two terms in parenthesis are suppressed compared to the first three terms by a factor of $\left(\frac{1}{D}\right)$. However, it turns out, that the temporal and the spatial frequencies are related by a factor of $\left(\frac{1}{\sqrt D}\right)$ even if we ignore the last two terms in equation \eqref{eq:eomPrelim}, mentioned above. This happens because in the scalar sector the divergence of the velocity fluctuation couples to the shape fluctuation (i.e., $\delta r$) and the coupling is through equation \eqref{delu} which involves a relative factor of $D$.\\
 This simply says that it is inconsistent to assume both the temporal and spatial frequencies to be of order ${\cal O}(1)$. Now we shall demand that the temporal frequency is of order ${\cal O}(1)$, but we shall not restrict the spatial frequencies. In that case the terms in the first  parenthesis in equation \eqref{eq:eomPrelim}  are certainly suppressed compared to the first three terms, but the terms in the last parenthesis need not be. Thus for our purpose $E_a^{tot}$ is given by
 \begin{equation}\label{eoma}
  -\partial_a \delta r - \partial_t\partial_a\delta r -\partial_t \delta u_a + \frac{1}{D}( \partial^b\partial_b \delta u_a + \partial^b\partial_b\partial_a \delta r) = 0
 \end{equation}
One might wonder that since we are considering the fluctuations with $k\sim\mathcal{O}(\sqrt{D})$, there might be instances where the subleading correction terms in the membrane equations of motion will contribute at the same order as the terms present in \eqref{eoma} and \eqref{delu}. But one can carefully think that this will not happen. The only type of potentially dangerous terms are where we have $(\hat{\nabla^2})^n$ (with $n>0$) in the numerator. But since action of each $\hat{\nabla^2}$ raises the order of the term by $D$, there would be a corresponding factor of $D^n$ in the denominator, and thus the order of this term will be still subleading compared to terms present in \eqref{eoma} and \eqref{delu}, even if $k\sim\mathcal{O}(\sqrt{D})$.

%
Now we find the scalar and vector QNMs of the membrane. Finding $\partial^aE_a^{tot}$ and substituting \eqref{delu} we get
\begin{equation}\label{diveq}
 -\partial^b\partial_b \delta r - \partial_t\partial^b\partial_b\delta r + D \partial^2_t \delta r - \partial_t \partial^b\partial_b \delta r + \frac{1}{D} \partial^a\partial_a\partial^b\partial_b \delta r = 0
\end{equation}
We can compare \eqref{diveq} with the analogous equation that can be derived for black p-brane in flat space as was done in \cite{yogesh2} and we note that the equation remains the same as in \cite{yogesh2} except that now the first term has negative sign. As we will see this sign difference implies that there is no instability in shape fluctuations unlike in the case of black p-brane in flat space.

Now we consider the plane wave expansion of the fluctuations as
\begin{equation}\label{rfl}
 \delta r = \delta r^0 e^{-i\omega t} e^{ik_ax^a}
\end{equation}
Thus substituting \eqref{rfl} into \eqref{diveq} and solving we get the scalar QNM frequencies
\begin{equation}
 \omega_s = \pm \frac{k}{\sqrt{D}}-\frac{ik^2}{D},~~~~\text{where}~~k^2=k_ak^a~~\text{and} ~~k = \sqrt{k^2}
\end{equation}
 Thus the most general solution to \eqref{diveq} is given by
\begin{equation}
 \delta r = \delta r^0_1 e^{-i\omega_1 t} e^{ik_ax^a} + \delta r^0_2 e^{-i\omega_2 t} e^{ik_ax^a}
\end{equation}
where,
\begin{equation}
  \omega_1 = \frac{k}{\sqrt{D}}-\frac{ik^2}{D},\quad \omega_2=-\frac{k}{\sqrt{D}}-\frac{ik^2}{D}
\end{equation}

Now we can take the form of the most general solution of $\delta u_a$ which solves \eqref{delu} and \eqref{eoma} as (Note there is only one vector QNM frequency as \eqref{eoma} has at max one time derivative acting on $\delta u_a$)

\begin{equation}\label{gsu}
  \delta u_a = \delta r^0_1 V_a^1 e^{-i\omega_1 t} e^{ik_ax^a} + \delta r^0_2 V_a^2 e^{-i\omega_2 t} e^{ik_ax^a}+ v_a e^{-i\omega_v t} e^{ik_ax^a}
\end{equation}
where $V_a^1$ and $V_a^2$ are vectors in the direction of $k_a$ and  $v_a$ is any vector such that $v_ak^a=0$.\\
Putting \eqref{gsu} into \eqref{delu} and \eqref{eoma} we get
\begin{equation}
   \quad \omega_v = -\frac{ik^2}{D}, \quad V_a^1 = \left( -i+\frac{\sqrt{D}}{k} \right)k_a, \quad V_a^2 = \left( -i-\frac{\sqrt{D}}{k} \right)k_a
\end{equation}
Thus we see that there is no instability in AdS case.
We write again the most general solution to the equations \eqref{eoma} and \eqref{delu} 
\begin{equation}\label{gensol}
 \begin{split}
  \delta r &= \delta r^0_1 e^{-i\omega_1 t} e^{ik_ax^a} + \delta r^0_2 e^{-i\omega_2 t} e^{ik_ax^a} \\
  \delta u_a &= \delta r^0_1 V_a^1 e^{-i\omega_1 t} e^{ik_ax^a} + \delta r^0_2 V_a^2 e^{-i\omega_2 t} e^{ik_ax^a}+ v_a e^{-i\omega_3 t} e^{ik_ax^a}
 \end{split}
\end{equation}
where, 
\begin{equation}\label{eq:paralin}
 \begin{split}
  \omega_1 &= \frac{k}{\sqrt{D}}-\frac{ik^2}{D},\quad \omega_2=-\frac{k}{\sqrt{D}}-\frac{ik^2}{D}, \quad \omega_3 = -\frac{ik^2}{D}, \quad v_ak^a=0 \\
  V_a^1 &= \left( -i+\frac{\sqrt{D}}{k} \right)k_a, \quad V_a^2 = \left( -i-\frac{\sqrt{D}}{k} \right)k_a
 \end{split}
\end{equation}

From \eqref{gensol} and \eqref{eq:paralin} we see that the interesting length scale along the $x^a$ directions is $\frac{1}{\sqrt{D}}$, rather than the scale of order unity we take. Next we work in the scaled limit adapted to capture the physics at length scale $\frac{1}{\sqrt{D}}$.

\subsubsection{Scaled nonlinear analysis and derivation of effective equations}
Now, similar to previous subsection, here we will consider a membrane configuration which has fluctuations about a uniform planar membrane in AdS. Taking hints from the linear analysis of the previous subsection, we consider a particular scaling limit of our membrane equations, like \cite{EmparanHydro}. Like \cite{EmparanHydro}, we will consider fluctuations that depend only on time and $p$ number of brane directions (with $p\sim\mathcal{O}(1)$), in the sense that the shape and velocity fluctuations will be function of time and some $p$ spacial coordinates. And also the velocity fluctuations will be only along time and the same $p$ directions. By considering this setup, we then reduce our membrane equations to a pair of nonlinear effective equations, which we then match with those of \cite{EmparanHydro}.

 We rewrite the background metric \eqref{ppmet} in the following form
 \begin{equation}\label{ppmet2}
   ds^2_{(bgd)} = g_{AB} dX^AdX^B = -r^2 dt^2 + \frac{dr^2}{r^2} + r^2 (dx^adx_a + dx^i dx_i)
  \end{equation}
Where the indices $a,b$ now take $p$ number of values. We will only consider fluctuations in these directions as mentioned above. The indices $i,j$ take $D-p-2$ number of values.

From the analysis in the previous subsection we could see that if we want the frequency along the time direction to be of order ${\cal O}(1)$, then the spatial frequency $k$ has to be very high, of the order of ${\cal O}\left(\sqrt{D}\right)$.
To zoom into this regime, in this subsection  we work with the scaled spatial coordinates $x^a\rightarrow y^a =\sqrt{D}~ x^a$. The spatial frequencies in the new coordinate will scale as 
$\tilde k_a = \frac{k_a}{\sqrt{D}}$. Therefore in the regime of interest the frequency along the new space coordinates will be of order one $\tilde k^a\sim{\cal O}(1)$.
So, the AdS Poincare patch metric \eqref{ppmet2} takes the form
\begin{equation}\label{pps}
 ds^2= -r^2 dt^2 + \frac{dr^2}{r^2} + r^2 (\frac{dy^ady_a }{D}+ dx^i dx_i)
\end{equation}

Now suppose we repeat the linearized analysis of the previous subsection in these new coordinates. 
 We shall normalize the fluctuations such that 
the components of the velocity vector field in the directions of $\partial_{y^a}$ are of order ${\cal O }(1)$. It follows that $\left[u\cdot dy^a\right] \sim {\cal O}\left(\frac{1}{D}\right)$. In old $x^a$ coordinate we already know the solution of $\left[u\cdot dx^a\right]$ (see equation \eqref{gsu} and the second equation of \eqref{gensol}). Solution in new coordinates will simply be the coordinate transform of the old solution.
In other words if we expand the velocity field as
$$u = u_0~ dt + \left(\frac{1}{D}\right) u_1(t,y^b) ~dt + \left(\frac{1}{D}\right)U_a(t,y^b) ~dy^a $$
then it follows that
$$\left(\frac{1}{D}\right)U_a =\frac{1}{\sqrt{D}}\left( \delta r_1^0~ V^1_a e^{-i\omega_1 t} e^{i\tilde k_a y^a} +\delta r_2^0~ V^2_a e^{-i\omega_2 t} e^{i\tilde k_a y^a} + v_a e^{-i\omega_v t} e^{i\tilde k_a y^a} \right)$$
In the RHS of the previous equation the extra factor of $\left(\frac{1}{\sqrt{D}}\right)$ comes as a consequence of coordinate transformation. From equation \eqref{eq:paralin} we could see $V^1_a$ and $V^2_a$ are of order ${\cal O}\left(\sqrt{D}\right)$. It follows that in these coordinates if we normalize our fluctuations such that $u\cdot\partial_{y^a}$ is of order ${\cal O}(1)$, then for consistency we must have $\delta r_1^0$ and $\delta r_2^0$ and therefore $\delta r$ to be of order ${\cal O}\left(\frac{1}{D}\right)$. This further implies that $ u_1\sim {\cal O}(1)$ (see equation \eqref{eq:utexp}).\\
Now we shall scale the radial coordinate so that in the new coordinate the amplitude of the radial perturbation is of order ${\cal O}(1)$.
\begin{equation}
 r=1+\frac{\rho}{D}
\end{equation}
With this coordinate redefinition 
the metric \eqref{pps} becomes 
\begin{equation}\label{ppsc}
 ds^2_{(bgd)} =g_{AB}dX^A dX^B = -\left( 1+\frac{\rho}{D} \right)^2 dt^2 + \frac{d\rho^2}{D^2\left( 1+\frac{\rho}{D} \right)^2} + \left( 1+\frac{\rho}{D} \right)^2 (\frac{dy^ady_a }{D}+ dx^i dx_i)
\end{equation}
%
%
%
In this scaled coordinate system the fluctuations are of the form
\begin{equation}\label{flu}
  \rho = Y(t,y^b), \quad u = u_0 dt + \frac{1}{D} u_1(t,y^b) dt + \frac{1}{D}U_a(t,y^b) dy^a 
\end{equation}
where $Y$, $u_1$ and $U_a$ are all of order ${\cal O}(1)$ in terms of $\left(1/D\right)$ expansion.
 Now we shall substitute these fluctuations in our `membrane equation' and evaluate it at very leading order in $\left(1/D\right)$ expansion. However now we shall  not consider any linearization with respect to the amplitude of the fluctuations.\\
 The procedure for evaluating the equation of motion is very similar as in the previous two subsections. So we shall be very brief here.

The metric induced on the membrane worldvolume is given by
\begin{equation}\label{ppindnl}
\begin{split}
 ds^2_{(ind)}&=g_{\mu\nu}^{(ind)}dz^{\mu}dz^{\nu}\\
 &= -\left( 1+\frac{Y}{D} \right)^2 dt^2  + \left( 1+\frac{Y}{D} \right)^2 (\frac{dy^ady_a }{D}+ dx^i dx_i)\\
 & \text{where } z^{\mu}\equiv \text{Coordinates along the membrane}\equiv\{t,y^a,x^i\}
\end{split}
\end{equation}
We use the notation $\tilde{\nabla}$ for covariant derivative with respect metric \eqref{ppindnl} and $\bar{\nabla}$ for covariant derivative constructed from metric \eqref{ppsc}.
In this notation the membrane equation is given by
\begin{equation}\label{membrane2}
\begin{split}
&{\cal{P}}^{\nu}_\mu\left\{\frac{\tilde{\nabla}^2 u_{\nu}}{\cal{K}}-\frac{\tilde{\nabla}_{\nu}{\cal{K}}}{\cal{K}}+u^{\alpha}{\cal K}_{\alpha\nu}-(u\cdot\tilde{\nabla})u_{\nu}\right\}=0~~~~\text{and}~~~~\tilde{\nabla}\cdot u=0\\
&\text{where}~~~{\cal{P}}_{\mu\nu}=g_{\mu\nu}^{(ind)}+u_{\mu}u_{\nu}
\end{split}
\end{equation}
 As before ${\cal K}_{\mu\nu}$ is given by the pullback of the extrinsic curvature tensor $K_{MN}$ expressed as a tensor in the full background metric.
  \begin{equation}\label{eq:nnex}
 \begin{split}
& {\cal K}_{\mu\nu} = \left( \frac{\partial X^M}{\partial z^\mu} \right) \left( \frac{\partial X^N}{\partial z^\nu} \right) K_{MN} \\
\text{where}&~K_{AB}=\Pi_{AC} \bar{\nabla}_C n_B\\
\text{where } \Pi_{AC}&=g_{AC} -n_A n_C
\end{split}
 \end{equation} 
 Here we have denoted the set $(\rho,t,a,i)$ by $X^M$ and the set $(t,a,i)$  by $z^\mu$.
 Now from explicit calculation we see that $K_{\rho N}\sim\mathcal{O}(D^{-2})$ (see appendix \ref{BB Calcula}).  Hence here also ${\cal K}_{\mu\nu}$ is just the `truncation' of the $K_{MN}$ evaluated on the membrane surface and the nonzero components are given by
  \begin{equation}
 \begin{split}
 {\cal K}_{tt} &= -\frac{\partial_t^2Y}{D} -\left( 1+\frac{2Y}{D} - \frac{\partial_aY\partial^aY}{2D} \right), \quad
 {\cal K}_{ta} = - \frac{\partial_t\partial_aY}{D} \\
 {\cal K}_{ab} &=  -\frac{\partial_a\partial_bY}{D}+\frac{\delta_{ab}}{D}, \quad  {\cal K}_{ij} = \delta_{ij}\left( 1+\frac{2Y}{D} - \frac{\partial_aY\partial^aY}{2D} \right)\\
  \mathcal{K} &= -\partial^b\partial_bY + (D-1) - \frac{\partial_a Y \partial^a Y}{2}
 \end{split}
 \end{equation}
 where the rest of the components are zero.

%
%
%
Normalization fixes the time component of the velocity field in terms of its space-component.
 \begin{equation}
  u_t = u_0 +\frac{u_1}{D} = -\left(1+\frac{Y}{D}+ \frac{U^aU_a}{2D}\right)
 \end{equation}
 The answer for the membrane projector $\mathcal{P}^\mu_\nu = \delta^\mu_\nu+u^{\mu} u_{\nu}$ is given by
 \begin{equation}\label{eq:effproj}
\begin{split}
\mathcal{P}^t_t &= -\frac{U^aU_a}{D}, \quad \mathcal{P}^t_a = \frac{U_a}{D}, \quad \mathcal{P}^a_t = -U^a\left( 1+\frac{Y}{D} \right)^{-1} -\frac{U^aU^bU_b}{2D}, \\
\mathcal{P}^a_b &= \delta^a_b -\frac{U_aU_b}{D}, \quad \mathcal{P}^i_j = \delta^i_j, \quad  \mathcal{P}^i_t = \mathcal{P}^t_i =\mathcal{O}(D^{-2}), \quad  \mathcal{P}^i_a =\mathcal{P}^a_i = \mathcal{O}(D^{-2})
\end{split}
\end{equation}
  Now, we denote the vector membrane equation (the 1st equation in \eqref{membrane2}) as
 \begin{equation*}
 \begin{split}
& E^{tot}_{\mu} \equiv \mathcal{P}^\nu_\mu E_{\nu}\\
\text{where}~~~&E_\mu \equiv \frac{\tilde{\nabla}^2u_\mu}{\mathcal{K}}-\frac{\tilde{\nabla}_\mu {\mathcal K}}{\mathcal{K}}+u^\nu K_{\nu\mu}-u^\nu \tilde{\nabla}_{\nu}u_{\mu} 
\end{split}
\end{equation*}
Note that $E_i$ and $E^{tot}_i$ won't contribute in the leading order analysis as both the background and the fluctuations satisfy translational symmetry along $x^i$ directions. Also we know from explicit computation that the leading terms in $E_a$ are of order ${\cal O}\left(\frac{1}{D}\right)$. Thus it is easy to see that we only need to evaluate the quantities $E_t$ to order $\mathcal{O}(1)$. 

At leading order it is easy to see that $E_t^{tot} = -U^aE_a^{tot}$. Thus the only independent components of the membrane equation are along $y^a$ directions. 
%
The relevant terms are given by (keeping terms of leading order in $1/D$)
\begin{equation}
 \frac{\tilde{\nabla}_t\mathcal{K}}{\mathcal{K}} = \mathcal{O}(D^{-1}), \quad \frac{\tilde{\nabla}^2u_t}{\mathcal{K}} = \mathcal{O}(D^{-1}), \quad (u\cdot{\cal K})_t = -1, \quad u.\tilde{\nabla} u_t = \mathcal{O}(D^{-1})
\end{equation}
 \begin{equation}
  \begin{split}
   \frac{\tilde{\nabla}_a\mathcal{K}}{\mathcal{K}} &= -\frac{\partial_a \partial^b\partial_bY}{D}- \frac{\partial^bY\partial_a\partial_bY}{D}, \quad \frac{\tilde{\nabla}^2u_a}{\mathcal{K}} = \frac{\partial^b\partial_b U_a}{D} + \frac{\partial^b Y \partial_b U_a}{D}, \\ (u\cdot{\cal K})_a &= -\frac{\partial_t \partial_a Y}{D}- \frac{U^b\partial_b\partial_a Y}{D}+ \frac{U_a}{D}, \quad u.\tilde{\nabla} u_a = \frac{\partial_t U_a}{D} + \frac{\partial_a Y}{D}+ \frac{U^b\partial_b U_a}{D}
  \end{split}
 \end{equation}
Thus we now can evaluate $E^{tot}_a$ and is given by
 \begin{equation}\label{mmef}
  \begin{split}
  &E^v_a \equiv \partial^b\partial_b U_a-\partial_a Y-U^b\partial_b U_a+\partial^b Y \partial_b U_a-U^b\partial_b\partial_a Y+\partial^bY\partial_a\partial_bY \\ &+\partial_a \partial^b\partial_b Y-\partial_t U_a-\partial_t \partial_a Y = 0
  \end{split}
 \end{equation}
 Also the equation $\tilde{\nabla}.u$ evaluates to
 \begin{equation}\label{duef}
  E^s \equiv \partial_bU^b + \partial_t Y + U_b\partial^b Y = 0
 \end{equation}
 Equations \eqref{mmef} and \eqref{duef} are the two effective equations for the membrane variables (membrane's shape and the velocity field on it) at leading order. Now we shall perform a variable redefinition which will recast these equation exactly as given in \cite{EmparanHydro}.
Consider the change of variables of the form
  \begin{equation}
  \begin{split}
   &Y(t,a)= \log{m(t,a)} \\
   &U_a(t,a) = \frac{p_a(t,a)-\partial_a m(t,a)}{m(t,a)}
  \end{split}
 \end{equation}
After performing this change in the effective equations \eqref{mmef} and \eqref{duef}, it can be shown that the linear combinations

\begin{equation}
E_t \equiv m(t,a)E^s \quad E_a \equiv p_a(t,a)E^s-m(t,a)E^v_a
\end{equation}

of the equations \eqref{mmef} and \eqref{duef} are just the effective equations derived in \cite{EmparanHydro} which are (in the notations used in this paper)
  
 \begin{equation}
  \begin{split}
   &\partial_t m - \partial^b\partial_b m + \partial_a p^a = 0 \\
   &\partial_t p_a - \partial^b\partial_b p_a + \partial_a m + \partial^b \left( \frac{p_ap_b}{m} \right) = 0
  \end{split}
 \end{equation}

 Thus we see from this procedure that the effective equations derived in \cite{EmparanHydro} are just a particular scaling limit of our membrane equations.  Note that our leading order membrane equations were derived in systematic expansion in $1/D$. But in this section we have rescaled our membrane equation to consider the length scales of order $\frac{1}{\sqrt{D}}$. Strictly speaking, this is beyond the regime of validity of our membrane equations. But it can be argued that this scaling is a consistent thing to do, in particular, the most general terms that can be written as subleading corrections to our membrane equations do not get so much enhanced so as they become comparable to the leading order terms in the membrane equations. The argument goes just like given in the Discussion of \cite{yogesh2} for black p-brane in flat spacetime and we won't rewrite here.


%

\section{Conclusion and Future Directions}\label{sec:conclude}
In this paper, we have used `large $D$' techniques to find new dynamical `black hole' solution to pure Einstein equations in presence of cosmological constant. The solutions are determined in an expansion in $\left(\frac{1}{D}\right)$ and are in `one-to-one' correspondence with a dynamical membrane (characterized by its shape and a velocity field on it) embedded in the asymptotic geometry (which could be AdS or dS). \\
The method we have used is manifestly covariant with respect to this asymptotic geometry (which we have referred to as `background'). We do not need to choose any coordinate system for the background geometry at any point of our derivation. The same calculation works  for both global AdS and Poincare patch. The form of the final answer also remains invariant. However,  they are different solutions with different asymptotic geometries and horizon topologies and this fact is encoded in the various covariant derivatives that appear in the final solution. These covariant derivatives are always defined with respect to the background.\\
We have applied this method to calculate the metric and the governing equation for the dual dynamical membrane upto the first subleading correction. Then we have performed several checks for our universal coordinate independent answer, by specializing to different coordinate systems. 
\begin{itemize}
\item We matched them against the known exact and static solutions -  Schwarzschild black hole/brane and Myers-Perry black holes for both asymptotically AdS and dS spaces.
\item We have linearized our membrane equations and matched them against the known spectrum of black hole/brane QNMs in AdS space and black hole QNMs in dS space.
\item We have taken a special scaling limit of our equations and recovered the dual effective hydrodynamic equations that was determined in \cite{EmparanHydro} for the AdS black-branes in large number of dimensions.
\end{itemize}
One immediate future direction would be to extend this calculation to the second subleading order. As described in \cite{radiation}, this is the order where we expect the leading entropy production for dynamical black holes. Also it would be easy  (though tedious)  to generalize this analysis to Einstein-Maxwell system in presence of cosmological constant. It would be interesting to calculate the stress tensor and the entropy current for this system following the method developed in \cite{radiation}.\\
For AdS space we know there exists another perturbative technique to construct new gravity solution that are dual to fluid dynamics in one lower dimension \cite{nonlinfluid}. This duality works in any dimension \cite{arbitrarydim} and therefore in particular large number dimensions where we can also apply $\left(\frac{1}{D}\right)$ expansion. It would be very interesting to explore how these two perturbative techniques could be patched together.

\section*{Acknowledgement}

It is a great pleasure to thank Shiraz Minwalla for initiating discussions on this topic and for his numerous suggestions throughout the course of this work.  We would also like to thank Arunabha Saha for collaboration at the initial stage.\\
We would  like to thank Amos Yarom, R. Loganayagam, Suman Kundu, Poulomi Nandi, Jyotirmoy Bhattacharya, Amitabh Virmani, Milan Patra for illuminating discussions. 
B.C. would like to acknowledge the hospitality of ICTP Trieste, IIT Kanpur, IIT Gandhinagar, TIFR Mumbai, IISER-Pune and ICTS-TIFR Bengaluru while this work was in progress. 
Y.D. would like to thank ICTP-Trieste, IISER-Pune, IOP-Bhubaneswar and especially IIT-Kanpur for hospitality while this work was in progress. P.B. would like to acknowledge the hospitality of IISER-Pune while this work was in progress.
 The work of S.B. was supported by an India Israel (ISF/UGC) joint research grant (UGC/PHY/2014236). The work of BC is partly supported by the D.S.T.-Max Planck Partner Group
 ``Quantum Black Holes'' between I.O.P. Bhubaneswar and A.E.I. Golm. The work of Y.D. was supported by the Infosys Endowment for the study of the Quantum Structure of Spacetime, as well as an Indo Israel (UGC/ISF) grant.  We  would  also like to acknowledge our debt to the people of India for their steady and generous support to research in the basic sciences.

\appendix
\section{Calculation of the homogeneous part -  $H_{AB}$}\label{app:homo}
In this section we shall give details of the computation for \eqref{eq:homoscalar}, \eqref{eq:homovector}, \eqref{eq:homotensor}, \eqref{eq:homoTrace} and their decoupled form as described in equations \eqref{eq:homot}, \eqref{eq:homov}, \eqref{eq:homof} and \eqref{eq:homoh}.
As mentioned before, we can determine the metric upto ${\cal O}\left(\frac{1}{D}\right)$  by solving the gravity equation \eqref{eq:eom2} upto order ${\cal O}\left(D\right)$. At this order $G^{(1)}_{AB}$ contributes simply as a linear fluctuation over the zeroth order metric $G^{[0]}_{AB} = g_{AB} + G^{(0)}_{AB}$. So here we shall first compute the form of the gravity equation \eqref{eq:eom2}, linearized about $G^{[0]}_{AB}$.\\
Let us denote the perturbed metric as 
$${\mathfrak g}_{AB} = G^{[0]}_{AB} + \frac{1}{D} G^{(1)}_{AB} =g_{AB} + \psi^{-D} O_A O_B +  \frac{1}{D} G^{(1)}_{AB}$$
Also, as it is clear from our discussion, in this linearized calculation we need to compute only the leading $D$ piece.\\
The linearized variation of the Christoffel symbols and the Ricci Tensor take the form
\begin{equation}\label{eq:deltaRab}
\begin{split}
&\delta\Gamma^A_{BC} =\frac{1}{D}\left( \frac{g^{AM} -\psi^{-D} O^A O^M}{2}\right)\left(D_B~ G^{(1)}_{MC} +D_C~ G^{(1)}_{MB}-D_M~ G^{(1)}_{BC}\right)\\
&\delta R_{AB} =\left( D_C ~\delta\Gamma^C_{AB}- D_B ~\delta\Gamma^C_{A C}\right)
\end{split}
\end{equation}

In equation \eqref{eq:deltaRab}, $D_A$ denotes the covariant derivative w.r.t $G^{[0]}_{AB}$. Now we  can easily convert  $D_A$ to $\nabla_A$ (i.e. the covariant derivative w.r.t $g_{AB}$) by introducing some new terms to account for the correction to Christoffel symbols generated from the  extra piece $(\psi^{-D} O_A O_B)$ .

\begin{equation}\label{eq:deltaRabSimp}
\begin{split}
&\delta R_{AB} =\underbrace{\nabla_C ~\delta\Gamma^C_{AB}}_{\text{Term}1}- \underbrace{\nabla_B ~\delta\Gamma^C_{A C}}_{\text{Term}2}-\underbrace{\left(\tilde\Gamma^M_{CA}\delta\Gamma^C_{MB}+\tilde\Gamma^M_{CB}\delta\Gamma^C_{MA}
\right)}_{\text{Term}3}+\underbrace{\tilde\Gamma^M_{AB}\delta\Gamma^C_{MC}}_{\text{Term}4}\\
&\text{where}\\
&\delta\Gamma^A_{BC} =\frac{1}{D}\left(\frac{ g^{AM} -\psi^{-D} O^A O^M}{2}\right)\bigg(\nabla_B~ G^{(1)}_{MC} +\nabla_C~ G^{(1)}_{MB}\\
&~~~~~~~~~~~~~~~~~~~~~~~~~~~~~~~~~~~~~~~~~~~~~~~~~~-\nabla_M~ G^{(1)}_{BC}
-2\tilde\Gamma^{M'}_{BC}~ G^{(1)}_{MM'}\bigg)\\
&\tilde\Gamma^A_{BC}=-\psi^{-D}\left(\frac{DN}{2\psi}\right)\left[O^A\left(n_B O_C + n_C O_B\right) - n^AO_B O_C + \psi^{-D} O^AO_BO_C\right]
\end{split}
\end{equation}
%


\subsection{Scalar sector}
In this subsection we shall compute $H_{AB}^{scalar}$. The relevant part of $\delta G_{AB}$ has the following form.
\begin{equation}\label{eq:ScalarPert}
\begin{split}
& G^{(1)}_{AB}|_{scalar} =O_A O_B\sum_{n} f_n(R)~ {\mathfrak s}_n\\
&R=D(\psi-1)
\end{split}
\end{equation}
To compute $H_{AB}^{scalar}$ we have to substitute equation \eqref{eq:deltaRabSimp} in \eqref{eq:ScalarPert} and compute only the leading $D$ piece.
\begin{equation}\label{eq:deltagmScalar}
\begin{split}
\delta\Gamma^C_{AB} =&~ \left(\frac{N}{2}\right)\sum_n \bigg[\left(O^Cn_AO_B + O^C n_B O_A -n^CO_A O_B\right)f_n'\\
&~~~~~~~~~~~~~~~~~~~ +(\psi^{-D} O^CO_AO_B)\left(f_n'-f_n\right)\bigg]{\mathfrak s}_n
+\text{Subleading terms}\\
\delta\Gamma^C_{AC} =&~ 0
\end{split}
\end{equation}
Now we shall substitute equation \eqref{eq:deltagmScalar} in each of the four terms in equation \eqref{eq:deltaRabSimp}.
\begin{equation}\label{eq:TermScalar}
\begin{split}
\text{Term}1 =&~\nabla_C ~\delta\Gamma^C_{AB}\\
=&~ \left(\frac{D N^2}{2}\right)\sum_n{\mathfrak s}_n\bigg[\left(f''_n+ f'_n\right)\left(n_B O_A + n_A O_B -O_BO_A\right)\\ 
&~~~~~~~~~~~~~~~~~~~~~~~+\psi^{-D}\left(f''_n - f'_n\right)O_BO_A\bigg]
+\text{Subleading terms}\\
\text{Term}2 =&~\nabla_B ~\delta\Gamma^C_{AC}
=0\\
\\
\text{Term}3 =&~\left(\tilde\Gamma^M_{CA}\delta\Gamma^C_{MB}+\tilde\Gamma^M_{CB}\delta\Gamma^C_{MA}
\right)\\
=&~-\left(D N^2\right)\psi^{-D}\sum_n\bigg[f'_n~O_B O_A\bigg]{\mathfrak s}_n+\text{Subleading terms}\\
\text{Term}4 =&~\tilde\Gamma^M_{AB}\delta\Gamma^C_{MC}=0
\end{split}
\end{equation}

So finally
\begin{equation}\label{eq:finHabScalar}
\begin{split}
H^{scalar}_{AB} &= \text{Term}1 -\text{Term}3\\
&=\left(\frac{D N^2}{2}\right)\sum_n{\mathfrak s}_n\left(f''_n+ f'_n\right)\bigg[n_B O_A + n_A O_B -\left(1-\psi^{-D}\right)O_BO_A\bigg]
\end{split}
\end{equation}

\subsection{Vector sector}
In this subsection we shall compute $H_{AB}^{vector}$. The relevant part of $\delta G_{AB}$ has the following form.
\begin{equation}\label{eq:VectorPert}
\begin{split}
 G^{(1)}_{AB}|_{vector} = \sum_n v_n(R)\bigg([{\mathfrak v}_n]_A O_B +[{\mathfrak v}_n]_B O_A\bigg),~~~R \equiv D(\psi-1)
\end{split}
\end{equation}
Now we shall substitute equation \eqref{eq:deltagmVector} in each of the four terms in equation \eqref{eq:deltaRabSimp}.
\begin{equation}\label{eq:deltagmVector}
\begin{split}
\delta\Gamma^A_{BC} =&~ \left(\frac{N}{2}\right)\sum_n \bigg\{ O^A\left(n_B [{\mathfrak v}_n]_C +n_C [{\mathfrak v}_n]_B\right)v_n'\\
&~~~~~~~~~~~~~~~- \left(n^A-\psi^{-D}O^A\right)\left(O_B [{\mathfrak v}_n]_C +O_C [{\mathfrak v}_n]_B\right)v_n'\\
&~~~~~~~~~~~~~~~+\left[v_n' (n_B O_C + n_C O_B) - v_n\left(\psi^{-D} O_B O_C\right)\right] [{\mathfrak v}_n]^A\bigg\}\\
&~~~~~~~~~~+\text{Subleading terms}\\
\delta\Gamma^C_{AC} =&~ 0\\
\end{split}
\end{equation}

\begin{equation}\label{eq:TermVector}
\begin{split}
\text{Term}1 =&~\nabla_C ~\delta\Gamma^C_{AB}\\
=&~ \left(\frac{D N^2}{2}\right)\sum_n\bigg(u_B\left[{\mathfrak v}_n\right]_A+u_A\left[{\mathfrak v}_n\right]_B\bigg)\left(v''_n+ v'_n\right)\\
&+ \left(\frac{D N^2}{2}\right)\sum_n\psi^{-D}\bigg(O_B\left[{\mathfrak v}_n\right]_A+O_A\left[{\mathfrak v}_n\right]_B\bigg)v''_n\\
&+\left(\frac{N}{2}\right)\sum_n \left(\nabla\cdot{\mathfrak v}_n\right)\bigg[v_n'\left(n_AO_B + n_B O_A\right) -\psi^{-D}v_nO_BO_A\bigg]\\
&+\text{Subleading terms}\\
\\
\text{Term}2 =&~\nabla_B ~\delta\Gamma^C_{AC}
=0\\
\\
\text{Term}3 =&~\left(\tilde\Gamma^M_{CA}\delta\Gamma^C_{MB}+\tilde\Gamma^M_{CB}\delta\Gamma^C_{MA}
\right)\\
=&~-\left(\frac{D N^2}{2}\right)\psi^{-D}\sum_nv_n'\bigg(O_B\left[{\mathfrak v}_n\right]_A+O_A\left[{\mathfrak v}_n\right]_B\bigg)+\text{Subleading terms}\\
\\
\text{Term}4 =&~\tilde\Gamma^M_{AB}\delta\Gamma^C_{MC}=0
\end{split}
\end{equation}

So finally
\begin{equation}\label{eq:finHabVector}
\begin{split}
H^{vector}_{AB} &= \text{Term}1 -\text{Term}3\\
&=
\left(\frac{N}{2}\right)\sum_n \left(\nabla\cdot{\mathfrak v}_n\right)\bigg[v_n'\left(n_AO_B + n_B O_A\right) -\psi^{-D}v_nO_BO_A\bigg]\\
&~+\left(\frac{D N^2}{2}\right)\sum_n\left(v''_n+ v'_n\right)\bigg\{\bigg(u_B\left[{\mathfrak v}_n\right]_A+u_A\left[{\mathfrak v}_n\right]_B\bigg)\\
&~~~~~~~~~~~~~~~~~~~~~~~~~~~~~~~~~~~~+\psi^{-D}\bigg(O_B\left[{\mathfrak v}_n\right]_A+O_A\left[{\mathfrak v}_n\right]_B\bigg)\bigg\}\\
\end{split}
\end{equation}

\subsection{Tensor sector}
In this subsection we shall compute $H_{AB}^{tensor}$. The relevant part of $\delta G_{AB}$ has the following form.
\begin{equation}\label{eq:TensorPert}
\begin{split}
 G^{(1)}_{AB}|_{tensor} = \sum_n t_n(R)~[{\mathfrak t}_n]_{AB},~~~R \equiv D(\psi-1)
\end{split}
\end{equation}

\begin{equation}\label{eq:deltagmTensor}
\begin{split}
\delta\Gamma^A_{BC} =&~ \left(\frac{N}{2}\right)\sum_n t_n'\bigg\{ \left[{\mathfrak t}_n\right]^A_C n_B+\left[{\mathfrak t}_n\right]^A_B n_C-\left(n^A-\psi^{-D} O^A\right)\left[\mathfrak t_n\right]_{BC}\bigg\}\\
&~~~~~~~~~~~+\text{Subleading terms}\\
\delta\Gamma^C_{AC} =&~ 0\\
\end{split}
\end{equation}
Now we shall substitute equation \eqref{eq:deltagmTensor} in each of the four terms in equation \eqref{eq:deltaRabSimp}.
\begin{equation}\label{eq:TermTensor}
\begin{split}
\text{Term}1 =&~\nabla_C ~\delta\Gamma^C_{AB}\\
=&~- \left(\frac{D N^2}{2}\right)\sum_n\left[t_n''(1-\psi^{-D}) + t_n'\right]\left[{\mathfrak t}_n\right]_{AB}\\
&+\left(\frac{N}{2}\right)\sum_n t_n'\bigg( n_B~(\nabla_C\left[{\mathfrak t}_n\right]^C_A)+n_A~(\nabla_C\left[{\mathfrak t}_n\right]^C_B)\bigg)\\
&+\text{Subleading terms}\\
\\
\text{Term}2 =&~\nabla_B ~\delta\Gamma^C_{AC}
=0\\
\text{Term}3 =&~\left(\tilde\Gamma^M_{CA}\delta\Gamma^C_{MB}+\tilde\Gamma^M_{CB}\delta\Gamma^C_{MA}
\right)=\text{Subleading terms}\\
\text{Term}4 =&~\tilde\Gamma^M_{AB}\delta\Gamma^C_{MC}=0
\end{split}
\end{equation}

So finally
\begin{equation}\label{eq:finHabTensor}
\begin{split}
H^{tensor}_{AB} &= \text{Term}1 \\
=&~- \left(\frac{D N^2}{2}\right)\sum_n\left[t_n''(1-\psi^{-D}) + t_n'\right]\left[{\mathfrak t}_n\right]_{AB}\\
&+\left(\frac{N}{2}\right)\sum_n t_n'\bigg( n_B~(\nabla_C\left[{\mathfrak t}_n\right]^C_A)+n_A~(\nabla_C\left[{\mathfrak t}_n\right]^C_B)\bigg)\\
\end{split}
\end{equation}

\subsection{Trace sector}
In this subsection we shall compute $H_{AB}^{trace}$. The relevant part of $\delta G_{AB}$ has the following form.
\begin{equation}\label{eq:TracePert}
\begin{split}
G^{(1)}_{AB}|_{trace} =\left(\frac{1}{D}\right) P_{AB}\sum_n h_n(R) {\mathfrak s}_n,~~~R \equiv D(\psi-1)
\end{split}
\end{equation}
As explained in section (\ref{sec:strategy}), we have an extra factor of $\left(\frac{1}{D}\right) $ compared to the expressions of $\delta G_{AB}$ in tensor, vector and the scalar sector. 

\begin{equation}\label{eq:deltagmTrace}
\begin{split}
\delta\Gamma^C_{AB} =&~ \left(\frac{N}{2D}\right)\sum_n h'_n \bigg[\left(n_AP^C_B + n_BP^C_A - n^CP_{AB}\right)
 +\left(\psi^{-D} O^CP_{AB}\right)\bigg]{\mathfrak s}_n\\
~&+\text{Subleading terms}\\
\delta\Gamma^C_{AC} =&~ \left(\frac{N}{2} \right)n_A\sum_nh'_n~{\mathfrak s}_n+\text{Subleading terms}\\
\end{split}
\end{equation}
Now we shall substitute equation \eqref{eq:deltagmTrace} in each of the four terms in equation \eqref{eq:deltaRabSimp}.
\begin{equation}\label{eq:TermTrace}
\begin{split}
\text{Term}1 =&~\nabla_C ~\delta\Gamma^C_{AB}\\
=&~- \left(\frac{N^2}{2}\right)\sum_n{\mathfrak s}_n\bigg\{\left[\left(1-\psi^{-D}\right)h_n'' + h_n'\right]P_{AB} + 2h'_n~ n_A n_B\bigg\}\\
&~~+\text{Subleading terms}\\
\\
\text{Term}2 =&~\nabla_B ~\delta\Gamma^C_{AC}\\
=&~\left(\frac{D N^2}{2}\right)\sum_n{\mathfrak s}_n\bigg[h_n''~n_A n_B\bigg]+\text{Subleading terms}\\
\\
\text{Term}3 =&~\left(\tilde\Gamma^M_{CA}\delta\Gamma^C_{MB}+\tilde\Gamma^M_{CB}\delta\Gamma^C_{MA}
\right)=\text{Subleading terms}\\
\\
\text{Term}4 =&~\tilde\Gamma^M_{AB}\delta\Gamma^C_{MC}\\
=&-\left(\frac{D N^2}{4}\right)\psi^{-D}\sum_n{\mathfrak s}_nh_n'\bigg[n_B O_A + n_A O_B -\left(1-\psi^{-D}\right)O_BO_A\bigg]\\
&~~+\text{Subleading terms}\\
\end{split}
\end{equation}

So finally
\begin{equation}
\begin{split}
H^{trace}_{AB} &= \text{Term}1 -\text{Term}2 +\text{Term}4\\
&=-\left(\frac{D N^2}{4}\right)\sum_n{\mathfrak s}_n\bigg\{2 h''_n ~n_A n_B +h'_n \left[n_B O_A +n_A O_B -\left(1-\psi^{-D}\right)O_BO_A\right]\psi^{-D}\bigg\}\\
&~~~-\left(\frac{ N^2}{2}\right)\sum_n{\mathfrak s}_n\bigg\{2 h'_n~n_A n_B +\left[h'_n +\left(1-\psi^{-D}\right)h''_n\right]P_{AB}\bigg\}+{\cal O}\left(\frac{1}{D}\right)\\
\end{split}
\end{equation}

Note that in the above equation, the second line is of order ${\cal O}(1)$. Since in our calculation we are only interested upto order ${\cal O}(D)$, we could ignore the second line. For our purpose
\begin{equation}
\begin{split}
H^{trace}_{AB} &= \text{Term}1 -\text{Term}2 +\text{Term}4\\
&=-\left(\frac{D N^2}{4}\right)\sum_n{\mathfrak s}_n\bigg\{2 h''_n ~n_A n_B +h'_n \left[n_B O_A +n_A O_B -\left(1-\psi^{-D}\right)O_BO_A\right]\psi^{-D}\bigg\}\\
\end{split}
\end{equation}
%

\section{Calculation of the sources - $S_{AB}$}\label{app:calsource}

In this section we shall give details of calculation of $S_{AB}$. As mentioned in subsection (\ref{subsec:source}) we have to evaluate ${\cal{E}}_{AB}$ on $G^{[0]}_{AB}$.
\begin{equation}\label{eq:sourceabs}
\begin{split}
{\cal{E}}_{AB}&=R_{AB}|_{G^{[0]}_{AB}}-(D-1)\lambda G^{[0]}_{AB}\\
&=\bar R_{AB}+\delta R_{AB}-(D-1)\lambda G^{[0]}_{AB}\\
&=(D-1)\lambda g_{AB} +\delta R_{AB}-(D-1)\lambda\left(g_{AB}+G^{(0)}_{AB}\right)\\
&=\delta R_{AB}-D\lambda~ G^{(0)}_{AB}+\text{Subleading Terms}
\end{split}
\end{equation}
Where $\bar R_{AB}$ is the Ricci tensor evaluated on the background metric $g_{AB}$ and 
$\delta R_{AB}$ is simply the difference between the Ricci tensor evaluated on $G^{[0]}_{AB}$ and Ricci tensor evaluated on $g_{AB}$.\\
Using this notation
\begin{equation}\label{source}
S_{AB}=\delta R_{AB}-D\lambda~ G^{(0)}_{AB}
\end{equation}
Now for our case,
\begin{equation}\label{Lead anstz}
\begin{split}
G_{AB}^{[0]}=g_{AB}+\psi^{-D} O_A O_B,~~~G_{AB}^{(0)} = \psi^{-D} O_A O_B
\end{split}
\end{equation}
As the one form field `O' is null, the inverse of the above metric \eqref{Lead anstz} becomes very simple.
\begin{equation}
\begin{split}
G^{[0] AB}=g^{AB}-\psi^{-D} O^A O^B
\end{split}
\end{equation}
Substituting lead ansatz in equation \eqref{eq:gamma2} we find
\begin{equation}\label{eq:deltagamma}
\begin{split}
\delta\Gamma^A_{BC}=\left[\frac{g^{AM}-\psi^{-D} O^A O^M}{2}\right]\bigg[\nabla_B\left(\psi^{-D} O_CO_M\right)&+\nabla_C \left(\psi^{-D} O_BO_M\right)\\
&-\nabla_M \left(\psi^{-D} O_CO_B\right)\bigg]
\end{split}
\end{equation}
Here $\nabla$ is covariant derivative with respect to the background metric $g_{AB}$.\\
For the convenience of computation we shall decompose $\delta\Gamma^A_{BC}$ in two parts
\begin{equation}\label{eq:gammadecomp}
\begin{split}
\delta\Gamma^A_{BC}=\delta\Gamma^A_{BC}|_{\text{lin.}}+\delta\Gamma^A_{BC}|_{\text{non-lin.}}
\end{split}
\end{equation}
where
\begin{equation}\label{eq:gammadecomp2}
\begin{split}
&\delta\Gamma^A_{BC}|_{\text{lin.}}=\frac{1}{2}\left\{\nabla_B(\psi^{-D}O_C O^A)+\nabla_C(\psi^{-D}O_B O^A)-\nabla^A(\psi^{-D}O_B O_C)\right\}\\
&\delta\Gamma^A_{BC}|_{\text{non-linear}}=\frac{1}{2}\psi^{-D}O^A(O\cdot \nabla)(\psi^{-D} O_B O_C)
\end{split}
\end{equation}
From \eqref{eq:covricci} we know that Ricci tensor can be written as
\begin{equation}
\begin{split}
R_{AB} &= \bar R_{AB} + \underbrace{\nabla_C\left[\delta\Gamma^C_{AB}\right] - \nabla_B\left[\delta\Gamma^C_{CA}\right] + \left[\delta\Gamma^C_{CE}\right]\left[\delta\Gamma^E_{AB}\right] - \left[\delta\Gamma^C_{BE}\right]\left[\delta\Gamma^E_{AC}\right]}_{\delta R_{AB}}
\end{split}
\end{equation}
From equation \eqref{eq:deltagamma} it follows
\begin{equation}\label{eq:stepsimp1}
\begin{split}
\delta\Gamma^C_{CA}&
=\frac{1}{2}\left\{\nabla_C(\psi^{-D} O_A O^C)+\nabla_A(\psi^{-D} O_C O^C)-\nabla^C(\psi^{-D} O_C O_A) \right\}\\
&~~~~~~~~~~~~+\frac{1}{2}\psi^{-D}O^C(O\cdot \nabla)(\psi^{-D}O_C O_A)\\
&=0
\end{split}
\end{equation}
The expression for $\delta R_{AB}$ simplifies once we substitute equation \eqref{eq:deltagamma}
\begin{equation}
\begin{split}
\delta R_{AB}&=\nabla_C\left[\delta\Gamma^C_{AB}\right]  - \left[\delta\Gamma^C_{BE}\right]\left[\delta\Gamma^E_{AC}\right]\\
&=\underbrace{\nabla_C\left[\delta\Gamma^C_{AB}|_{\text{linear}}\right]}_{\delta R_{AB}|{\text{linear}}} + \underbrace{\nabla_C\left[\delta\Gamma^C_{AB}|_{\text{non-linear}}\right]}_{\delta R^{(1)}_{AB}|{\text{non-linear}}}\underbrace{- \left[\delta\Gamma^C_{BE}\right]\left[\delta\Gamma^E_{AC}\right]}_{\delta R^{(2)}_{AB}|{\text{non-linear}}}
\end{split}
\end{equation}

At first we present the calculation of $\delta R^{(2)}_{AB}|_{\text{non-linear}}$
\begin{equation}
\begin{split}
\delta R^{(2)}_{AB}|_{\text{non-lin.}}&=\underbrace{-\left[\delta\Gamma^C_{BE}|_{\text{lin.}}\right]\left[\delta\Gamma^E_{AC}|_{\text{lin.}}\right]}_{\text{Term-1}}\underbrace{-\left[\delta\Gamma^C_{BE}|_{\text{lin.}}\right]\left[\delta\Gamma^E_{AC}|_{\text{non-lin.}}\right]}_{\text{Term-2}}\\
&\underbrace{-\left[\delta\Gamma^C_{BE}|_{\text{non-lin.}}\right]\left[\delta\Gamma^E_{AC}|_{\text{lin.}}\right]}_{\text{Term-3}}\underbrace{-\left[\delta\Gamma^C_{BE}|_{\text{non-lin.}}\right]\left[\delta\Gamma^E_{AC}|_{\text{non-lin.}}\right]}_{\text{Term-4}}
\end{split}
\end{equation}

\begin{equation}
\begin{split}
\text{Term-4}&\equiv-\left[\delta\Gamma^C_{BE}|_{\text{non-lin.}}\right]\left[\delta\Gamma^E_{AC}|_{\text{non-lin.}}\right]\\
&=-\left\{\frac{1}{2}\psi^{-D}O^C(O\cdot\nabla)(\psi^{-D}O_B O_E)\right\}\left\{\frac{1}{2}\psi^{-D}O^E(O\cdot\nabla)(\psi^{-D}O_C O_A)\right\}\\
&=0
\end{split}
\end{equation}

\begin{equation}
\begin{split}
\text{Term-3}&\equiv-\left[\delta\Gamma^C_{BE}|_{\text{non-lin.}}\right]\left[\delta\Gamma^E_{AC}|_{\text{lin.}}\right]\\
&=-\left\{\frac{1}{2}\psi^{-D}O^C (O\cdot\nabla)(\psi^{-D}O_B O_E)\right\}\\
&~~~~~~~~~~~~~~~~~~~\frac{1}{2}\bigg\{\nabla_C(\psi^{-D}O_A O^E)+\nabla_A(\psi^{-D}O_C O^E)-\nabla^E(\psi^{-D}O_C O_A)\bigg\}\\
&=-\frac{1}{4}\psi^{-D}\left\{(O\cdot \nabla)(\psi^{-D}O_B O_E)\right\}\left\{(O\cdot \nabla)(\psi^{-D}O_A O^E)\right\}\\
&=-\frac{1}{4}\psi^{-3D}\{(O\cdot\nabla)O_E\}\{(O\cdot\nabla)O^E\}O_B O_A\\
&=0
\end{split}
\end{equation}
In the last step we have used \eqref{eq:SubsidiaryO3}.\\
Similarly,
\begin{equation}
\text{Term-2}=0
\end{equation}
Now we shall compute Term-1, which is non-zero and a bit complicated.
\begin{equation}\label{eq:Term-1St1}
\begin{split}
\delta R^{(2)}_{AB}|_{\text{non-lin.}}=&-\left[\delta\Gamma^C_{BE}|_{\text{lin.}}\right]\left[\delta\Gamma^E_{AC}|_{\text{lin.}}\right]\\
=&-\frac{1}{2}\bigg\{\nabla_B(\psi^{-D}O_E O^C)+\nabla_E(\psi^{-D}O_B O^C)-\nabla^C(\psi^{-D}O_B O_E)\bigg\}\\
&~~~~~~~~~~~~~~~~~\frac{1}{2}\bigg\{\nabla_C(\psi^{-D}O_A O^E)+\nabla_A(\psi^{-D}O_C O^E)-\nabla^E(\psi^{-D}O_C O_A)\bigg\}\\
=&-\frac{1}{4}\bigg\{\cancel{\psi^{-2D}(\nabla_B O_E)(\nabla_C O^E)O^C O_A}-\cancel{\psi^{-2D}O_E(\nabla_B O^C)O_A(\nabla^E O_C)}\\
&+\nabla_E(\psi^{-D}O_B O^C)\nabla_C(\psi^{-D}O_A O^E)+\bcancel{\psi^{-2D}O_B(\nabla_E O^C)O^E(\nabla_A O_C)}\\[10pt]
&-\psi^{-2D}O_B(\nabla_E O^C)(\nabla^E O_C)O_A-\psi^{-2D}O_B(\nabla^C O_E)O_A(\nabla_C O^E)\\
&-\bcancel{\psi^{-2D}O_B(\nabla^C O_E)O_C(\nabla_A O^E)}+\nabla^C(\psi^{-D}O_B O_E)\nabla^E(\psi^{-D}O_C O_A)\bigg\}\\
=&~\frac{1}{2}\psi^{-2D}(\nabla_E O^C)(\nabla^E O_C)O_B O_A-\frac{1}{2}\nabla_E(\psi^{-D}O_B O^C)\nabla_C(\psi^{-D}O_A O^E)\\
=&-\frac{1}{2}~\left[(O\cdot\nabla)\left(\psi^{-D}O_B\right)\right]~\left[(O\cdot\nabla)\left(\psi^{-D}O_A\right)\right]\\
&+\psi^{-2D}\left(\frac{DN}{2\psi}\right)2\left[n^E (O\cdot\nabla) O_E\right]O_BO_A\\
&+\left(\frac{\psi^{-2D}}{2}\right)\left[\left(\nabla_E O_C\right)\left(\nabla^E O^C -\nabla^C O^E\right)\right]O_BO_A + {\cal O}(1)\\
\end{split}
\end{equation}
Now using the fact that
$$\left(\nabla_E O_C\right)\left(\nabla^E O^C\right)=\left(\nabla_E O_C\right)\left(\nabla^C O^E\right)=\frac{K^2}{D} + {\cal O}(1)$$
we finally find
\begin{equation}\label{eq:Term-1St2}
\begin{split}
&\delta R^{(2)}_{AB}|_{\text{non-lin.}}\\
=&-\frac{1}{2}~(O\cdot\nabla)\left[\psi^{-D}O_B\right]~~(O\cdot\nabla)\left[\psi^{-D}O_A\right]
+\psi^{-2D}\left(\frac{DN}{\psi}\right)\left[n^E (O\cdot\nabla) O_E\right]O_BO_A\\
=&-\frac{1}{2}~(O\cdot\nabla)\left[\psi^{-D}O_B\right]~~(O\cdot\nabla)\left[\psi^{-D}O_A\right]
+\psi^{-2D} K\left[u^C (O\cdot\nabla) n_C\right]O_BO_A + {\cal O}(1)\\
\end{split}
\end{equation}
In the last line we have used the fact that 
$$\left(\frac{DN}{\psi}\right) = K +{\cal O}(1)$$

Now we proceed to the calculation of $\delta R^{(1)}_{AB}|_{\text{non-lin.}}$
\begin{equation}\label{eq:RNL-1}
\begin{split}
\delta R^{(1)}_{AB}|_{\text{non-lin.}}
&=\nabla_C\left[\delta\Gamma^C_{AB}|_{\text{non-linear}}\right]
=\nabla_C\left\{\frac{1}{2}\psi^{-D}O^C(O\cdot\nabla)(\psi^{-D} O_B O_A)\right\}\\
&=\left(\frac{\psi^{-D}}{2}\right) \bigg[(\nabla\cdot O)~ (O\cdot\nabla)\left(\psi^{-D} O_BO_A\right)+O_A (O\cdot\nabla)\left[(O\cdot\nabla)(\psi^{-D}O_B)\right]\bigg] \\
&~+ \frac{1}{2}\left[(O\cdot\nabla)\left(\psi^{-D} O_A\right)\right]~\left[(O\cdot\nabla)\left(\psi^{-D} O_B\right)\right]
+ \frac{1}{2}(O\cdot\nabla)\left[\psi^{-2D}O_B(O\cdot\nabla)O_A\right]\\
\\
&=\left(\frac{\psi^{-2D}}{2}\right) \left(\frac{DN}{\psi} -\nabla\cdot O\right)\left[\left(\frac{DN}{\psi} \right)O_BO_A- (O\cdot\nabla)(O_BO_A)\right]\\
&~~~~~+ \frac{1}{2}\left[(O\cdot\nabla)\left(\psi^{-D} O_A\right)\right]~\left[(O\cdot\nabla)\left(\psi^{-D} O_B\right)\right]\\
&~~~~~-\left(\frac{\psi^{-2D}}{2}\right)\left(O\cdot\nabla\right) \left[\frac{DN}{\psi} O_A O_B\right] + {\cal O}(1)\\
\end{split}
\end{equation}
%
%
We can use identity\eqref{I0} to simplify \eqref{eq:RNL-1}
\begin{equation}\label{eq:RNL-1st2}
\begin{split}
&\delta R^{(1)}_{AB}|_{\text{non-lin.}}\\
&=\left(\frac{\psi^{-2D}}{2}\right) \left[(n\cdot\nabla){K}+{K}(\nabla\cdot u)\right] O_BO_A
-\left(\frac{\psi^{-2D}}{2}\right)\left(O\cdot\nabla\right) \left[{K}~ O_A O_B\right]\\
& + \frac{1}{2}\left[(O\cdot\nabla)\left(\psi^{-D} O_A\right)\right]~\left[(O\cdot\nabla)\left(\psi^{-D} O_B\right)\right]+ {\cal O}(1)\\
&=\left(\frac{\psi^{-2D}}{2}\right) \left[(u\cdot\nabla){K}+{K}(\nabla\cdot u)\right] O_BO_A
-\psi^{-2D}~K\left[u^C(O\cdot\nabla) n_C\right] O_AO_B\\
& + \frac{1}{2}\left[(O\cdot\nabla)\left(\psi^{-D} O_A\right)\right]~\left[(O\cdot\nabla)\left(\psi^{-D} O_B\right)\right]+ {\cal O}(1)
\end{split}
\end{equation}
In the last step we have used the subsidiary condition on $O_A$.
\begin{equation}
(O\cdot\nabla) O_A = \left[u^C(O\cdot\nabla) n_C\right] O_A
\end{equation}
Adding \eqref{eq:Term-1St2} and \eqref{eq:RNL-1st2} we get the desired expression for $\delta R_{AB}|_{\text{non-lin.}}$

\begin{equation}
\begin{split}
&\delta R_{AB}|_{\text{non-lin.}}
=\left(\frac{\psi^{-2D}}{2}\right) \left[(u\cdot\nabla){K}+{K}(\nabla\cdot u)\right] O_BO_A+{\cal O}(1)\\
\end{split}
\end{equation}

Finally, $\delta R_{AB}|_{\text{non-lin.}}$ becomes
\begin{equation}\label{eq:RNLAB}
\begin{split}
&\delta R_{AB}|_{\text{non-lin.}}\\
&=\left(\frac{\psi^{-2D}}{2}\right)\left[\frac{DN}{\psi}(\nabla\cdot u)+(u\cdot\nabla)\left(\frac{DN}{\psi}\right)\right]O_A O_B\\
&=\left(\frac{\psi^{-2D}}{2}\right)\left(\frac{DN}{\psi}(\tilde{\nabla}\cdot u)\right)~O_A O_B
\end{split}
\end{equation}
Where, $\tilde{\nabla}$ is defined as follows, for any general  tensor with $n$ indices $W_{A_1A_2\cdots A_n}$
\begin{equation}\label{tildedef}
\tilde{\nabla}_A W_{A_1A_2\cdots A_n}=\Pi_A^C~\Pi_{A_1}^{C_1}\Pi_{A_2}^{C_2}\cdots \Pi_{A_n}^{C_n}\left(\nabla_C W_{C_1C_2\cdots C_n}\right)
\end{equation}\\
Now, we shall calculate the linear terms in Ricci tensor
\begin{equation}
\begin{split}
\delta R_{AB}|_{\text{lin.}}&=\nabla_C\left[\delta\Gamma^C_{BA}|_{\text{lin.}}\right]\\
&=\underbrace{\frac{1}{2}\nabla_C\left\{\nabla_B\left(\psi^{-D}O_A O^C\right)\right\}}_{T_1}+\underbrace{\frac{1}{2}\nabla_C\left\{\nabla_A\left(\psi^{-D}O_B O^C\right)\right\}}_{T_2}\underbrace{-\frac{1}{2}\nabla_C\left\{\nabla^C\left(\psi^{-D}O_A O_B\right)\right\}}_{T_3}
\end{split}
\end{equation}
\begin{equation}\label{T_1}
\begin{split}
T_1&=\frac{1}{2}\nabla_C\left\{\nabla_B\left(\psi^{-D}O_A O^C\right)\right\}\\
&=\frac{1}{2}[\nabla_C,\nabla_B]\left(\psi^{-D}O_A O^C\right) +\frac{1}{2}\nabla_B\nabla_C \left(\psi^{-D}O_A O^C\right)\\
&=\left(\frac{\psi^{-D}}{2}\right)\bar R_{EB} O^E O_A- \frac{1}{2}\nabla_B \left[\psi^{-D}\left\{\left(\frac{DN}{\psi}-\nabla\cdot O\right)O_A - \left(O\cdot\nabla\right)O_A\right\}\right]
+ {\cal O}(1)\\
&=\left(\frac{D\lambda}{2}\right)\psi^{-D}O_AO_B +\left(\frac{DN}{2\psi}\right)\psi^{-D}\left[\frac{DN}{\psi} - \nabla\cdot O- u^C(O\cdot\nabla)n_C\right]n_B O_A\\
&~~~~~~~~~~~~~~~~~~~~~-\frac{\psi^{-D}}{2}O_A\nabla_B\left(\tilde{\nabla}\cdot u\right) + {\cal O}(1)\\
\end{split}
\end{equation}
In the last step we have used subsidiary condition on $O$ and also the fact that 
$$\left(\frac{DN}{\psi} - \nabla\cdot O\right)=\left(\frac{DN}{\psi} -{K}\right) + {\cal O}(1)\sim {\cal O}(1)$$
Similarly, 
\begin{equation}\label{T_2}
\begin{split}
T_2&=\left(\frac{D\lambda}{2}\right)\psi^{-D}O_AO_B +\left(\frac{DN}{2\psi}\right)\psi^{-D}\left[\frac{DN}{\psi} - \nabla\cdot O- u^C(O\cdot\nabla)n_C\right]n_A O_B\\
&~~~~~~~~~~~~~~~~~~~~~-\frac{\psi^{-D}}{2}O_B\nabla_A\left(\tilde{\nabla}\cdot u\right) + {\cal O}(1)\\
\end{split}
\end{equation}

\begin{equation}\label{T_3}
\begin{split}
T_3&=-\frac{1}{2}\nabla_C\nabla^C(\psi^{-D}O_B O_A)\\
&=-\frac{1}{2}\left(\nabla^2\psi^{_D}\right)O_AO_B -\left(\nabla_C \psi^{-D}\right) \left(\nabla^CO_A O_B\right) - \frac{\psi^{-D}}{2}\nabla^2(O_A O_B)\\
&=\psi^{-D}\bigg[\left(\frac{DN}{\psi}\right)(n\cdot\nabla)\left( O_A O_B\right) - \frac{1}{2}\nabla^2(O_A O_B)\bigg]
\end{split}
\end{equation}\\ 
Adding \eqref{T_1},\eqref{T_2},\eqref{T_3} we get the expression for ${[R_L]}_{AB}$
\begin{equation}\label{eq:RLAB}
\begin{split}
&\delta R_{AB}|_{\text{lin.}}\\
=&~\psi^{-D}\bigg[\left(\frac{DN}{\psi}\right)(n\cdot\nabla)\left( O_A O_B\right)  +D\lambda ~O_AO_B-\frac{1}{2}(O_A\nabla_B+O_B\nabla_A)\left(\tilde{\nabla}\cdot u\right)\\
&~~~~~~~~- \frac{1}{2}\nabla^2(O_A O_B)+\left(\frac{DN}{2\psi}\right)\left(\frac{DN}{\psi} - \nabla\cdot O- u^C(O\cdot\nabla)n_C\right)\left(n_B O_A+n_A O_B\right) \bigg]
+ {\cal O}(1)\\
=&~\psi^{-D}\bigg[K~(n\cdot\nabla)\left( O_A O_B\right) - \frac{1}{2}\nabla^2(O_A O_B) +D\lambda ~O_AO_B-\frac{1}{2}(O_A\nabla_B+O_B\nabla_A)\left(\tilde{\nabla}\cdot u\right)\\
&~~~~~~~~+\frac{K}{2}\left(\frac{(n\cdot\nabla)K}{K} + \nabla\cdot u- u^C(O\cdot\nabla)n_C\right)\left(n_B O_A+n_A O_B\right) \bigg]
+ {\cal O}(1)\\
\end{split}
\end{equation}\\
Using, the following identities
\begin{equation}
(n\cdot\nabla)(O_A O_B)=2 [u^C(n\cdot\nabla)n_C] O_A O_B+\left(O_A P^C_B+O_B P^C_A\right)[(u\cdot\nabla)O_C]\\
\end{equation}
\begin{equation}\label{eq:ccii}
\begin{split}
O_B \nabla^2 O_A+O_A\nabla^2 O_B&=2\left[K[u^D(n\cdot\nabla)n_D]+(u\cdot\nabla)K\right]O_A O_B+ (O_B P^C_A+O_A P^C_B )\nabla^2O_C\\
&~~~~~~~~-[\left(\nabla^C O_D\right)(\nabla_C O^D)][n_A O_B+n_B O_A ]
\end{split}
\end{equation}
We have used the identity \eqref{I1} for the derivation of the above equation.\\
\begin{equation}
\begin{split}
(O_A\nabla_B+O_B\nabla_A)\left(\tilde{\nabla}\cdot u\right)&=\left(P^E_A O_B+P^E_B O_A\right)\tilde{\nabla}_E\left(\tilde{\nabla}\cdot u\right)+2~O_A O_B(u\cdot\nabla)(\tilde{\nabla}\cdot u)\\
&~~~~+(n_A O_B+n_B O_A)(O\cdot\nabla)(\tilde{\nabla}\cdot u)\\
&=(n_A O_B+n_B O_A)(n\cdot\nabla)(\tilde{\nabla}\cdot u)+{\cal O}(1)
\end{split}
\end{equation}\\
The expression of $\delta R_{AB}|_{\text{lin.}}$ becomes
\begin{equation}\label{delrlin}
\begin{split}
&\delta R_{AB}|_{\text{lin.}}\\
&=\psi^{-D}\Bigg[D~\lambda ~(O_A O_B)+\left(O_B P^C_A+O_A P^C_B\right)\left(K(u\cdot\nabla)O_C-\frac{1}{2}\nabla^2 O_C\right)\\
&+(n_A O_B+N_B O_A)\bigg\{\frac{K}{2}\left(\frac{(n\cdot\nabla)K}{K} + \nabla\cdot u- u^C(O\cdot\nabla)n_C\right)+\frac{1}{2}K_{CD} K^{CD}-\frac{1}{2}(n\cdot\nabla)\left(\tilde{\nabla}\cdot u\right)\bigg\}\Bigg]
\end{split}
\end{equation}\\
Substituting \eqref{delrlin} and \eqref{eq:RNLAB} in \eqref{source} we get the source term $S_{AB}^{(-1)}$
\begin{equation}\label{delrAB}
\begin{split}
&S_{AB}=\delta R_{AB}-D\lambda G^{(0)}_{AB}=\delta R_{AB}|_{\text{lin.}}+\delta R_{AB}|_{\text{non-lin.}}-D\lambda ~ \psi^{-D} O_A O_B\\
&=\psi^{-D}\Bigg[\psi^{-D} \left(\frac{K}{2}\right)\left(\tilde{\nabla}\cdot u\right) O_BO_A+\left(O_B P^C_A+O_A P^C_B\right)\left(K(u\cdot\nabla)O_C-\frac{1}{2}\nabla^2 O_C\right)\\
&~~~+(n_A O_B+N_B O_A)\bigg\{\frac{K}{2}\left(\frac{(n\cdot\nabla)K}{K} + \nabla\cdot u- u^C(O\cdot\nabla)n_C\right)+\frac{1}{2}K_{CD} K^{CD}-\frac{1}{2}(n\cdot\nabla)(\tilde{\nabla}\cdot u)\bigg\}\Bigg]\\
&=\psi^{-D}\left(\frac{K}{2}\right)\Bigg\{\psi^{-D} \left(\tilde{\nabla}\cdot u\right) O_BO_A+\left(O_BP^C_A+O_AP^C_B\right)\left[\frac{\tilde{\nabla}^2 u_C}{K}-\frac{\nabla_C K}{K}+u^D K_{DC}-(u\cdot\nabla)u_C\right]\\
&+(n_A O_B+n_B O_A)\bigg[\frac{1}{K}K_{CD} K^{CD}-\frac{1}{K}(n\cdot\nabla)(\tilde{\nabla}\cdot u)+\frac{(n\cdot\nabla)K}{K}+\tilde{\nabla}\cdot u-2~\frac{(u\cdot\nabla)K}{K}+u\cdot K\cdot u\bigg]\Bigg\}\\
\end{split}
\end{equation}\\
In the last line we have used the following identity (see appendix \ref{app:Identity} for derivation)
\begin{equation}\label{eq:I4}
\begin{split}
&P^C_B\nabla^2 O_C=P^C_B~\bigg[\nabla_C K-\tilde{\nabla}^2 u_C+K\left(u^D K_{DC}-(u\cdot\nabla)u_C\right)\bigg]+{\cal{O}}(1)
\end{split}
\end{equation}\\
Now,
\begin{equation}
\begin{split}
S_{AB}&=\psi^{-D}\left(\frac{K}{2}\right)\Bigg\{\psi^{-D} (\tilde{\nabla}\cdot u) O_BO_A+\left(O_BP^C_A+O_AP^C_B\right)\left[\frac{\tilde{\nabla}^2 u_C}{K}-\frac{\nabla_C K}{K}+u^D K_{DC}-(u\cdot\nabla)u_C\right]\\
&+(n_A O_B+n_B O_A)\bigg[\frac{1}{K}K_{CD} K^{CD}-\frac{\tilde{\nabla}^2 K}{K^2}+2~\frac{u\cdot\nabla K}{K}-u\cdot K\cdot u-\frac{1}{K}u^D\bar{R}_{DE} u^E+\frac{(n\cdot\nabla)K}{K}\\
&~~~~~~~~~~~~~~~~~~~~~~~~~+\tilde{\nabla}\cdot u-2~\frac{(u\cdot\nabla)K}{K}+(u\cdot K\cdot u)\bigg]\Bigg\}\\
\end{split}
\end{equation}
In the last line we have used the following identity
\begin{equation}
(n\cdot\nabla)\left(\tilde{\nabla}\cdot u\right)=\left(\frac{\tilde{\nabla}^2 K}{K}-2~(u\cdot\nabla)K+K~(u\cdot K\cdot u)+u^D\bar{R}_{DE} u^E\right)
\end{equation}
Where $\left(\tilde{\nabla}\cdot u\right)$ is given in appendix \ref{deluderi}.\\ 
Using the identity \eqref{eq:impi} and \eqref{urunrn} we get $S_{AB}$
\begin{equation}\label{source3}
\begin{split}
S_{AB}&=\psi^{-D}\left(\frac{K}{2}\right)\Bigg[\psi^{-D} \left(\tilde{\nabla}\cdot u\right) O_BO_A+(n_A O_B+n_B O_A)\left(\tilde{\nabla}\cdot u\right)\\
&+(O_BP^C_A+O_AP^C_B)\left(\frac{\tilde{\nabla}^2 u_C}{K}-\frac{\nabla_C K}{K}+u^D K_{DC}-(u\cdot\nabla)u_C\right)\Bigg]\\
\end{split}
\end{equation}
Let us note the presence of  `$K(\tilde\nabla\cdot u)$' term in $S_{AB}$. From the leading order calculation it follows that it is of order ${\cal O}(D)$ on $\psi=1$ hypersurface(see eq \eqref{sclrcon}). This is sort of `anomalous',  since naive order counting suggests that this term should be of order ${\cal O}(D^2)$ and this may not be the case once we are away from the membrane.

Now for any generic term, which is of order ${\cal O}(1)$  when evaluated on  $(\psi =1)$ hypersurface, will have corrections of order ${\cal O}\left(1\over D\right)$ (or further suppressed) as one goes away from $\psi=1$. But, for `anomalous' term like $K(\tilde\nabla\cdot u)$ that is not the case. Below, we shall examine this term in more detail.
We can expand $(\tilde{\nabla}\cdot u)$ in $\left[\psi -1 = {R\over D}\right]$ as follows
\begin{equation}\label{nabuoff3}
\begin{split}
\tilde{\nabla}\cdot u &=\left(\tilde{\nabla}\cdot u\right)_{\psi=1}+\frac{\psi-1}{N}(n\cdot\nabla)\left(\tilde{\nabla}\cdot u\right)_{\psi=1}\\
&=\left(\tilde{\nabla}\cdot u\right)_{R=0}+\left(\frac{R}{K}\right)\left(\frac{\tilde{\nabla}^2 K}{K}-2~(u\cdot\nabla)K+K~(u\cdot K\cdot u)+u^D\bar{R}_{DE} u^E\right)_{R=0}\\
&=\left(\tilde{\nabla}\cdot u\right)_{R=0}-\frac{R}{K}\left(\tilde{\nabla}\cdot E\right)_{R=0}
\end{split}
\end{equation}
We don't need to expand any other term since $\tilde{\nabla}\cdot u$ is the only `anomalous' term in this order. 
Substituting \eqref{nabuoff3} in \eqref{source3} we get the final expression for $S_{AB}$
\begin{equation}\label{source2}
\begin{split}
S_{AB}&=e^{-R}\left(\frac{K}{2}\right)\Bigg[e^{-R} \left(\left(\tilde{\nabla}\cdot u\right)_{R=0}-\frac{R}{K}\left(\tilde{\nabla}\cdot E\right)_{R=0}\right) O_BO_A\\
&~~~~~~~+(n_A O_B+n_B O_A)\left(\left(\tilde{\nabla}\cdot u\right)_{R=0}-\frac{R}{K}\left(\tilde{\nabla}\cdot E\right)\right)_{R=0}\\
&+(O_BP^C_A+O_AP^C_B)\left(\frac{\tilde{\nabla}^2 u_C}{K}-\frac{\nabla_C K}{K}+u^D K_{DC}-(u\cdot\nabla)u_C\right)_{R=0}\Bigg]\\
\end{split}
\end{equation}

\section{Intermediate steps for matching with AdS Black Hole}\label{app:AdsBHMatch}
Since we know that the horizon is not at $r=1$, this implies $\psi(r=1) \neq 1$.\\
 We shall assume the following expansion of $\psi$ around $r=1$.
 \begin{equation}\label{exp}
 \begin{split}
 \psi(r) &= 1 +\frac{X_1}{D} + \frac{X_2}{D^2} + \left(a_{10} +\frac{a_{11}}{D}\right)(r-1) + a_{20} (r-1)^2 + {\cal O}\left(\frac{1}{D}\right)^3\\
 \text{where}&~ ~X_1, ~X_2,~ a_{10},~a_{11},~a_{20}~~\text{are constants and }~~(r-1)\sim{\cal O}\left(\frac{1}{D}\right)\\
 \end{split}
 \end{equation}
 
 Substituting equation \eqref{exp} in equation \eqref{subpsi} and solving it order by order in $\left(\frac{1}{D}\right)$  we find the following solutions for the coefficients.
 \begin{equation}\label{sol1}
 a_{10}=1,~~~a_{11} = X_1  -2,~~~a_{20}=0
 \end{equation}
 
To fix $X_1$ and $X_2$ we have to use the fact that $\psi=1$ correspond to horizon.
We can determine the horizon of Schwarzschild-AdS black hole $r_0$ order by order in $\left(\frac{1}{D}\right)$.
\begin{equation}\label{r0det}
r_0 = 1-\frac{\log 2}{D} +\left(\frac{1}{D}\right)^2\left[-2 \log 2 +\frac{(\log 2)^2}{2}\right] + {\cal O}\left(\frac{1}{D}\right)^3
\end{equation}
 Noe setting $\psi(r_0) =1$ we find
 $$X_1 = \log 2,~~X_2= \frac{(\log 2)^2}{2}$$
 
 So finally we found 
 \begin{equation}\label{finpsi}
 \begin{split}
 \psi(r)& = 1 +\frac{\log 2}{D} +\left(\frac{1}{D}\right)^2\left[\frac{(\log 2)^2}{2}\right] \\
 &+ \left[1 +\left( \frac{\log 2-2}{D}\right)\right] (r-1) + {\cal O}\left(\frac{1}{D}\right)^3\\
 \end{split}
 \end{equation}

\section{ Linearized analysis: Details of the calculation}\label{app:QNMpoincare}
\subsection{QNM for AdS/dS Schwarzschild Black hole}\label{QNM AdS Global}

The Christoffel symbols (here we report only nonzero components) calculated for the metric \eqref{globmet} are given by (with $\bar g_{ab}$ and $\bar\Gamma^a_{bc}$ as metric and Christoffel symbol on unit sphere )
\begin{equation}
 \begin{split}
  \Gamma^r_{ab} &= -r \left( 1-\frac{\sigma r^2}{L^2} \right)\bar{g}_{ab},~~~
  \Gamma^a_{rb} = \frac{1}{r} \delta^a_b,~~~
  \Gamma^r_{tt} = -r \left( 1-\frac{\sigma r^2}{L^2} \right)\frac{\sigma}{L^2} \\
    \Gamma^t_{rt} &= -r  \left( 1-\frac{\sigma r^2}{L^2} \right)^{-1} \frac{\sigma}{L^2},~~~
  \Gamma^r_{rr} = r  \left( 1-\frac{\sigma r^2}{L^2} \right)^{-1} \frac{\sigma}{L^2},~~~\Gamma^a_{bc}=\bar\Gamma^a_{bc} \\   
 \end{split}
\end{equation}
 The normal vector to membrane surface is given by
 \begin{equation}
  \begin{split}
   n_r &=  \left( 1-\frac{\sigma r^2}{L^2} \right)^{-\frac{1}{2}},~~
   n_t =  \left( 1-\frac{\sigma r^2}{L^2} \right)^{-\frac{1}{2}} (-\epsilon \partial_t \delta r) ,~~
   n_a =  \left( 1-\frac{\sigma r^2}{L^2} \right)^{-\frac{1}{2}} (-\epsilon \bar\nabla_a \delta r) 
  \end{split}
 \end{equation}
 The answer for $\nabla_An_B$ is given by
\begin{equation}
 \begin{split}
  \nabla_rn_r &= 0,~ ~~~
  \nabla_rn_t = \left( 1-\frac{\sigma r^2}{L^2} \right)^{-\frac{3}{2}}\frac{2\sigma r}{L^2}(-\epsilon \partial_t \delta r) \\
  \nabla_tn_r &= \left( 1-\frac{\sigma r^2}{L^2} \right)^{-\frac{3}{2}}\frac{\sigma r}{L^2}(-\epsilon \partial_t \delta r),~~~\nabla_an_t = (-\epsilon \partial_t \bar\nabla_a \delta r)\left( 1-\frac{\sigma r^2}{L^2} \right)^{-\frac{1}{2}}  \\
  \nabla_tn_t &= \left( 1-\frac{\sigma r^2}{L^2} \right)^{-\frac{1}{2}}(-\epsilon \partial^2_t \delta r) + \left( 1-\frac{\sigma r^2}{L^2} \right)^{\frac{1}{2}}\frac{\sigma r}{L^2} \\
  \nabla_rn_a &= (-\epsilon \bar\nabla_a \delta r)\left[ \frac{\sigma r}{L^2}\left( 1-\frac{\sigma r^2}{L^2} \right)^{-\frac{3}{2}}-\frac{1}{r}\left( 1-\frac{\sigma r^2}{L^2} \right)^{-\frac{1}{2}} \right] \\
  \nabla_an_r &= (\epsilon \bar\nabla_a \delta r)\frac{1}{r}\left( 1-\frac{\sigma r^2}{L^2} \right)^{-\frac{1}{2}},~~~
  \nabla_tn_a = (-\epsilon \partial_t \bar\nabla_a \delta r)\left( 1-\frac{\sigma r^2}{L^2} \right)^{-\frac{1}{2}} \\
\nabla_an_b &= \left( 1-\frac{\sigma r^2}{L^2} \right)^{-\frac{1}{2}}(-\epsilon \bar\nabla_a \bar\nabla_b \delta r) + r\left( 1-\frac{\sigma r^2}{L^2} \right)^{\frac{1}{2}}\bar{g}_{ab}
 \end{split}
\end{equation}
The answer for `spacetime' projector $P_{B}^{A}=\delta^A_B - n^A n_B $ is
\begin{equation}
\begin{split}
 & P^r_r = 0,\quad P^t_t = 1,\quad P^a_b = \delta^a_b,\quad P^t_a = 0,\quad P^a_t = 0,\\
 & P^r_t = \epsilon \partial_t \delta r,\quad P^t_r = \left( 1-\frac{\sigma r^2}{L^2} \right)^{-2} (-\epsilon \partial_t \delta r), \\
 & P^r_a = \epsilon \bar\nabla_a \delta r,\quad P^a_r = \frac{1}{r^2}\left( 1-\frac{\sigma r^2}{L^2} \right)^{-1} (\epsilon \bar\nabla^a \delta r)
\end{split}
\end{equation}
The answer for the spacetime form of the Extrinsic curvature $K_{MN}$ is 
\begin{equation}
 \begin{split}
  K_{rr} &= 0,~~~ 
  K_{rt} = \left( 1-\frac{\sigma r^2}{L^2} \right)^{-\frac{3}{2}} \frac{\sigma r}{L^2} (-\epsilon \partial_t \delta r),~~~
  K_{ra} = \frac{1}{r} \left( 1-\frac{\sigma r^2}{L^2} \right)^{-\frac{1}{2}}(\epsilon \bar\nabla_a \delta r) \\
K_{ta} &= \left( 1-\frac{\sigma r^2}{L^2} \right)^{-\frac{1}{2}}(-\epsilon \partial_t \bar\nabla_a \delta r),~~~
  K_{tt} = \left( 1-\frac{\sigma r^2}{L^2} \right)^{-\frac{1}{2}}(-\epsilon \partial^2_t \delta r)+\left( 1-\frac{\sigma r^2}{L^2} \right)^{\frac{1}{2}}\frac{\sigma r}{L^2} \\ 
  K_{ab} &= \left( 1-\frac{\sigma r^2}{L^2} \right)^{-\frac{1}{2}}(-\epsilon \bar\nabla_a\bar\nabla_b \delta r)+r\left( 1-\frac{\sigma r^2}{L^2} \right)^{\frac{1}{2}}\bar{g}_{ab} 
 \end{split}
\end{equation}
Nonzero Christoffel symbol components for the metric \eqref{lind} is given by
\begin{equation}
 \begin{split}
  \Gamma^t_{tt} &= -\left( 1-\frac{\sigma }{L^2} \right)^{-1}\frac{\sigma}{L^2}(\epsilon \partial_t \delta r),~~~
  \Gamma^a_{tt} = -\frac{\sigma}{L^2}(\epsilon \bar\nabla^a \delta r) \\
  \Gamma^t_{at} &= -\left( 1-\frac{\sigma }{L^2} \right)^{-1}\frac{\sigma}{L^2}(\epsilon \bar\nabla_a \delta r),~~~
  \Gamma^t_{ab} = \left( 1-\frac{\sigma }{L^2} \right)^{-1}(\epsilon \partial_t \delta r)\bar{g}_{ab} \\
  \Gamma^a_{tb} &= (\epsilon \partial_t \delta r)\delta^a_b,~~~
  \Gamma^a_{bc} = \bar{\Gamma}^a_{bc} + \epsilon(\bar\nabla_b \delta r \delta^a_c+\bar\nabla_c \delta r \delta^a_b-\bar\nabla^a \delta r \bar{g}_{bc})
 \end{split}
\end{equation}
The answer for $\hat{\nabla}_\mu u_\nu$ is given by
\begin{equation}
\begin{split}
\hat{\nabla}_t u_t &= 0,~~~
 \hat{\nabla}_t u_a = \epsilon \partial_t \delta u_a - \left( 1-\frac{\sigma}{L^2} \right)^{-\frac{1}{2}}\left(\frac{\sigma}{L^2}\right) (\epsilon \bar\nabla_a \delta r) \\
 \hat{\nabla}_a u_t &= 0,~~~
 \hat{\nabla}_a u_b = \epsilon \bar\nabla_a \delta u_b + \left( 1-\frac{\sigma}{L^2} \right)^{-\frac{1}{2}} (\epsilon \partial_t \delta r) \bar{g}_{ab}
 \end{split}
\end{equation}

\subsection{Effective equations for AdS black brane from scaled membrane equations}

\subsubsection{Linearized fluctuation analysis}\label{QNM Poincare}

 The nonzero Christoffel symbols for the metric \eqref{ppmet} are
 \begin{equation}
   \Gamma^r_{rr} = \frac{-1}{r},\quad
   \Gamma^r_{ab} = -r^3 \delta_{ab},\quad
   \Gamma^a_{rb} = \frac{1}{r} \delta^{a}_{b}, \quad
   \Gamma^r_{tt} = r^3,\quad
   \Gamma^t_{rt} = \frac{1}{r}\quad
 \end{equation}
The normal to membrane surface is
 \begin{equation}
  n_r = \frac{1}{r},\quad n_a = \frac{-\epsilon \partial_a \delta r}{r},\quad n_t = \frac{-\epsilon \partial_t \delta r}{r} 
 \end{equation}
 The answer for $\nabla_Mn_N$ is given by
\begin{equation}
\begin{split}
  \nabla_rn_r &= 0, \quad  \nabla_rn_t = \frac{2\epsilon \partial_t \delta r}{r^2}, \quad \nabla_tn_r = \frac{\epsilon \partial_t \delta r}{r^2},  \quad \nabla_tn_t = -\frac{\epsilon \partial^2_t \delta r}{r} -r^2, \quad \\
  \nabla_rn_a &= \frac{2\epsilon \partial_a \delta r}{r^2}, \quad \nabla_an_r = \frac{\epsilon \partial_a \delta r}{r^2}, \quad \nabla_tn_a = \frac{-\epsilon \partial_t \partial_a \delta r}{r}, \\
  \nabla_an_t &= \frac{-\epsilon \partial_t \partial_a \delta r}{r}, \quad \nabla_an_b = \frac{-\epsilon \partial_a \partial_b \delta r}{r} + r^2 \delta_{ab}
\end{split}
\end{equation}
where the rest of the components are zero.\\
The answer for the projector $P_A^B=\delta_A^B - n_An^B$ is given by
\begin{equation}
 \begin{split}
  P^r_r &= 0,\quad P_t^t = 1,\quad P^a_b = \delta^a_b,\quad P^a_t = 0, \quad P^t_a = 0, \\
  P^r_t &= \epsilon \partial_t \delta r, \quad P^t_r = \frac{-\epsilon \partial_t \delta r}{r^4}, \quad \quad P^r_a = \epsilon \partial_a \delta r, \quad \quad P^a_r = \frac{\epsilon \partial^a \delta r}{r^4} 
 \end{split}
\end{equation}

The spacetime Extrinsic curvature $K_{MN}$ is given by

\begin{equation}
\begin{split}
 K_{rr} &= 0, \quad K_{rt}  = \frac{\epsilon \partial_t \delta r}{r^2} , \quad K_{ra} = \frac{\epsilon \partial_a \delta r}{r^2} \\ K_{tt} &= \frac{-\epsilon \partial^2_t \delta r}{r} - r^2, \quad K_{ta}  =\frac{-\epsilon \partial_t \partial_a \delta r}{r}, \quad K_{ab} = \frac{-\epsilon \partial_a \partial_b \delta r}{r} + r^2 \delta_{ab}
 \end{split}
\end{equation}
where the rest of the components are zero.\\
The nonzero Christoffel symbols for the metric \eqref{ppind} are
\begin{equation}
 \begin{split}
  \Gamma^t_{tt} &= \epsilon \partial_t \delta r, \quad  \Gamma^a_{tt} = \epsilon \partial^a \delta r, \quad \Gamma^t_{at} = \epsilon \partial_a \delta r, \quad \Gamma^t_{ab} = \epsilon \partial_t \delta r\delta_{ab} \\
  \Gamma^a_{tb} &= \epsilon \partial_t \delta r\delta^a_b, \quad \Gamma^a_{bc} = \epsilon  ( \partial_b \delta r \delta^a_c + \partial_c \delta r \delta^a_b - \partial^a \delta r \delta_{bc})
 \end{split}
\end{equation}
The answer for $\hat{\nabla}_\mu u_\nu$ is given by
\begin{equation}
 \hat{\nabla}_t u_t = 0, \quad \hat{\nabla}_t u_a = \epsilon \partial_t \delta u_a + \epsilon \partial_a \delta r, \quad \hat{\nabla}_a u_t = 0, \quad
 \hat{\nabla}_a u_b = \epsilon \partial_a \delta u_b + \epsilon \partial_t \delta r \delta_{ab}
\end{equation}
where the rest of the components are zero.

\subsubsection{Scaled nonlinear analysis}\label{BB Calcula}
The nonzero Christoffel symbols for the metric \eqref{ppsc} are
\begin{equation}
 \begin{split}
  \Gamma^\rho_{\rho\rho} &= -\frac{1}{D} \left( 1+\frac{\rho}{D} \right)^{-1}, \quad \Gamma^\rho_{ab} =-\left( 1+\frac{\rho}{D} \right)^{3}\delta_{ab}, \quad \Gamma^\rho_{ij} =-D\left( 1+\frac{\rho}{D} \right)^{3}\delta_{ij}, \\
  \Gamma^a_{\rho b} &= \frac{1}{D}\left( 1+\frac{\rho}{D} \right)^{-1}\delta^a_b,\quad  \Gamma^i_{\rho j} = \frac{1}{D}\left( 1+\frac{\rho}{D} \right)^{-1}\delta^i_j,\\
  \Gamma^\rho_{tt} &= D\left( 1+\frac{\rho}{D} \right)^{3}, \quad  \Gamma^t_{\rho t} = \frac{1}{D}\left( 1+\frac{\rho}{D} \right)^{-1}
 \end{split}
\end{equation}
The normal to the membrane surface is given by
\begin{equation}
 \begin{split}
  n_\rho &= \frac{1}{D}\left( 1-\frac{\rho}{D}-\frac{\partial^aY\partial_a Y}{2D} \right) \\
  n_t &= -\frac{\partial_t Y}{D}\left( 1-\frac{\rho}{D}-\frac{\partial^aY\partial_a Y}{2D} \right) \\
  n_a &= -\frac{\partial_a Y}{D}\left( 1-\frac{\rho}{D}-\frac{\partial^aY\partial_a Y}{2D} \right) 
 \end{split}
\end{equation}
The projector $P^M_N=\delta^M_N-n^Mn_N$ is given by
\begin{equation}
 \begin{split}
  P^\rho_\rho &= \frac{\partial^aY\partial_a Y}{D}, \quad P^t_t = 1, \quad P^i_j = \delta^i_j, \quad P^a_b =\delta^a_b, \quad P^t_a = \mathcal{O}(n^{-2}), \quad P^a_t = -\frac{\partial^aY\partial_t Y}{D}, \\
  P^\rho_t &= \partial_t Y, \quad P^t_\rho =\mathcal{O}(n^{-2}), \quad P^\rho_a = \partial_a Y, \quad P^a_\rho = \frac{\partial^aY}{D}
 \end{split}
\end{equation}
where the rest of the components are zero.\\
The answer for $\bar{\nabla}_Mn_N$ is given by
\begin{equation}
\begin{split}
 \bar{\nabla}_\rho n_\rho &= \mathcal{O}(D^{-2}), \quad \bar{\nabla}_\rho n_t = \mathcal{O}(D^{-2}), \quad \bar{\nabla}_t n_\rho = \mathcal{O}(D^{-2}), \\  \bar{\nabla}_t n_t &= -\frac{\partial^2_t Y}{D} - \left( 1 + \frac{2\rho}{D} - \frac{\partial_aY\partial^aY}{2D}\right), \quad \bar{\nabla}_a n_\rho = \mathcal{O}(D^{-2}), \quad \bar{\nabla}_\rho n_a = \mathcal{O}(D^{-2}), \\ \bar{\nabla}_t n_a &= -\frac{\partial_t \partial_aY}{D}, \quad  \bar{\nabla}_a n_t = -\frac{\partial_t \partial_aY}{D}, \quad \bar{\nabla}_a n_b = -\frac{\partial_a \partial_bY}{D} +\frac{\delta_{ab}}{D}, \\ \quad \bar{\nabla}_i n_j &= \delta_{ij}\left( 1+ \frac{2\rho}{D} - \frac{\partial_aY\partial^aY}{2D} \right),
\end{split}
\end{equation}
where the rest of the components are zero.\\
The answer for spacetime form of the Extrinsic curvature $K_{MN}$ is
\begin{equation}
 \begin{split}
  K_{\rho\rho} &= \mathcal{O}(D^{-2}), \quad  K_{\rho t} = \mathcal{O}(D^{-2}), \quad  K_{\rho a} = \mathcal{O}(D^{-2}),\quad K_{\rho i} = \mathcal{O}(D^{-2}), \\
  K_{tt} &= -\frac{\partial_t^2Y}{D} -\left( 1+\frac{2\rho}{D} - \frac{\partial_aY\partial^aY}{2D} \right), \quad K_{ta} = -\frac{\partial_a\partial_tY}{D}, \quad K_{ti} = \mathcal{O}(D^{-2}), \\ K_{ab} &= -\frac{\partial_a\partial_bY}{D}+\frac{\delta_{ab}}{D}, \quad K_{ij} = \delta_{ij}\left( 1+\frac{2\rho}{D} - \frac{\partial_aY\partial^aY}{2D} \right)
 \end{split}
\end{equation}
where the rest of the components are zero.\\
 Nonzero Christoffel symbols components for the induced metric \eqref{ppindnl} are given by
 \begin{equation}
  \begin{split}
   \Gamma^t_{tt} &= \frac{\partial_t Y}{D}, \quad \Gamma^a_{tt} = \partial^a Y, \quad  \Gamma^t_{at} = \frac{\partial_a Y}{D}, \quad \Gamma^t_{ab} = \frac{\partial_t Y}{D^2}\delta_{ab}, \quad \Gamma^t_{ij} = \frac{\partial_t Y}{D^2}\delta_{ij}, \\
   \Gamma^a_{tb} &= \frac{\partial_t Y}{D}\delta^a_b, \quad \Gamma^i_{tj} = \frac{\partial_t Y}{D}\delta^i_j, \quad \Gamma^a_{bc} = \frac{1}{D}(\partial_bY \delta^a_c + \partial_cY \delta^a_b-\partial^aY \delta_{bc}), \\ \Gamma^a_{jk} &= -\partial^a Y \delta_{jk}, \quad \Gamma^i_{ja} = \frac{\partial_a Y}{D} \delta^i_j
  \end{split}
 \end{equation}
 The answer for $\hat{\nabla}_\mu u_\nu$ is given by
 \begin{equation}
  \begin{split}
   \tilde{\nabla}_tu_t &= -\frac{\partial_t(U_aU^a)}{2D}, \quad  \tilde{\nabla}_tu_a = \frac{\partial_t U_a}{D} + \frac{\partial_a Y}{D}, \quad \tilde{\nabla}_au_t = -\frac{\partial_a(U_bU^b)}{2D}, \\ \tilde{\nabla}_au_b &= \frac{\partial_a U_b}{D}, \quad \tilde{\nabla}_iu_j = \frac{\partial_t Y}{D}\delta_{ij} + \frac{U^a\partial_a Y}{D}\delta_{ij}
  \end{split}
 \end{equation}
 where the rest of the components are zero.

\section{The derivation of $(\nabla\cdot u)$}\label{deluderi}
Note that to compute the full space-time divergence of $u^A$ we also need to know the normal derivative of $u^A$ away from the membrane.
\begin{equation}\label{eq:deriu}
\begin{split}
\nabla\cdot u=&~P^{AB} \nabla_A u_B +n^B (n\cdot \nabla) u_B\\
=&~~P^{AB} \nabla_A u_B -u^B (n\cdot \nabla) n_B\\
=&~~P^{AB} \nabla_A u_B -\frac{(u\cdot\nabla) K}{K}\\
\end{split}
\end{equation}
In the last line we have used the identity $\bigg[(n\cdot\nabla) n_A = \frac{\Pi^C_A\nabla_C K}{K}\bigg]$.\\
We know that the first term in equation \eqref{eq:deriu} is of order ${\cal O}(1)$ on the membrane. It follows from the equation of motion at zeroth order. However, to determine the source term we need to know this expression even away from the $(\psi =1)$ hypersurface. Below we shall determine this term in an expansion in $(\psi-1) $ and we shall see that the coefficient of the linear term is also of order ${\cal O}(1)$.\\
Consider the expansion  of $u^A$ from $(\psi =1)$  hypersurface.
 \begin{equation}
\begin{split}
u^A=u^A|_{\psi=1}+\frac{\psi-1}{N}[(n\cdot \nabla)u^A]|_{\psi=1}+\cdots
\end{split}
\end{equation}
 Substituting this expansion in first term of the equation \eqref{eq:deriu} we find 
 
 \begin{equation}\label{eq:usimpl1}
\begin{split}
&~~~~(P^{AB}\nabla_A u_B)\\
&=P^{AB}\nabla_A\bigg(u_B|_{\psi=1}+\frac{\psi-1}{N}[(n\cdot \nabla)u_B]|_{\psi=1}+\cdots\bigg)\\
&=P^{AB}\nabla_A u_B|_{\psi=1}+P^{AB}\nabla_A\bigg(\frac{\psi-1}{N}[(n\cdot \nabla)u_B]|_{\psi=1}+\cdots\bigg)\\
&=P^{AB}\nabla_A u_B|_{\psi=1}+\bigg(\frac{\psi-1}{N}\bigg)P^{AB}\nabla_A[(n\cdot \nabla)u_B]|_{\psi=1}-P^{AB}\bigg(\frac{\psi-1}{N^2}\bigg)(\nabla_A N)[(n\cdot\nabla)u_B]|_{\psi=1}\\
&=P^{AB}\nabla_A u_B|_{\psi=1}+\bigg(\frac{\psi-1}{N}\bigg)P^{AB}\nabla_A[(n\cdot \nabla)u_B]|_{\psi=1}+{\cal{O}}\bigg(\frac{1}{D}\bigg)
\end{split}
\end{equation}
Now we shall process the coefficient of $(\psi-1)$.
 \begin{equation}\label{eq:usimpl2}
\begin{split}
P^{AB}\nabla_A [(n\cdot\nabla)u_B]&=P^{AB}\nabla_A\left[-n_B\frac{(u\cdot\nabla)K}{K}+P^D_B[(n\cdot\nabla)n_D-(u\cdot\nabla)O_D]\right]\\
&=-(u\cdot\nabla)K+K[n^D(u\cdot\nabla)O_D]+P^{AD}\nabla_A[(n\cdot\nabla)n_D-(u\cdot\nabla)O_D]\\
&=-u^D\bar{R}_{DE} n^E+K~(u\cdot K\cdot u)-2~(u\cdot\nabla)K+\frac{\tilde{\nabla}^2 K}{K}+u^D\bar{R}_{DE} u^E\\
&=\frac{\tilde{\nabla}^2 K}{K}-2~(u\cdot\nabla)K+K~(u\cdot K\cdot u)+u^D\bar{R}_{DE} u^E
\end{split}
\end{equation}
Note that $(\psi-1)$ is also of order ${\cal O}\left(\frac{1}{D}\right)$. Therefore combining  equations \eqref{eq:usimpl1} and \eqref{eq:usimpl2} we find 
$$\nabla\cdot u = \left(\tilde{\nabla}\cdot u\right)\bigg{|}_{\psi=1}-\frac{(u\cdot\nabla )K}{K}+\frac{\psi-1}{N}\left[\frac{\tilde{\nabla}^2 K}{K}-2~(u\cdot\nabla)K+K~(u\cdot K\cdot u)+u^D\bar{R}_{DE} u^E\right] + {\cal O}\left(\frac{1}{D}\right)$$
 \section{The divergence of the vector constraint equation at 1st order}
 The membrane  equation at 1st order is given in equation \eqref{eq:vecconst1}.  For convenience we are quoting the equation here again. 
 \begin{equation}\label{eq:constraintrep}
\begin{split}
&P^A_B\left[ \frac{\tilde\nabla^2 u_A}{K} -\frac{\nabla_A{K}}{K}+  u_C {K}^C_A - (u\cdot\nabla)u_A \right] = {\cal O}\left(\frac{1}{D}\right)
\end{split}
\end{equation}
 We could compute the divergence of each of the term separately.

\begin{equation}
\text{Divergence} \equiv \underbrace{\nabla^B\left(P^A_B \frac{\tilde\nabla^2 u_A}{K}\right)}_{Term-1} -\underbrace{\nabla^B\left(P^A_B \frac{\nabla_A{K}}{K}\right)}_{Term-2}+ \underbrace{\nabla^B\left(P^A_B~u_C {K}^C_A\right)}_{Term-3} - \underbrace{\nabla^B\left(P^A_B~(u\cdot\nabla)u_A\right)}_{Term-4}
\end{equation}

\begin{equation}\label{eq:proces1f}
\begin{split}
\text{Term-1}&\equiv\nabla^B\left(P^A_B \frac{\tilde\nabla^2 u_A}{K}\right)\\
&=- n^A[{\nabla}^2 u_A-K(n\cdot\nabla)u_A]+\frac{1}{K}P^A_B \nabla^B\left[\nabla^2 u_A-K(n\cdot\nabla)u_A\right]\\
&=(u\cdot\nabla)K+\frac{1}{K}P^{AB}\left[-\bar{R}_{BD}(\nabla^D u_A)+\bar{R}_{BEAD}(\nabla^E u^D)+\nabla^E(\bar{R}_{BEAD}~u^D)+\nabla^2(\nabla_B u_A)\right]\\
&~~~~~~~~~~~~~~~~~~~~~~~~-P^{AB}\nabla_B[(n\cdot\nabla)u_A]\\
&=(u\cdot\nabla)K+\frac{1}{K}\nabla^2(\nabla\cdot u)-P^{AB}\nabla_B[(n\cdot\nabla)u_A]\\
&=(u\cdot\nabla)K
\end{split}
\end{equation}
In the last line we have used \eqref{eq:usimpl1} for the expression of $(\nabla\cdot u)$

\begin{equation}\label{eq:proces2f}
\begin{split}
\text{Term-2}&\equiv \nabla^B\left(P^A_B \frac{\nabla_A{K}}{K}\right)\\
&=\frac{\nabla^2 {K}}{{K}}-(n\cdot\nabla){K}\\
&=\frac{\tilde{\nabla}^2{K}}{{K}}
\end{split}
\end{equation}

\begin{equation}\label{eq:proces3f}
\begin{split}
\text{Term-3}&\equiv \nabla^B\left(P^A_B~u_C {K}^C_A\right)\\
&=P^A_B ~u^C\nabla^B(\Pi^D_A\nabla_D n_C)\\
&=-Ku^C(n\cdot\nabla)n_C+u^C\nabla^2 n_C\\
&=(u\cdot\nabla)K
\end{split}
\end{equation} 

\begin{equation}\label{eq:proces4f}
\begin{split}
\text{Term-4}&\equiv \nabla^B\left(P^A_B~(u\cdot\nabla)u_A\right)\\
&=-K~n^A(u\cdot\nabla)u_A+P^{AB}(\nabla_B u^E)(\nabla_E u_A)+P^{AB}u^E(\nabla_B \nabla_E u_A)\\
&=K~(u\cdot K\cdot u)+P^{AB} u^E\bar{R}_{BEAD}u^D\\
&=K~(u\cdot K\cdot u)+u^E\bar{R}_{ED}u^D
\end{split}
\end{equation}


Adding equations \eqref{eq:proces1f}, \eqref{eq:proces2f}, \eqref{eq:proces3f} and \eqref{eq:proces4f} we get the divergence of the vector constraint equation as
\begin{equation}\label{vecdiver}
\text{Divergence}= -u^C \bar R_{DC} u^D - \frac{\tilde\nabla^2K }{K} + 2 (u\cdot \nabla) K - K(u^A K_{AB} u^B)\approx 0
\end{equation}
But this is still not in the form of equation \eqref{eq:vecdiv} and we need to do few more manipulations. Note that because $\bar R_{DC}$ (Ricci tensor evaluated on the background) is proportional to the background metric $\bar g_{DC}$ and both $u$ and $n$ are normalized time-like and space-like vectors respectively,

\begin{equation}\label{urunrn}
u^C \bar R_{DC} ~u^D = -n^C \bar R_{DC} ~n^D
\end{equation}\\
Now we could use the following identity
\begin{equation}\label{eq:impi}
\begin{split}
(n\cdot\nabla K) &= -n^A \bar R_{AD} ~n^D + \frac{\tilde\nabla^2 K}{K} -K_{AB} K^{AB}\\
\text{Proof}:&\\
(n\cdot \nabla) {\cal K}&=(n^A \nabla_A)(\nabla_B n^B)\\
&=n^A\big[\nabla_A,\nabla_B\big]n^B+n^A \nabla_B(\nabla_A n^B)\\
&=-n^A \bar{R}_{ABD}^{~~~~~B}~ n^D+\nabla_B\big[(n\cdot\nabla)n^B\big]-K_{AB} K^{AB}\\
&=-n^A \bar{R}_{AD}n^D+\nabla_B\left[\frac{\Pi^{BA}\nabla_A{K}}{{K}}\right]-K_{AB} K^{AB}\\ 
&=-n^A \bar R_{AD} ~n^D + \frac{\tilde\nabla^2 K}{K} -K_{AB} K^{AB}\\
\end{split}
\end{equation}
After substituting equation \eqref{eq:impi} in the expression of the divergence we get the following equation
\begin{equation}
(n\cdot\nabla)K+K_{AB} K^{AB}-2~(u\cdot\nabla)K+K~(u\cdot K\cdot u)={\cal{O}}(1)
\end{equation}

\section{Identities}\label{app:Identity}
In this appendix we shall prove some identities that we have used for our computation.
\subsection{Proof of \eqref{eq:leadeq2} from \eqref{eq:SubsidiaryPsi}}
\begin{equation}\label{I0}
\begin{split}
\nabla^2(\psi^{-D})&=0\\
\Rightarrow \frac{DN}{\psi}- K &=\frac{(n\cdot\nabla)N}{N}-\frac{N}{\psi}\\
&=\frac{(n\cdot\nabla)(\psi~K)}{\psi~K}-\frac{N}{\psi} + {\cal O}\left(\frac{1}{D}\right)\\
&= \frac{(n\cdot\nabla)K}{ K}+ {\cal O}\left(\frac{1}{D}\right)\\
\end{split}
\end{equation}
\subsection{Proof of equations \eqref{eq:ccii}}
We have used the following identity for derivation of \eqref{eq:ccii}
\begin{equation}\label{I1}
\begin{split}
u^A\nabla^2n_A &=u^A\nabla_C\left[n^C(n\cdot\nabla)n_A + K^C_A\right]\\
&=K\left[u^A(n\cdot\nabla)n_A \right]+u^A\nabla_C K^C_A+{\cal O}(1)\\
&=K\left[u^A(n\cdot\nabla)n_A \right]+u^A\nabla_C K^C_A+{\cal O}(1)\\
&=K\left[u^A(n\cdot\nabla)n_A \right]+(u\cdot\nabla)K+{\cal O}(1)\\
&= 2(u\cdot\nabla)K+{\cal O}(1)
\end{split}
\end{equation}

In deriving equation \eqref{I1} we have used the following identity
\begin{equation}\label{I3}
(n\cdot\nabla)n_A=\Pi^C_A\left[\frac{\nabla_C K}{K}\right]+{\cal{O}}\left(\frac{1}{D}\right)
\end{equation}

Proof of \eqref{I3}
\begin{equation}
\begin{split}
\nabla_A N^2 &= \nabla_A [(\nabla_B\psi) (\nabla^B\psi)]\\
\Rightarrow2 N \nabla_A N &= 2(\nabla^B\psi )(\nabla_A \nabla_B \psi)\\
\Rightarrow~ N \nabla_A N &= (\nabla^B\psi )(\nabla_B \nabla_A \psi)\\
\Rightarrow~ N \nabla_A N &= Nn^B\nabla_B (Nn_A)\\
\Rightarrow~~~~  \nabla_A N &= (n\cdot\nabla)(Nn_A)\\
\Rightarrow(n\cdot\nabla)n_A &= \Pi^C_A\left(\frac{\nabla_C N}{N}\right)\\
&=\Pi^C_A\left[\frac{\nabla_C K}{K}\right]+{\cal{O}}\left(\frac{1}{D}\right)
\end{split}
\end{equation}
\subsection{Proof of \eqref{eq:I4}}

\begin{equation}
P^C_B\nabla^2 O_C=P^C_B(\nabla^2 n_C-\nabla^2 u_C)
\end{equation}

\begin{equation}\label{del2n}
\begin{split}
P^C_B \nabla^2 n_C&=P^C_B~ \nabla^D\nabla_D n_C\\
&=P^C_B~ \nabla^D\nabla_D\left( \frac{\nabla_C\psi}{N}\right)\\
&=P^C_B~ \nabla^D\left(\frac{\nabla_D\nabla_C\psi}{N}-\frac{1}{N^2}(\nabla_D N)(\nabla_C\psi)\right)\\
&=P^C_B~\left(\frac{\nabla^D\nabla_C\nabla_D\psi}{N}-\frac{2}{N^2}(\nabla_D N)(\nabla^D\nabla_C\psi)\right)\\
&=\frac{1}{N}P^C_B\bigg([\nabla_D,\nabla_C]\nabla^D\psi+\nabla_C\nabla_D\nabla^D\psi\bigg)+{\cal{O}}(1)\\
&=\frac{1}{N}P^C_B\bigg(\bar{R}_{DCE}^{~~~~~~D}~\nabla^E\psi+\nabla_C\nabla_D(N n^D)\bigg)+{\cal{O}}(1)\\
&=\frac{1}{N}P^C_B\nabla_C\bigg(n_D \nabla^D N + N \nabla^D n_D\bigg) +{\cal{O}}(1)\\
&=P^C_B \left(\frac{N\nabla_C \nabla^D n_D}{N}+\frac{(\nabla_C N)(\nabla^D n_D)}{N}\right) +{\cal{O}}(1)\\
&= P^C_B\left(\nabla_C K+\frac{\nabla_C K}{K} K \right) +{\cal{O}}(1)\\
&= 2P^C_B \nabla_C K+{\cal{O}}(1)
\end{split}
\end{equation}

\begin{equation}\label{del2u}
\begin{split}
P^C_B(\nabla^2u_C)&=P^C_B~\nabla^D\nabla_D u_C\\
&=P^C_B\nabla^D\bigg(\Pi^E_C~\nabla_Du_E+n_C n^E\nabla_D u_E\bigg)\\
&=P^C_B\nabla^D\bigg(\Pi^E_C~\nabla_Du_E\bigg)+{\cal{O}}(1)\\
&=P^C_B\nabla^D\bigg(\Pi^E_C~\Pi^F_D~\nabla_F u_E+\Pi^E_C~n_D(n\cdot\nabla)u_E\bigg)+{\cal{O}}(1)\\
&=P^C_B\nabla^D\bigg(\Pi^E_C~\Pi^F_D~\nabla_F u_E\bigg)+P^C_B K(n\cdot\nabla)u_C+{\cal{O}}(1)\\
&=P^C_B\Pi^D_N\nabla^N\bigg(\Pi^E_C~\Pi^F_D~\nabla_F u_E\bigg)+P^C_B n^D(n\cdot\nabla)\bigg(\Pi^E_C~\Pi^F_D~\nabla_F u_E\bigg)\\
&~~~~~~~~~~~~~~~~~~~~~~~~~~~~~~~~+P^C_B K(n\cdot\nabla)u_C+{\cal{O}}(1)\\
&=P^C_B\Pi^D_N\nabla^N\bigg(\Pi^E_C~\Pi^F_D~\nabla_F u_E\bigg)+P^C_B K(n\cdot\nabla)u_C+{\cal{O}}(1)\\
\end{split}
\end{equation}

Adding\eqref{del2n} and \eqref{del2u} we get the expression for $P^C_B\nabla^2 O_C$

\begin{equation}\label{del2O}
\begin{split}
P^C_B\nabla^2 O_C&=P^C_B~[2\nabla_C K-\Pi^D_N\nabla^N\bigg(\Pi^E_C~\Pi^F_D~\nabla_F u_E\bigg)-K(n\cdot\nabla)u_C]+{\cal{O}}(1)
\end{split}
\end{equation}

Now from our subsidiary condition,
\begin{equation}\label{subsi}
\begin{split}
&~~~~~~~~~P^C_B(O\cdot\nabla)O_C=0\\
\Rightarrow &P^C_B(n\cdot\nabla)u_C=P^C_B~[(n\cdot\nabla)n_C-(u\cdot\nabla)n_C+(u\cdot\nabla)u_C]
\end{split}
\end{equation}

Substituting \eqref{subsi} in \eqref{del2O} we get,
\begin{equation}
\begin{split}
P^C_B\nabla^2 O_C&=P^C_B~\bigg\{2\nabla_C K-\Pi^D_N\nabla^N\left(\Pi^E_C~\Pi^F_D~\nabla_F u_E\right)\\
&~~~~~~~~~~~~~~~- K\left[(n\cdot\nabla)n_C-(u\cdot\nabla)n_C+(u\cdot\nabla)u_C\right]\bigg\}+{\cal{O}}(1)\\
&=P^C_B~\bigg\{\nabla_C K-\Pi^D_N\nabla^N\bigg(\Pi^E_C~\Pi^F_D~\nabla_F u_E\bigg)+K\left[u^D K_{DC}-(u\cdot\nabla)u_C\right]\bigg\}+{\cal{O}}(1)\\
&=P^C_B~\bigg\{\nabla_C K-\tilde{\nabla}^2 u_C+K\left[u^D K_{DC}-(u\cdot\nabla)u_C\right]\bigg\}+{\cal{O}}(1)
\end{split}
\end{equation}
Where $\tilde{\nabla}$ is defined in eq \eqref{tildedef}.
\bibliography{larged}
\bibliographystyle{JHEP}

\end{document}